\documentclass[12pt]{article}

\usepackage[T1]{fontenc}
\usepackage[utf8]{inputenc}
\usepackage{lmodern}

\usepackage{tikz}
\usetikzlibrary{shapes.geometric}
\usepackage{xcolor}
\usepackage{booktabs}
\usepackage{array}
\usepackage{subcaption}

\definecolor{lightblue}{RGB}{207,226,243}

\usepackage[hyphens]{url}
\usepackage{algorithm}
\usepackage{algpseudocode}
\usepackage[most]{tcolorbox}
\usepackage{multirow}
\usepackage{threeparttable}
\usepackage{amsmath}
\usepackage{amssymb} 
\usepackage{mathtools}

\usepackage{amsthm}
\usepackage{thmtools}
\usepackage{thm-restate}

\usepackage{booktabs}
\usepackage{multirow}
\usepackage{threeparttable}
\usepackage{tabularx}

\usepackage[margin=1in]{geometry}
\usepackage{setspace}
\usepackage{booktabs}
\usepackage[labelsep=period,font=small,labelfont=bf,margin=5ex]{caption}
\usepackage{adjustbox}
\usepackage{float}
\usepackage{subcaption} 
\usepackage{graphicx}
\usepackage{enumitem}
\usepackage{tabularray}
\usepackage{multibib}
\usepackage[sort]{natbib}
\usepackage[bookmarks=false]{hyperref}

\usepackage[utf8]{inputenc}
\usepackage{booktabs}
\usepackage{array}
\usepackage{tabularx}

\usepackage{quiver} 

\usepackage{algorithm}
\usepackage{algpseudocode}

\usepackage{ragged2e}
\usepackage{amssymb}
\usepackage{pifont}  

\newcolumntype{P}[1]{>{\RaggedRight\arraybackslash}p{#1}} 

\usepackage{makecell}

\newcolumntype{Y}{>{\RaggedRight\arraybackslash}X} 

\newcolumntype{L}[1]{>{\RaggedRight\arraybackslash}p{#1}}
\newcolumntype{C}[1]{>{\centering\arraybackslash}p{#1}}

\usepackage{threeparttable}

\usepackage{subcaption}

\usepackage{xr} 

\hypersetup{
  colorlinks=true,
  citecolor=black,
  linkcolor=black,
  urlcolor=black
}

\bibpunct{(}{)}{;}{a}{}{,}  
\renewcommand{\harvardurl}[1]{\textbf{URL:} \url{#1}}  
\newcites{app}{Additional References}

\usepackage{multibib}
\usepackage[sort]{natbib}
\usepackage[bookmarks=false]{hyperref}

\title{Modeling Bipartite Dynamic Networks: An Additive and Multiplicative Effects Model}

\author{Jing Luo\thanks{Ph.D. Candidate in the Department of Political Science at The Ohio State University.  Email: \texttt{luo.1405@osu.edu}.}}

\begin{document}


\maketitle

\frenchspacing
\begin{abstract}

\noindent 
Researchers frequently study interactions between two distinct types of actors, represented as bipartite networks. These networks exhibit dependence patterns that differ from those in one-mode networks and therefore require models tailored to their structure. This paper develops an additive and multiplicative effects (AME) framework for longitudinal bipartite data. First, I distinguish the dependence structure and specify the corresponding modeling assumptions. Second, I introduce the bipartite dynamic AME model and develop an estimation procedure based on block coordinate descent. Third, I incorporate a squared iterative method to improve computational efficiency. Using simulations and an application to global production networks, I show that the model improves coefficient estimation, more accurately recovers the data-generating process, and better captures the multiplicative latent structure. The model reveals evolving patterns in countries’ global production engagement that are not captured by observed covariates or static specifications. I provide an R package, \href{https://github.com/nanajing7/RAMEN}{RAMEN}, to facilitate implementation.
\end{abstract}

\begin{flushright}
Abstract word count:  149\\ 
Total word count: 5,869
\end{flushright}
 
\doublespacing
\allowdisplaybreaks


\newpage
\section{Introduction}

The additive and multiplicative effects (AME) model is widely used in political science research to analyze relational data, including interstate conflict \citep{minhas2024decomposing}, UN voting \citep{kim2023dynamic}, armed group interactions during civil war \citep{dorff2020networks}, and defense cooperation \citep{kinne2024network}. This model is well suited to these settings because relations are rarely independent: ties involving the same actors may be correlated, and relations may exhibit triadic and other higher-order dependence. Ignoring such dependence may lead to biased parameter estimates and misleading inference about quantities of interest \citep{minhas2019inferential, cranmer2016critique, hoff2021additive}. The AME model accounts for these dependence using additive random effects and multiplicative latent factors.

Many political relationships, however, involve interactions between two distinct sets of actors. Such bipartite relational data are common, including in studies of government–rebel group interactions \citep{wang2022resilience}, politicians–political claims \citep{leifeld2017discourse}, legislator–bill sponsorship \citep{cranmer2011inferential}, and politicians–special interest groups \citep{desmarais2015fates}. These relationships often evolve over time, forming longitudinal bipartite relational data.

\begin{figure}[!htbp]
\centering
\caption{Dependence Structures in One-Mode and Bipartite Networks}
\label{fig:bipartite_dependence}
\begin{minipage}[t]{0.48\textwidth}
\vspace{0pt}
\centering
\begin{tikzpicture}[x=1cm,y=1cm,every node/.style={font=\small},
    actor/.style={circle,draw=black,line width=0.5pt,minimum size=6mm,inner sep=0pt},
    normaledge/.style={-{Latex[length=1.8mm,width=1.2mm]},line width=0.55pt,draw=black!60},
    highlight/.style={line width=1.6pt,draw=black!80},
    note/.style={font=\footnotesize,align=center}]
\node[font=\bfseries] at (2.8,5.2) {(a) One-Mode Network};
\node[actor,fill=red!80] (a1) at (0.9,3.45) {};
\node[actor,fill=red!80] (a2) at (2.1,4.40) {};
\node[actor,fill=red!80] (a3) at (3.7,4.00) {};
\node[actor,fill=red!80] (a4) at (1.7,2.10) {};
\node[actor,fill=red!80,font=\small\bfseries,text=white] (a5) at (3.2,1.85) {$a$};
\node[actor,fill=red!80] (a6) at (4.55,2.80) {};
\draw[normaledge] (a1) -- (a2);
\draw[normaledge] (a1) -- (a3);
\draw[normaledge] (a1) -- (a4);
\draw[normaledge] (a1) -- (a5);
\draw[normaledge] (a2) -- (a3);
\draw[normaledge] (a2) -- (a4);
\draw[normaledge] (a2) -- (a5);
\draw[normaledge] (a3) -- (a4);
\draw[normaledge] (a3) -- (a5);
\draw[normaledge] (a4) -- (a5);
\draw[normaledge] (a5) -- (a6);
\draw[-{Latex[length=2mm,width=1.3mm]},line width=1.5pt,bend left=17] (a3) to (a6);
\draw[-{Latex[length=2mm,width=1.3mm]},line width=1.5pt,bend left=17] (a6) to (a3);
\node[note,anchor=west] at (4.85,3.55) {Dyadic\\reciprocity};
\draw[highlight] (a1) -- (a2);
\draw[highlight] (a2) -- (a4);
\draw[highlight] (a4) -- (a1);
\node[note,anchor=east] at (0.60,2.35) {Triadic\\closure};
\node[note,anchor=north] at (5.7,2.0) {Sender--receiver covariance};
\end{tikzpicture}
\end{minipage}
\hfill
\begin{minipage}[t]{0.48\textwidth}
\vspace{0pt}
\centering
\begin{tikzpicture}[x=1cm,y=1cm,every node/.style={font=\small},
    rownode/.style={circle,draw=black,line width=0.5pt,minimum size=6mm,inner sep=0pt},
    colnode/.style={regular polygon,regular polygon sides=3,draw=black,line width=0.5pt,minimum size=7mm,inner sep=0pt},
    bipedge/.style={line width=0.55pt,draw=black!60},
    highlight/.style={line width=1.6pt,draw=black!80},
    note/.style={font=\footnotesize,align=center}]
\node[font=\bfseries] at (2.8,5.2) {(b) Bipartite Network};
\node[font=\small] at (1.2,4.45) {Row nodes};
\node[font=\small] at (4.1,4.45) {Column nodes};
\node[rownode,fill=red!80] (r1) at (1.2,3.6) {};
\node[rownode,fill=red!80] (r2) at (1.2,2.5) {};
\node[rownode,fill=red!80] (r3) at (1.2,1.4) {};
\node[colnode,fill=blue!65] (c1) at (4.1,3.7) {};
\node[colnode,fill=blue!65] (c2) at (4.1,2.8) {};
\node[colnode,fill=blue!65] (c3) at (4.1,1.9) {};
\node[colnode,fill=blue!65] (c4) at (4.1,1.0) {};
\draw[bipedge] (r1) -- (c1);
\draw[bipedge] (r1) -- (c2);
\draw[bipedge] (r1) -- (c3);
\draw[bipedge] (r2) -- (c1);
\draw[bipedge] (r2) -- (c2);
\draw[bipedge] (r2) -- (c4);
\draw[bipedge] (r3) -- (c2);
\draw[bipedge] (r3) -- (c3);
\draw[bipedge] (r3) -- (c4);
\draw[highlight] (r1) -- (c1);
\draw[highlight] (r1) -- (c2);
\draw[highlight] (r2) -- (c1);
\draw[highlight] (r2) -- (c2);
\node[note,anchor=west] at (4.55,3.2) {Four-cycle};
\end{tikzpicture}
\end{minipage}
\begin{minipage}{0.94\textwidth}
\footnotesize
\vspace*{0.4em}
\textit{Note:} In one-mode networks, node $a$ can act as both a sender and a receiver, allowing sender--receiver covariance. The same structure also permits dyadic reciprocity and triadic closure. In bipartite networks, sender and receiver roles are assigned to distinct node sets, so sender--receiver covariance, reciprocity and triangle-based closure are absent. Higher-order dependence instead arises through structures such as four-cycles.
\end{minipage}
\end{figure}
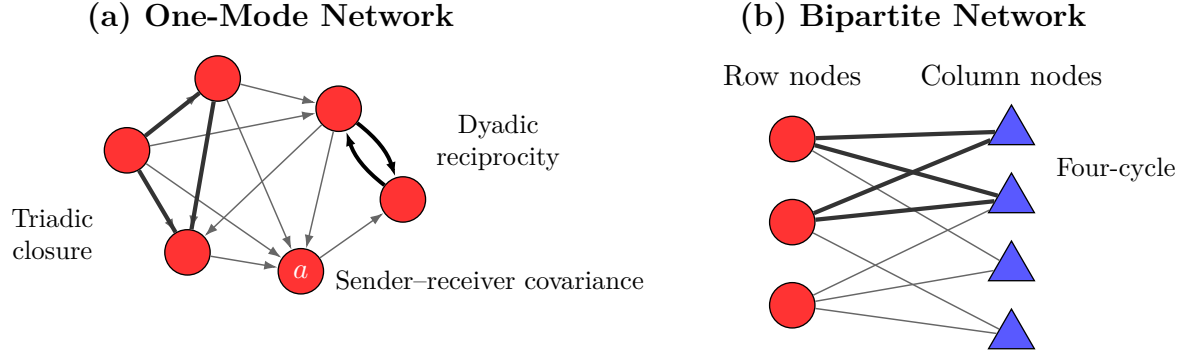

Longitudinal bipartite relational data exhibit a different configuration of dependence from one-mode networks and therefore require an AME framework tailored to this structure. Figure~\ref{fig:bipartite_dependence} illustrates some examples within a single network realization. In the one-mode network, actors can occupy both sender and receiver roles, permitting sender--receiver covariance, dyadic reciprocity, and triadic closure. In the bipartite network, row and column nodes belong to distinct sets and ties occur only across them. Reciprocity and triangle-based closure are therefore absent. Higher-order dependence instead arises through structures such as four-cycles, in which two row nodes are connected to the same two column nodes. A bipartite AME framework must accordingly account for the absence of these one-mode structures while capturing higher-order dependence specific to bipartite networks.

Moving to the longitudinal bipartite setting, the model must account for dependence over time. Ignoring this temporal dependence may cause persistence from prior periods to be attributed to observed covariates or to other model components, leading the model to misrepresent the underlying data-generating process (DGP). Temporal dependence also complicates estimation because additive effects and latent factors may evolve differently over time, increasing the dimensionality of the parameter space. These challenges are especially pronounced in large longitudinal bipartite networks, such as legislator–bill relationships, country-UN resolution interactions, and country–product trade relationships observed across multiple periods.

This article develops a formal framework for the bipartite dynamic AME model. First, I specify the dependence structure of bipartite relational data and the corresponding modeling assumptions. Second, I introduce the model and estimate it using a nested empirical Bayes maximum a posteriori (MAP) procedure based on block coordinate descent (BCD). I quantify estimation uncertainty using a conditional parametric bootstrap. Third, I incorporate SQUAREM fixed-point acceleration into the proposed estimation procedure to improve computational efficiency for high-dimensional longitudinal networks \citep{du2020squarem}. To facilitate implementation, I provide an accompanying R package, \texttt{RAMEN}.\footnote{Available at \url{https://github.com/nanajing7/RAMEN}.}

I apply the proposed approach to both simulated data and international trade data. In simulations, I show that the model (1) more accurately estimates covariate coefficients and achieves well-calibrated coverage than commonly used alternatives, (2) better captures the underlying DGP, yielding improved out-of-sample prediction, and (3) more accurately recovers the multiplicative latent structure. In the empirical application, I construct a bipartite global production network linking countries and industries, and examine both exogenous covariates and endogenous network factors that shape countries' engagement in global shipbuilding production. The model (1) provides interpretable estimates of covariate effects, (2) uncovers substantively meaningful patterns of cross-country heterogeneity, and (3) reveals evolving engagement patterns that are not captured by conventional regression approaches or static AME specifications. Together, these findings advance methodological tools for network analysis and deepen our understanding of countries’ engagement in global production.

This article makes three contributions. First, it extends the AME framework to longitudinal bipartite relational data by specifying the dependence structure of bipartite networks and allowing both additive and multiplicative components to evolve over time. Second, it develops a scalable estimation and inference framework that incorporates MAP estimation, uncertainty quantification, goodness-of-fit assessment, and computational acceleration. Third, it provides a general methodological framework for studying longitudinal bipartite networks across a wide range of political settings. Such data arise in legislative politics \citep{zelizer2019position, battaglini2020effectiveness, kim2021mapping, lo2026statistical}, conflict \citep{kinne2020guns, wang2022resilience}, political representation \citep{aaroe2020cognitive, boucher2019tariff}, and international political economy \citep{schoeneman2024bipartite}. Yet, they are often analyzed with methods designed for independent dyads or one-mode networks. This article clarifies how the dependence structure and modeling assumptions of bipartite networks differ from those of one-mode networks and thereby demonstrates the need for methods tailored to this class of relational data. In doing so, it expands the methodological toolkit for analyzing complex interdependent data structures in political science.


The rest of the paper proceeds as follows. Section 2 specifies the dependence structure of bipartite relational data and introduces the bipartite dynamic AME model. Section 3 presents the simulations and empirical application. Section 4 concludes with broader implications and directions for future research.

\section{Methods}

This section first discusses the dependence structure of bipartite relational data and the corresponding modeling assumptions, which motivate a bipartite extension of the AME framework. It then introduces the dynamic bipartite AME model and describes its specification and estimation.

\subsection{Dependence in Bipartite Networks}

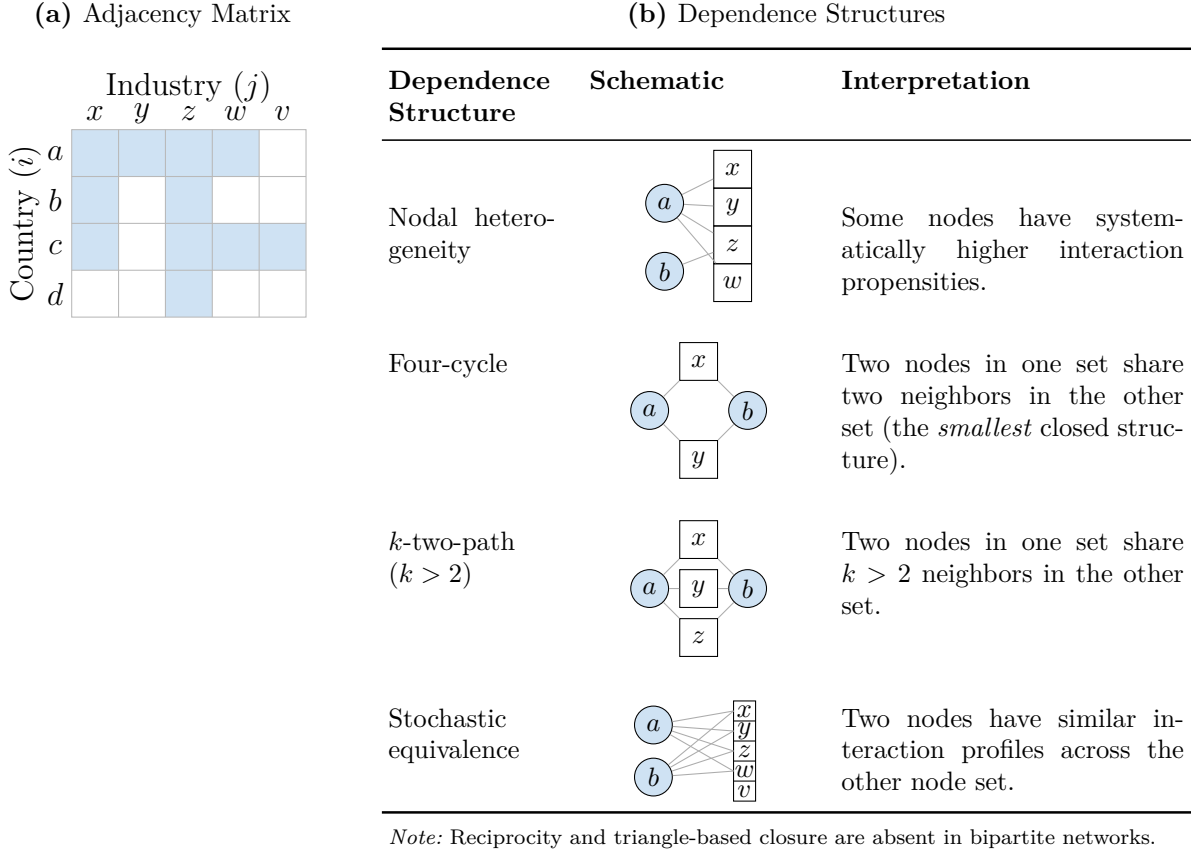
\begin{figure}[h]
\centering
\caption{Bipartite Relational Data and Dependence Structures}
\label{fig:bipartite_setting}
\begin{subfigure}[t]{0.29\textwidth}
\centering
\caption{Adjacency Matrix}
\label{fig:bipartite_matrix}
\vspace{0.18cm}
\hspace*{-0.40cm}
\begin{tikzpicture}[x=0.62cm,y=0.62cm]
\fill[lightblue] (0,3) rectangle (1,4);
\fill[lightblue] (1,3) rectangle (2,4);
\fill[lightblue] (2,3) rectangle (3,4);
\fill[lightblue] (3,3) rectangle (4,4);
\fill[lightblue] (0,2) rectangle (1,3);
\fill[lightblue] (2,2) rectangle (3,3);
\fill[lightblue] (0,1) rectangle (1,2);
\fill[lightblue] (2,1) rectangle (3,2);
\fill[lightblue] (3,1) rectangle (4,2);
\fill[lightblue] (4,1) rectangle (5,2);
\fill[lightblue] (2,0) rectangle (3,1);
\draw[gray!55] (0,0) rectangle (5,4);
\draw[gray!55] (1,0) -- (1,4);
\draw[gray!55] (2,0) -- (2,4);
\draw[gray!55] (3,0) -- (3,4);
\draw[gray!55] (4,0) -- (4,4);
\draw[gray!55] (0,1) -- (5,1);
\draw[gray!55] (0,2) -- (5,2);
\draw[gray!55] (0,3) -- (5,3);
\node at (0.5,4.35) {$x$};
\node at (1.5,4.35) {$y$};
\node at (2.5,4.35) {$z$};
\node at (3.5,4.35) {$w$};
\node at (4.5,4.35) {$v$};
\node at (-0.35,3.5) {$a$};
\node at (-0.35,2.5) {$b$};
\node at (-0.35,1.5) {$c$};
\node at (-0.35,0.5) {$d$};
\node at (2.5,4.9) {Industry ($j$)};
\node[rotate=90] at (-1.0,2) {Country ($i$)};
\end{tikzpicture}
\end{subfigure}%
\hspace{0.02\textwidth}%
\begin{subfigure}[t]{0.67\textwidth}
\centering
\caption{Dependence Structures}
\vspace{0pt}
\footnotesize
\renewcommand{\arraystretch}{1.35}
\setlength{\tabcolsep}{2.5pt}
\begin{tabular}{p{0.20\linewidth}p{0.31\linewidth}p{0.41\linewidth}}
\toprule
\textbf{Dependence Structure} & \textbf{\ \ Schematic} & \textbf{Interpretation} \\
\midrule
Nodal heterogeneity
&
\centering
\begin{tikzpicture}[scale=0.50,baseline=(current bounding box.center)]
\node[circle,draw,fill=lightblue,minimum size=5mm,inner sep=0pt] (a) at (0,0.3) {$a$};
\node[circle,draw,fill=lightblue,minimum size=5mm,inner sep=0pt] (b) at (0,-1.5) {$b$};
\node[rectangle,draw,minimum size=5mm,inner sep=1pt] (x) at (1.8,1.2) {$x$};
\node[rectangle,draw,minimum size=5mm,inner sep=1pt] (y) at (1.8,0.2) {$y$};
\node[rectangle,draw,minimum size=5mm,inner sep=1pt] (z) at (1.8,-0.8) {$z$};
\node[rectangle,draw,minimum size=5mm,inner sep=1pt] (w) at (1.8,-1.8) {$w$};
\draw[gray!65] (a) -- (x);
\draw[gray!65] (a) -- (y);
\draw[gray!65] (a) -- (z);
\draw[gray!65] (a) -- (w);
\draw[gray!65] (b) -- (z);
\end{tikzpicture}
&
Some nodes have systematically higher interaction propensities.
\\
\vspace{0.35em} Four-cycle
&
\centering
\vspace{0.35em}
\begin{tikzpicture}[scale=0.54,baseline=(current bounding box.center)]
\node[circle,draw,fill=lightblue,minimum size=5mm,inner sep=0pt] (a) at (0,0) {$a$};
\node[circle,draw,fill=lightblue,minimum size=5mm,inner sep=0pt] (b) at (2.4,0) {$b$};
\node[rectangle,draw,minimum size=5mm,inner sep=1pt] (x) at (1.2,1.2) {$x$};
\node[rectangle,draw,minimum size=5mm,inner sep=1pt] (y) at (1.2,-1.2) {$y$};
\draw[gray!65] (a) -- (x);
\draw[gray!65] (x) -- (b);
\draw[gray!65] (a) -- (y);
\draw[gray!65] (y) -- (b);
\end{tikzpicture}
&
\vspace{0.35em}
Two nodes in one set share two neighbors in the other set (the \textit{smallest} closed structure).
\\
\vspace{0.35em} $k$-two-path ($k>2$)
&
\centering
\vspace{0.35em}
\begin{tikzpicture}[scale=0.54,baseline=(current bounding box.center)]
\node[circle,draw,fill=lightblue,minimum size=5mm,inner sep=0pt] (a) at (0,0) {$a$};
\node[circle,draw,fill=lightblue,minimum size=5mm,inner sep=0pt] (b) at (2.4,0) {$b$};
\node[rectangle,draw,minimum size=5mm,inner sep=1pt] (x) at (1.2,1.2) {$x$};
\node[rectangle,draw,minimum size=5mm,inner sep=1pt] (y) at (1.2,0) {$y$};
\node[rectangle,draw,minimum size=5mm,inner sep=1pt] (z) at (1.2,-1.2) {$z$};
\draw[gray!65] (a) -- (x);
\draw[gray!65] (x) -- (b);
\draw[gray!65] (a) -- (y);
\draw[gray!65] (y) -- (b);
\draw[gray!65] (a) -- (z);
\draw[gray!65] (z) -- (b);
\end{tikzpicture}
&
\vspace{0.35em}
Two nodes in one set share $k>2$ neighbors in the other set.
\\
\vspace{0.35em} Stochastic equivalence
&
\centering
\vspace{0.35em}
\begin{tikzpicture}[scale=0.53,baseline=(current bounding box.center)]
\node[circle,draw,fill=lightblue,minimum size=5mm,inner sep=0pt] (a) at (0,0.65) {$a$};
\node[circle,draw,fill=lightblue,minimum size=5mm,inner sep=0pt] (b) at (0,-0.65) {$b$};
\draw (2.0,-1.25) rectangle (2.55,1.25);
\draw (2.0, 0.75) -- (2.55, 0.75);
\draw (2.0, 0.25) -- (2.55, 0.25);
\draw (2.0,-0.25) -- (2.55,-0.25);
\draw (2.0,-0.75) -- (2.55,-0.75);
\node at (2.275, 1.00) {$x$};
\node at (2.275, 0.50) {$y$};
\node at (2.275, 0.00) {$z$};
\node at (2.275,-0.50) {$w$};
\node at (2.275,-1.00) {$v$};
\draw[gray!65] (a) -- (2.0, 1.00);
\draw[gray!65] (a) -- (2.0, 0.50);
\draw[gray!65] (a) -- (2.0, 0.00);
\draw[gray!65] (a) -- (2.0,-0.50);
\draw[gray!65] (b) -- (2.0, 0.50);
\draw[gray!65] (b) -- (2.0, 0.00);
\draw[gray!65] (b) -- (2.0,-0.50);
\draw[gray!65] (b) -- (2.0, 1.00);
\end{tikzpicture}
&
\vspace{0.35em}
Two nodes have similar interaction profiles across the other node set.
\\
\bottomrule
\multicolumn{3}{l}{\scriptsize \textit{Note:} Reciprocity and triangle-based closure are absent in bipartite networks.}
\\
\end{tabular}
\end{subfigure}
\end{figure}

Figure~\ref{fig:bipartite_setting} (a) illustrates a bipartite relational structure using global production networks as an example. Countries form a row-node set, indexed by $i \in \{a,b, c, d \}$, and industries form a column-node set, indexed by $j \in \{x,y,z,w,v\}$. Each matrix entry, $Y_{ij}$, represents a country--industry dyad, with shaded cells indicating observed ties. Ties are defined only between the two node sets.

Figure~\ref{fig:bipartite_setting} (b) presents forms of dependence that can arise in bipartite relational data. Like one-mode networks, bipartite relational data exhibit nodal heterogeneity. Some countries are systematically more engaged across industries, and some industries attract systematically greater engagement across countries. Dyads sharing a country or an industry may therefore remain correlated after conditioning on observed covariates.

Not all forms of dependence found in one-mode networks carry over to bipartite data. Because ties are defined only across the two node sets, a country--industry tie $Y_{ij}$ has no reverse counterpart $Y_{ji}$ in the same relational matrix. Reciprocity is therefore not defined. Triangle-based transitivity and closure are likewise structurally impossible, since countries do not tie directly to other countries, nor do industries tie directly to other industries. The absence of these structures makes a direct application of the one-mode AME model inappropriate because its dependence assumptions do not match the bipartite data structure.


In bipartite networks, higher-order dependence can instead arise through configurations involving actors from both node sets. As shown in Figure~\ref{fig:bipartite_setting} (b), suppose that countries $a$ and $b$ both have interactions with industries $x$ and $y$. The four dyads $\{Y_{ax},Y_{ay},Y_{bx},Y_{by}\}$ form a four-cycle. Because bipartite networks contain no odd-length cycles, the four-cycle is the smallest closed configuration. K-two-path configurations extend the four-cycle to cases in which two nodes from one set share k>2 neighbors from the other set. For example, countries $a$ and $b$ both interact with industries $x$, $y$, and $z$. Such configurations indicate that similarity between two countries extends across a broader set of relationships. Stochastic equivalence generalizes this idea to nodes' overall interaction profiles. Two countries are stochastically equivalent when they exhibit similar patterns of interaction across the full set of industries.

In sum, bipartite relational data retain node-level heterogeneity, exclude reciprocity and triangle-based closure, and exhibit distinct forms of higher-order dependence. These structural differences have direct implications for model specification, inference, and fit. Specifically, conventional regression approaches assume independent dyads, while one-mode AME models rely on dependence assumptions that do not match the bipartite data structure. Applying either approach to bipartite data may bias coefficient estimates, miscalibrate uncertainty, and weaken predictive performance. Higher-order dependence is also important for model fit. Failing to represent higher-order dependence specific to bipartite networks can result in poor goodness of fit and convergence problems \citep{stivala2025improving, wang2009exponential}. Together, these considerations motivate a bipartite AME framework whose assumptions and model components align with the dependence structure of the data.

\subsection{Bipartite AME Model}

Let $Y_{ij}$ denote the observed interaction between a Type 1 (row) actor $i$ and a Type 2 (column) actor $j$, and let $\mathbf{x}_{ij}$ denote a P-dimensional vector of observed covariates. The bipartite AME model accounts for node-level heterogeneity and higher-order dependence:

\begin{equation*}
Y_{ij} = \mathbf{x}_{ij}^{\top}\boldsymbol{\beta} + a_i + b_j + \mathbf{u}_i^{\top}\mathbf{v}_j + \epsilon_{ij}, \qquad \epsilon_{ij} \overset{\mathrm{iid}}{\sim} \mathcal{N}(0,\sigma_{\epsilon}^{2}),
\label{eq:bipartite_ame}
\end{equation*}

\noindent where $\boldsymbol{\beta}$ is the coefficient vector associated with $\mathbf{x}_{ij}$; $a_i$ and $b_j$ are additive random effects capturing unobserved heterogeneity among Type 1 and Type 2 actors; $\mathbf{u}_i$ and $\mathbf{v}_j$ are $K$-dimensional latent vectors; and $\epsilon_{ij}$ is an idiosyncratic error term. The multiplicative term $\mathbf{u}_i^\top \mathbf{v}_j$ represents higher-order dependence, with $K$ determining the rank and therefore the complexity of the multiplicative relational structure that the model can represent. A smaller $K$ imposes a more restrictive low-rank representation, whereas a larger $K$ allows the model to represent more complex patterns of dependence.

The coefficients, additive random effects, and multiplicative latent factors are assigned zero-centered Gaussian priors:

\begin{align*}
\boldsymbol{\beta} &\sim \mathcal{N}(\mathbf{0},\sigma_{\beta}^{2}\mathbf{I}_{P}), \\
a_i &\overset{\mathrm{iid}}{\sim} \mathcal{N}(0,\sigma_{a}^{2}), \qquad b_j \overset{\mathrm{iid}}{\sim} \mathcal{N}(0,\sigma_{b}^{2}), \\
\mathbf{u}_i &\overset{\mathrm{iid}}{\sim} \mathcal{N}(\mathbf{0},\sigma_{U}^{2}\mathbf{I}_{K}), \qquad \mathbf{v}_j \overset{\mathrm{iid}}{\sim} \mathcal{N}(\mathbf{0},\sigma_{V}^{2}\mathbf{I}_{K}),
\label{eq:bipartite_ame_priors}
\end{align*}

\noindent where \(\sigma_\beta^2\), \(\sigma_a^2\), \(\sigma_b^2\), \(\sigma_U^2\), and \(\sigma_V^2\) govern the prior dispersion of the coefficient, additive-effect, and latent-factor components, respectively.

\begin{table}[!htbp]
\centering
\caption{Model Components in One-Mode and Bipartite AME Models}
\label{tab:ame_comparison}
\footnotesize
\renewcommand{\arraystretch}{1.25}
\begin{tabularx}{\textwidth}{@{}>{\raggedright\arraybackslash}p{0.21\textwidth}>{\raggedright\arraybackslash}X>{\raggedright\arraybackslash}X@{}}
\toprule
\textbf{Model component} & \textbf{One-Mode AME} & \textbf{Bipartite AME} \\
\midrule
\multicolumn{3}{@{}l}{\textit{Additive component}}\\
Nodal heterogeneity
&
\(a_i\) and \(b_i\) capture actors' baseline interaction propensities.
&
\(a_i\) and \(b_j\) capture actors' baseline interaction propensities.
\\
\addlinespace[0.4em]
Sender--receiver covariance
&
\textbf{Same node set.} Actors can both send and receive ties, so \(\operatorname{Cov}(a_i,b_i)=\sigma_{ab}\).
&
\textbf{Distinct node sets.} Type 1 and Type 2 actors occupy separate roles, so \(\sigma_{ab}\) is not defined.
\\
\midrule
\multicolumn{3}{@{}l}{\textit{Residual component}}\\
Within-dyad residual covariance
&
\textbf{Reverse ties.} Reverse ties exist, so \(\operatorname{Cov}(\epsilon_{ij},\epsilon_{ji})=\rho\sigma_{\epsilon}^{2}\).
&
\textbf{No reverse ties.} \(Y_{ij}\) has no reverse counterpart, so \(\rho\) is not defined.
\\
\midrule
\multicolumn{3}{@{}l}{\textit{Multiplicative component}}\\
Higher-order dependence
&
\textbf{One-mode structure.} Captures triangle-based closure, stochastic equivalence, and other higher-order patterns.
&
\textbf{Bipartite structure.} Captures four-cycles, stochastic equivalence and other higher-order patterns. \\
\midrule
\multicolumn{3}{@{}l}{\textit{Covariate component}}\\
Covariate effects
&
Includes observed covariates.
&
 Includes observed covariates.
\\
\bottomrule
\end{tabularx}
\vspace{0.4em}
\end{table}

Table~\ref{tab:ame_comparison} summarizes the bipartite AME specification and highlights its differences from the one-mode model. Both models include additive random effects to capture heterogeneity in actors' baseline interaction propensities. The models differ in whether these additive effects can covary. In the one-mode model, the sender and receiver additive effects are modeled jointly as $\{(a_1, b_1), \cdots, (a_n, b_n) \} \overset{iid}{\sim} \mathcal{N}(0,\Sigma_{ab})$, $\Sigma_{ab} = \begin{pmatrix} \sigma_a^2 & \sigma_{ab} \\ \sigma_{ab} & \sigma_b^2 \end{pmatrix}$, where $\sigma_{ab}$ captures the covariance between an actor's propensity to send and receive ties. In the bipartite model, the additive effects for the two node sets are specified separately as $a_i\overset{\mathrm{iid}}{\sim}\mathcal{N}(0,\sigma_a^2)$ and $b_j\overset{\mathrm{iid}}{\sim}\mathcal{N}(0,\sigma_b^2)$. The covariance parameter $\sigma_{ab}$ is not defined.

The specification of the residual component also differs. The one-mode model allows the errors associated with reverse ties $Y_{ij}$ and $Y_{ji}$ to be correlated: $\{(\epsilon_{ij}, \epsilon_{ji}): i \neq j\} \overset{iid}{\sim} \mathcal{N}(0, \Sigma_{\epsilon})$, $\Sigma_{\epsilon} = \sigma_{\epsilon}^2 \begin{pmatrix} 1 & \rho \\ \rho & 1 \end{pmatrix}$, where $\rho$ captures residual reciprocity. Because bipartite data contain no reverse ties, $\rho$ is not defined, and the model specifies $\epsilon_{ij}\overset{\mathrm{iid}}{\sim}\mathcal{N}(0,\sigma_{\epsilon}^{2})$.

The multiplicative component in the bipartite AME retains the similar bilinear form but represents different dependence patterns. In the one-mode model, each actor $i$ has sender and receiver latent vectors, $\mathbf{u}_i$ and $\mathbf{v}_i$, shared across its outgoing and incoming ties. The latent terms in a transitive triad, $\mathbf{u}_i^\top\mathbf{v}_j$, $\mathbf{u}_j^\top\mathbf{v}_k$, and $\mathbf{u}_i^\top\mathbf{v}_k$, are therefore linked through common actor-level factors, allowing the multiplicative component to represent transitivity. In the bipartite model, each Type 1 actor $i$ has a latent vector $\mathbf{u}_i$, and each Type 2 actor $j$ has a latent vector $\mathbf{v}_j$. Their inner product, $\mathbf{u}_i^\top\mathbf{v}_j$, represents the latent contribution to the expected interaction between $i$ and $j$. The vector $\mathbf{u}_i$ enters all interactions involving Type 1 actor $i$, while $\mathbf{v}_j$ enters all interactions involving Type 2 actor $j$. Consequently, the latent terms in a four-cycle, $\mathbf{u}_i^\top\mathbf{v}_j$, $\mathbf{u}_i^\top\mathbf{v}_{j'}$, $\mathbf{u}_{i'}^\top\mathbf{v}_j$, and $\mathbf{u}_{i'}^\top\mathbf{v}_{j'}$, are linked through shared Type 1 and Type 2 factors. This structure allows the multiplicative component to represent four-cycle patterns and, more generally, other forms of higher-order dependence. Because actors with similar latent vectors generate similar interaction profiles across the opposite node set, the model also captures stochastic equivalence.

The covariate component has the same form in both models. Observed covariates enter through $\mathbf{x}_{ij}^\top\boldsymbol{\beta}$, where $\mathbf{x}_{ij}$ may include characteristics of Type 1 actors, Type 2 actors, and dyads. The corresponding elements of $\boldsymbol{\beta}$ estimate their conditional associations with $Y_{ij}$.

\subsection{Bipartite Dynamic AME Model}

I extend the bipartite AME model to a dynamic setting by allowing its coefficients, additive effects, and multiplicative latent factors to evolve over time. For Type 1 actors $i=1,\ldots,N$, Type 2 actors $j=1,\ldots,M$, and periods $t=1,\ldots,T$, the model is

\begin{equation*}
Y_{ijt}=\mathbf{x}_{ijt}^{\top}\boldsymbol{\beta}_t+a_{it}+b_{jt}+\mathbf{u}_{it}^{\top}\mathbf{v}_{jt}+\epsilon_{ijt},
\qquad
\epsilon_{ijt}\overset{\mathrm{iid}}{\sim}\mathcal{N}(0,\sigma_{\epsilon}^{2}),
\label{eq:dynamic_bipartite_ame}
\end{equation*}

\noindent where $Y_{ijt}$ denotes the observed interaction between $i$ and $j$ at time $t$; $\mathbf{x}_{ijt}$ contains observed node- and dyad-specific covariates; $\boldsymbol{\beta}_t$ is the corresponding time-varying coefficient vector; $a_{it}$ and $b_{jt}$ are additive random effects capturing time-varying nodal heterogeneity; $\mathbf{u}_{it}$ and $\mathbf{v}_{jt}$ are time-varying latent vectors; and the multiplicative term $\mathbf{u}_{it}^{\top} \mathbf{v}_{jt}$ captures higher-order dependence that evolves over time; and $\epsilon_{ijt}$ is an idiosyncratic error term assumed to be independently distributed across dyads and time.

All time-varying components follow first-order Gaussian random walks. For $t=2,\ldots,T$,

\begin{align*}
\boldsymbol{\beta}_t \mid \boldsymbol{\beta}_{t-1}
&\sim\mathcal{N}\left(\boldsymbol{\beta}_{t-1},\tau_{\beta}^{2}\mathbf{I}_{P}\right),
\nonumber\\
a_{it} \mid a_{i,t-1}
&\overset{\mathrm{}}{\sim}\mathcal{N}\left(a_{i,t-1},\tau_{a}^{2}\right),
\nonumber\\
b_{jt} \mid b_{j,t-1}
&\overset{\mathrm{}}{\sim}\mathcal{N}\left(b_{j,t-1},\tau_{b}^{2}\right),
\nonumber\\
\mathbf{u}_{it} \mid \mathbf{u}_{i,t-1}
&\overset{\mathrm{}}{\sim}\mathcal{N}\left(\mathbf{u}_{i,t-1},\tau_{UV}^{2}\mathbf{I}_{K}\right),
\nonumber\\
\mathbf{v}_{jt} \mid \mathbf{v}_{j,t-1}
&\overset{\mathrm{}}{\sim}\mathcal{N}\left(\mathbf{v}_{j,t-1},\tau_{UV}^{2}\mathbf{I}_{K}\right).
\end{align*}

\noindent Conditional on the previous-period states, the innovations are independent across actors, periods, and parameter blocks. The initial states have zero-centered Gaussian distributions: 

\begin{align*}
\boldsymbol{\beta}_1
&\sim\mathcal{N}\left(\mathbf{0},\sigma_{\beta}^{2}\mathbf{I}_{P}\right),
\nonumber\\
a_{i1}
&\overset{\mathrm{}}{\sim}\mathcal{N}\left(0,\sigma_{a}^{2}\right),
\nonumber\\
b_{j1}
&\overset{\mathrm{}}{\sim}\mathcal{N}\left(0,\sigma_{b}^{2}\right),
\nonumber\\
\mathbf{u}_{i1}
&\overset{\mathrm{}}{\sim}\mathcal{N}\left(\mathbf{0},\sigma_{U}^{2}\mathbf{I}_{K}\right),
\nonumber\\
\mathbf{v}_{j1}
&\overset{\mathrm{}}{\sim}\mathcal{N}\left(\mathbf{0},\sigma_{V}^{2}\mathbf{I}_{K}\right).
\label{eq:dynamic_bipartite_initial}
\end{align*}

\noindent The initial-state variances $\sigma_{\beta}^{2}$, $\sigma_{a}^{2}$, $\sigma_{b}^{2}$, $\sigma_{U}^{2}$, and $\sigma_{V}^{2}$ determine the dispersion of each component at the beginning of the observation period. The innovation variances $\tau_{\beta}^{2}$, $\tau_{a}^{2}$, $\tau_{b}^{2}$, and $\tau_{UV}^{2}$ govern the magnitude of their period-to-period changes. Smaller innovation variances imply smoother trajectories and stronger temporal continuity, whereas larger values allow more rapid adjustment. 

I assign a common innovation variance $\tau_{UV}^{2}$ to the Type 1 and Type 2 latent trajectories to address an identification issue discussed in the next subsection. Briefly, $\mathbf{U}_t$ and $\mathbf{V}_t$ enter the model only through their product $\mathbf{U}_t\mathbf{V}_t^\top$. For any nonzero constant $c$, replacing $\mathbf{U}_t$ with $c\mathbf{U}_t$ and $\mathbf{V}_t$ with $c^{-1}\mathbf{V}_t$ leaves the fitted multiplicative component unchanged. Separate innovation variances for the two latent factors would therefore depend on an arbitrary allocation of scale between $\mathbf{U}_t$ and $\mathbf{V}_t$. Using a common innovation variance avoids this arbitrary distinction and stabilizes estimation.

\subsection{Estimation}

I estimate the model using a nested empirical Bayes MAP procedure. Conditional on the variance components, an inner BCD loop sequentially updates the coefficients, additive effects, and latent factors. An outer loop updates the variance components, including the observation variance $\sigma_\epsilon^2$, the initial-state variances $\sigma_\beta^2,\sigma_a^2,\sigma_b^2,\sigma_U^2,\sigma_V^2$, and the innovation variances $\tau_\beta^2,\tau_a^2,\tau_b^2,\tau_{UV}^2$. These estimates determine the shrinkage and temporal-smoothing penalties used in the next inner loop.

For fixed variance components, the MAP estimates minimize the following penalized least-squares objective:

\begin{align*}
Q(\Theta)
={}&
\sum_{t=1}^{T} \sum_{i=1}^{N} \sum_{j=1}^{M} \left( Y_{ijt} - a_{it} - b_{jt} - \mathbf{x}_{ijt}^{\top}\boldsymbol{\beta}_t -\mathbf{u}_{it}^{\top}\mathbf{v}_{jt} \right)^2 \nonumber\\
&+ \lambda_\beta \left\| \boldsymbol{\beta}_1 \right\|_2^2 + \lambda_a \left\| \mathbf{a}_1 \right\|_2^2 + \lambda_b \left\| \mathbf{b}_1 \right\|_2^2 + \lambda_U \left\| \mathbf{U}_1 \right\|_F^2 + \lambda_V \left\| \mathbf{V}_1 \right\|_F^2 \nonumber\\
&+ \gamma_\beta \sum_{t=2}^{T} \left\| \boldsymbol{\beta}_t - \boldsymbol{\beta}_{t-1} \right\|_2^2 + \gamma_a \sum_{t=2}^{T} \left\| \mathbf{a}_t - \mathbf{a}_{t-1} \right\|_2^2 \nonumber\\
&+ \gamma_b \sum_{t=2}^{T} \left\| \mathbf{b}_t - \mathbf{b}_{t-1} \right\|_2^2 + \gamma_{UV} \sum_{t=2}^{T} \left[ \left\| \mathbf{U}_t - \mathbf{U}_{t-1} \right\|_F^2 + \left\| \mathbf{V}_t - \mathbf{V}_{t-1} \right\|_F^2 \right].
\end{align*}

\noindent The first term is the residual sum of squares. The penalties indexed by $\lambda$ arise from the initial-state priors, while those indexed by $\gamma$ arise from the random-walk priors. Their values are determined by the corresponding variance components:

\[ \lambda_{\beta}=\frac{\sigma_{\epsilon}^{2}}{\sigma_{\beta}^2},
\qquad \lambda_a=\frac{\sigma_{\epsilon}^{2}}{\sigma_a^2},
\qquad \lambda_b=\frac{\sigma_{\epsilon}^{2}}{\sigma_b^2}, \]

\[ \lambda_U=\frac{\sigma_{\epsilon}^{2}}{\sigma_U^2},
\qquad \lambda_V=\frac{\sigma_{\epsilon}^{2}}{\sigma_V^2}, \]

and

\[
\gamma_{\beta}=\frac{\sigma_{\epsilon}^{2}}{\tau_{\beta}^{2}},
\qquad
\gamma_a=\frac{\sigma_{\epsilon}^{2}}{\tau_a^{2}},
\qquad
\gamma_b=\frac{\sigma_{\epsilon}^{2}}{\tau_b^{2}},
\qquad
\gamma_{UV}=\frac{\sigma_{\epsilon}^{2}}{\tau_{UV}^{2}}.
\]

\noindent Larger values of $\lambda$ impose stronger shrinkage on the initial states, whereas larger values of $\gamma$ produce smoother trajectories. For the latent factors, these penalties also help regularize the relative scale of $\mathbf{U}_t$ and $\mathbf{V}_t$.


Because $\mathbf{U}_t$ and $\mathbf{V}_t$ enter the residual sum of squares only through their product, reciprocal rescaling leaves this term unchanged: $\mathbf{U}_t\mathbf{V}_t^\top = (c\mathbf{U}_t) \left(\frac{1}{c}\mathbf{V}_t\right)^\top, c>0.$ The data-fit term alone therefore cannot distinguish among these equivalent scales. For fixed $\lambda_U$ and $\lambda_V$, however, rescaling changes the initial-state penalty to $\lambda_Uc^2\|\mathbf{U}_1\|_F^2 + \lambda_Vc^{-2}\|\mathbf{V}_1\|_F^2.$ The initial-state and random-walk penalties therefore constrain the relative scale of $\mathbf{U}_t$ and $\mathbf{V}_t$.

\begin{table}[h]
\caption{Empirical Bayes MAP Estimation via BCD}
\label{alg:ebmap}
\centering
\begin{tcolorbox}[colback=gray!8,colframe=gray!35,boxrule=0.4pt,arc=3pt,left=8pt,right=8pt,top=8pt,bottom=8pt,width=\linewidth]
\begingroup
\small
\newcommand{\AlgHeading}[2]{%
  \par\smallskip\noindent\hspace*{#1}%
  \parbox[t]{\dimexpr\linewidth-#1\relax}{\raggedright #2\strut}%
}
\newcommand{\AlgBullet}[2]{%
  \par\smallskip\noindent\hspace*{#1}%
  \makebox[1em][l]{\(\bullet\)}%
  \parbox[t]{\dimexpr\linewidth-#1-1em\relax}{\raggedright #2\strut}%
}
\newcommand{\AlgFormula}[1]{%
  \par\smallskip\noindent\hspace*{2.25em}%
  \parbox[t]{\dimexpr\linewidth-2.25em\relax}{\centering #1\par}%
}
\begin{tabularx}{\linewidth}{@{}>{\raggedleft\arraybackslash}p{1.5em}@{\hspace{0.6em}}>{\raggedright\arraybackslash}X@{}}
1: &
\textbf{Initialization}
\AlgBullet{1.25em}{Initialize additive effects and set \(\boldsymbol{\beta}_t=\mathbf{0}\) for all \(t\).}
\AlgBullet{1.25em}{Initialize \(\mathbf{U}_t\) and \(\mathbf{V}_t\) using temporally aligned SVDs.}
\AlgBullet{1.25em}{Initialize all penalty parameters to one.}
\\[0.5em]
2: &
\textbf{Repeat}
\AlgHeading{1.25em}{\textbf{Inner BCD loop}}
\AlgBullet{2.5em}{Update the coefficient trajectory \(\{\boldsymbol{\beta}_t\}_{t=1}^{T}\).}
\AlgBullet{2.5em}{Update the additive-effect trajectories \(\{a_{it}\}_{t=1}^{T}\) for \(i=1,\ldots,N\) and \(\{b_{jt}\}_{t=1}^{T}\) for \(j=1,\ldots,M\).}
\AlgBullet{2.5em}{Update the latent-factor trajectories \(\{\mathbf{u}_{it}\}_{t=1}^{T}\) and \(\{\mathbf{v}_{jt}\}_{t=1}^{T}\).}
\AlgBullet{2.5em}{Repeat until the penalized objective stabilizes.}
\AlgHeading{1.25em}{\textbf{End inner BCD loop}}
\AlgBullet{1.25em}{Normalize the latent-factor trajectories by reciprocal scaling while preserving \(\mathbf{U}_t\mathbf{V}_t^\top\) for all \(t\).}
\AlgBullet{1.25em}{Apply a common orthogonal Procrustes rotation across periods while preserving \(\mathbf{U}_t\mathbf{V}_t^\top\) and the normalized scale.}
\AlgHeading{1.25em}{\textbf{Outer empirical Bayes loop}}
\AlgBullet{2.5em}{Update \(\sigma_\epsilon^2\) from the residual sum of squares.}
\AlgBullet{2.5em}{Update the initial-state and temporal innovation variances:}
\AlgFormula{%
  \(\sigma_\beta^2,\ 
    \sigma_a^2,\ 
    \sigma_b^2,\ 
    \sigma_U^2,\ 
    \sigma_V^2,\ 
    \tau_\beta^2,\ 
    \tau_a^2,\ 
    \tau_b^2,\ 
    \tau_{UV}^2\).
}
\AlgBullet{2.5em}{Convert the updated variance components into penalty parameters for the next inner BCD loop.}
\\[0.5em]
3: &
\textbf{Until} the variance components and fitted values converge.
\\
\\
\end{tabularx}
\endgroup
\end{tcolorbox}
\end{table}

Table \ref{alg:ebmap} summarizes the estimation procedure.\footnote{Mathematical details are reported in Appendix~\ref{sec:Estimation}.} The algorithm proceeds in three stages. First, the model parameters are initialized. The Type 1 additive effects are initialized using the row means of $\mathbf{Y}_t$, and the Type 2 additive effects are initialized using the column means after subtracting the period-specific overall mean. The coefficient vectors are initialized at zero. Initial values for $\mathbf{U}_t$ and $\mathbf{V}_t$ are obtained from balanced rank-$K$ singular-value decompositions (SVDs) of the residualized outcome matrices and sequentially aligned across periods.

Because the multiplicative term makes the objective function non-convex, estimation may be sensitive to the initial latent factors and may converge to different local minima. I therefore use a multiple-start initialization strategy and generate additional starting values by adding small Gaussian perturbations to the aligned SVD factors. The inner BCD loop is run from each candidate with all penalties initially set to one, and the solution with the lowest penalized objective is retained.

Second, conditional on fixed variance components, the inner BCD loop sequentially updates the complete temporal trajectories of the coefficients, additive effects, and latent factors while holding the remaining parameter blocks fixed. Because the random-walk penalties link adjacent periods, each conditional update is obtained by solving a tridiagonal or block-tridiagonal linear system. The updates continue until the inner-loop convergence criterion is met.

After the inner loop converges, I normalize and align the latent factors to obtain a stable and reproducible representation. A whole-trajectory reciprocal scale normalization fixes the relative scale of $\{\mathbf{U}_t\}_{t=1}^{T}$ and $\{\mathbf{V}_t\}_{t=1}^{T}$ without changing $\mathbf{U}_t\mathbf{V}_t^\top$. A common orthogonal Procrustes rotation then aligns the latent factors at all periods with the retained initialization. Using the same rotation across periods preserves their temporal alignment.

Third, the outer loop updates the variance components, including the observation variance, initial-state variances, and temporal innovation variances using conditional second moments for each parameter block. These estimates determine the penalties used in the next inner loop. The inner and outer loops continue until the overall convergence criteria are met.

Convergence is assessed separately for the two loops. The inner loop converges when the relative change in the penalized objective falls below the specified tolerance. The outer loop converges when the relative changes in both the variance components and fitted values fall below their respective tolerances.


\subsection{SQUAREM Acceleration}

At each outer iteration, the updated variance components determine a new set of penalty parameters. Because these parameters may update only gradually, the outer loop can converge slowly. I use the squared iterative method implemented in \texttt{SQUAREM} to accelerate this process \citep{du2020squarem}. The method first performs two ordinary outer updates. It then compares the changes from these successive updates and extrapolates a larger step in the estimated update direction. If the extrapolation improves convergence, the outer loop continues from the extrapolated penalty values; otherwise, it reverts to the ordinary updates. In either case, the loop continues until the original convergence criteria are satisfied.\footnote{Appendix Figure~\ref{fig:simulation_compute} presents a comparison of computation time between the standard estimation procedure and the SQUAREM-accelerated procedure.}

\subsection{Inference}

I quantify uncertainty using a conditional parametric bootstrap. The procedure conditions on the observed actors, periods, covariates, and latent dimension $K$, while preserving the fitted dependence structure. For each bootstrap replication $b=1,\ldots,B$, I hold the estimated latent relational structure $\widehat{\mathbf{U}}_t\widehat{\mathbf{V}}_t^\top$ fixed and draw new observation errors:
\[
\epsilon_{ijt}^{*(b)}
\sim \mathcal{N}\left(0,\hat{\sigma}_{\epsilon}^{2}\right).
\]
These errors are combined with the bootstrap coefficient and additive-effect trajectories described below to generate a new panel from the fitted observation model.

Because the MAP trajectories are smoothed by the temporal penalties, generating bootstrap panels from the fitted trajectories would understate uncertainty in the coefficient and additive-effect estimates. I therefore restore the missing variation implied by the estimated initial-state and innovation variances before generating each bootstrap panel. Specifically, consider a parameter block $\boldsymbol{\theta}_t \in \left\{ \boldsymbol{\beta}_t,\mathbf{a}_t,\mathbf{b}_t \right\},$ and let $q_\theta$ denote the number of scalar parameters in the block. The additional initial-state variance is
\[
s_{\theta,0}^{2} = \widehat{\sigma}_{\theta}^{2} - \frac{ \left\lVert \widehat{\boldsymbol{\theta}}_1 \right\rVert_2^2 }{q_\theta }.
\]

The additional innovation variance is
\[
s_{\theta,\Delta}^{2} = \widehat{\tau}_{\theta}^{2} - \frac{1}{q_\theta(T-1)} \sum_{t=2}^{T} \left\lVert \widehat{\boldsymbol{\theta}}_t - \widehat{\boldsymbol{\theta}}_{t-1} \right\rVert_2^2.
\]

For bootstrap replication $b$, the initial state is generated as

\[
\boldsymbol{\theta}_1^{*(b)} = \widehat{\boldsymbol{\theta}}_1 + \mathbf{e}_{\theta,1}^{*(b)},
\qquad
\mathbf{e}_{\theta,1}^{*(b)} \sim \mathcal{N}\left( \mathbf{0}, s_{\theta,0}^{2}\mathbf{I}_{q_\theta} \right),
\]

and, for $t=2,\ldots,T$, the trajectory evolves according to
\[
\boldsymbol{\theta}_t^{*(b)} = \boldsymbol{\theta}_{t-1}^{*(b)} + \left( \widehat{\boldsymbol{\theta}}_t - \widehat{\boldsymbol{\theta}}_{t-1} \right) + \mathbf{e}_{\theta,t}^{*(b)},
\qquad
\mathbf{e}_{\theta,t}^{*(b)} \sim \mathcal{N}\left( \mathbf{0}, s_{\theta,\Delta}^{2}\mathbf{I}_{q_\theta}
\right).
\]

\noindent This construction preserves the estimated period-to-period changes while restoring the additional variation implied by the fitted variance components.

I do not apply this dispersion adjustment to $\mathbf{U}_t$ and $\mathbf{V}_t$ because their estimated values depend on the scale and orientation imposed during identification. Adding dispersion separately to $\mathbf{U}_t$ and $\mathbf{V}_t$ would make the bootstrap distribution depend on the particular scale normalization and rotation used to identify the latent factors. I hold the fitted latent contribution $\widehat{\mathbf{U}}_t\widehat{\mathbf{V}}_t^\top$ fixed when generating each bootstrap panel. The bootstrap outcomes are therefore
\[
Y_{ijt}^{*(b)} = \mathbf{x}_{ijt}^{\top}\boldsymbol{\beta}_t^{*(b)} + a_{it}^{*(b)} + b_{jt}^{*(b)} + \widehat{\mathbf{u}}_{it}^{\top} \widehat{\mathbf{v}}_{jt} + \epsilon_{ijt}^{*(b)}.
\]

For each bootstrap realization, I re-estimate the model using the same estimation procedure. Each replication uses the multiple-start strategy described above to reduce sensitivity to local minima. The resulting bootstrap distributions provide standard errors and percentile confidence intervals for the coefficients, additive effects, and variance components.\footnote{The accompanying software also implements a full-model parametric bootstrap that evaluates estimator recovery against known generating values. Details are provided in Appendix~\ref{sec:inference}.}

For the multiplicative component, inference is based on the identified product $\mathbf{U}_{t}\mathbf{V}_{t}^\top$. Although $\widehat{\mathbf{U}}_t\widehat{\mathbf{V}}_t^\top$ is held fixed when generating each panel, the multiplicative component is re-estimated in every replication. The resulting bootstrap distribution therefore captures estimation uncertainty conditional on the fitted latent structure. For each dyad and period, I summarize the bootstrap estimates of $\mathbf{u}_{it}^{\top}\mathbf{v}_{jt}$ by their mean, standard deviation, and the proportion of estimates that are positive. A proportion near one or zero indicates a consistently positive or negative latent interaction, whereas a proportion near $0.5$ indicates instability about its sign.

\section{Empirical Evaluation}

This section evaluates the proposed model using simulations and an application to international trade data. The simulations assess model performance. The empirical application illustrates its ability to capture and interpret evolving network structure.

\subsection{Simulation Study}

The simulation study has three objectives. First, I assess whether the proposed model improves coefficient estimation and the empirical coverage of nominal 95\% confidence intervals relative to widely used alternatives. Second, I evaluate whether it more accurately recovers the DGP when temporal structure is present. Third, I examine whether it more accurately recovers the network structure represented by the multiplicative latent component.

The simulations use a bipartite dynamic DGP with time-varying coefficients, additive effects, and multiplicative latent factors. Unless otherwise varied, each simulated panel contains $N=80$ Type 1 actors, $M=100$ Type 2 actors, $T=5$ periods, and $P = 3$ covariates. The true multiplicative component has rank $K=2$. The coefficients, additive effects, and latent factors evolve according to Gaussian random walks, such as $\boldsymbol{\beta}_t\mid\boldsymbol{\beta}_{t-1}\sim\mathcal{N}\!\left(\boldsymbol{\beta}_{t-1},\tau_\beta^2\mathbf{I}_P\right)$. Unless otherwise varied, the initial state variances of the additive effects and latent factors are set to one, and the innovation variances are $\tau_\beta^2=0.02^2, \ \tau_a^2=\tau_b^2=0.10^2, \tau_U^2=\tau_V^2=0.07^2.$ Innovations are independent across periods and parameter blocks and, for the additive effects and latent factors, across actors. The covariates are constructed to distinguish the consequences of additive and multiplicative dependence: the first covariate is correlated with the multiplicative component, the second is correlated with the additive component, and the third is independent of both. These correlations are generated through actor attributes shared by the covariates and the corresponding unobserved components. The baseline correlation is $\rho_x=0.40$. The observation error variance is $\sigma_\epsilon^2=4$.\footnote{Appendix~\ref{sec:Simulation} provides the complete DGP and parameter settings.}

The first simulation focuses on coefficient estimation. The initial coefficient vector is $\boldsymbol{\beta}_1=(0.8,-0.6,0.4)^\top$, and the subsequent coefficient vectors follow the random-walk process described above. I compare the bipartite dynamic AME estimator with pooled OLS, row-and-column random effects, two-way fixed effects, a lagged-outcome model, and a static bipartite AME model fitted separately to each period. These benchmarks account for different features of the DGP: the random- and fixed-effects models incorporate additive actor heterogeneity, the lagged-outcome model incorporates temporal dependence, and the static AME model captures additive and multiplicative dependence without linking networks across periods. I also include the accelerated bipartite dynamic AME estimator to verify that acceleration does not substantively change the estimates.

\begin{figure}[!htbp]
    \centering
    \caption{Coefficient Recovery: Bipartite Dynamic AME versus Alternative Models}
    \includegraphics[width=16cm]{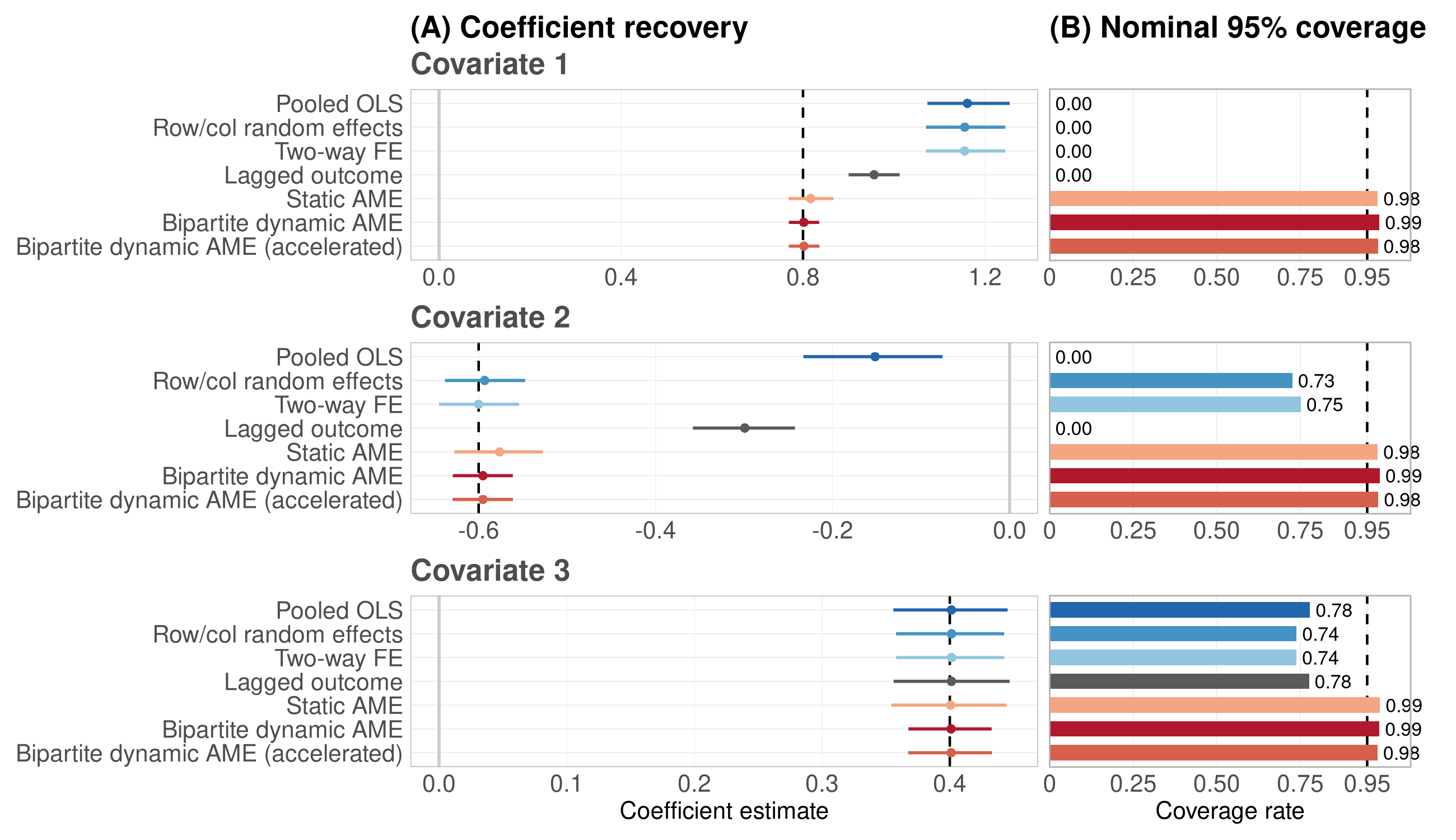}
    \label{fig:sim_coefficient_recovery}
\end{figure}

I conduct 1,000 Monte Carlo replications. Figure~\ref{fig:sim_coefficient_recovery} (a) reports the mean coefficient estimate for each method together with the Monte Carlo distribution of estimation error. The dashed vertical lines indicate the true coefficient values. Figure~\ref{fig:sim_coefficient_recovery} (b) reports the empirical coverage of nominal 95\% confidence intervals. The AME intervals are obtained using the conditional parametric bootstrap, while the remaining methods use their corresponding analytic intervals.

For the first covariate, which is correlated with the multiplicative component, the four non-AME alternatives produce biased estimates and zero coverage. The static and dynamic AME estimators recover the coefficient and attain coverage rates between 0.98 and 0.99. For the second covariate, which is correlated with the additive component, pooled OLS and the lagged-outcome model remain substantially biased and have zero coverage. The random-effects and two-way fixed-effects estimates are closer to the true value, but their coverage rates are only 0.73 and 0.75, respectively. The AME estimators again attain coverage rates between 0.98 and 0.99. For the third covariate, which is independent of both dependence components, all methods recover the coefficient reasonably well, but coverage among the non-AME alternatives ranges from 0.74 to 0.78, compared with 0.98 to 0.99 for the AME estimators. The ordinary and accelerated versions of the bipartite dynamic AME estimator produce nearly identical results. Overall, the first simulation shows that the bipartite dynamic AME estimator more accurately recovers the coefficients and provides well-calibrated confidence intervals in the presence of unobserved network dependence.\footnote{Additional performance statistics for this simulation are reported in Appendix Table~\ref{tab:sim1_additional_stats}.}

The second simulation focuses on temporal dependence. I examine whether accounting for temporal dependence improves the model's ability to recover the DGP, which is evaluated through out-of-sample prediction. I use the same DGP as in the first simulation to generate longitudinal bipartite networks. In each replication, I randomly hold out a subset of dyads in the final period as a test set and use all earlier periods and the remaining observed dyads in the final period for estimation. Predictive performance is evaluated using out-of-sample root mean squared error (RMSE).

I conduct three sets of experiments. 
First, I vary the temporal drift scale $s\in\{0,0.25,0.5,1,2,4\}$ to assess the performance of the bipartite dynamic AME estimator as the network evolves at different rates. For each time-varying component $\theta\in\{\beta,a,b,U,V\}$, the innovation standard deviation is set to $\tau_\theta(s)=s\tau_\theta,$ where $\tau_\theta$ is its baseline value. Thus, $s=0$ produces parameter trajectories that remain constant, $s=1$ corresponds to the baseline DGP, and larger values imply greater changes between adjacent periods. The number of periods is fixed at $T=5$, and the holdout fraction is fixed at $30\%$. Second, I vary the number of periods, $T\in\{2,3,5,7,10\}$, to assess the performance of the bipartite dynamic AME estimator as the time horizon increases. The temporal drift scale is fixed at $s=1$, and the final-period holdout fraction is fixed at $30\%$. Third, I vary the final-period holdout fraction over $\{10\%,30\%,50\%,80\%\}$ to assess the performance of the bipartite dynamic AME estimator as the final-period network becomes more sparsely observed. The temporal drift scale is fixed at $s=1$, and the number of periods is fixed at $T=5$.

\begin{figure}[ht]
    \centering
    \caption{Out-of-Sample Prediction under Temporal Dependence}
    \includegraphics[width=16cm]{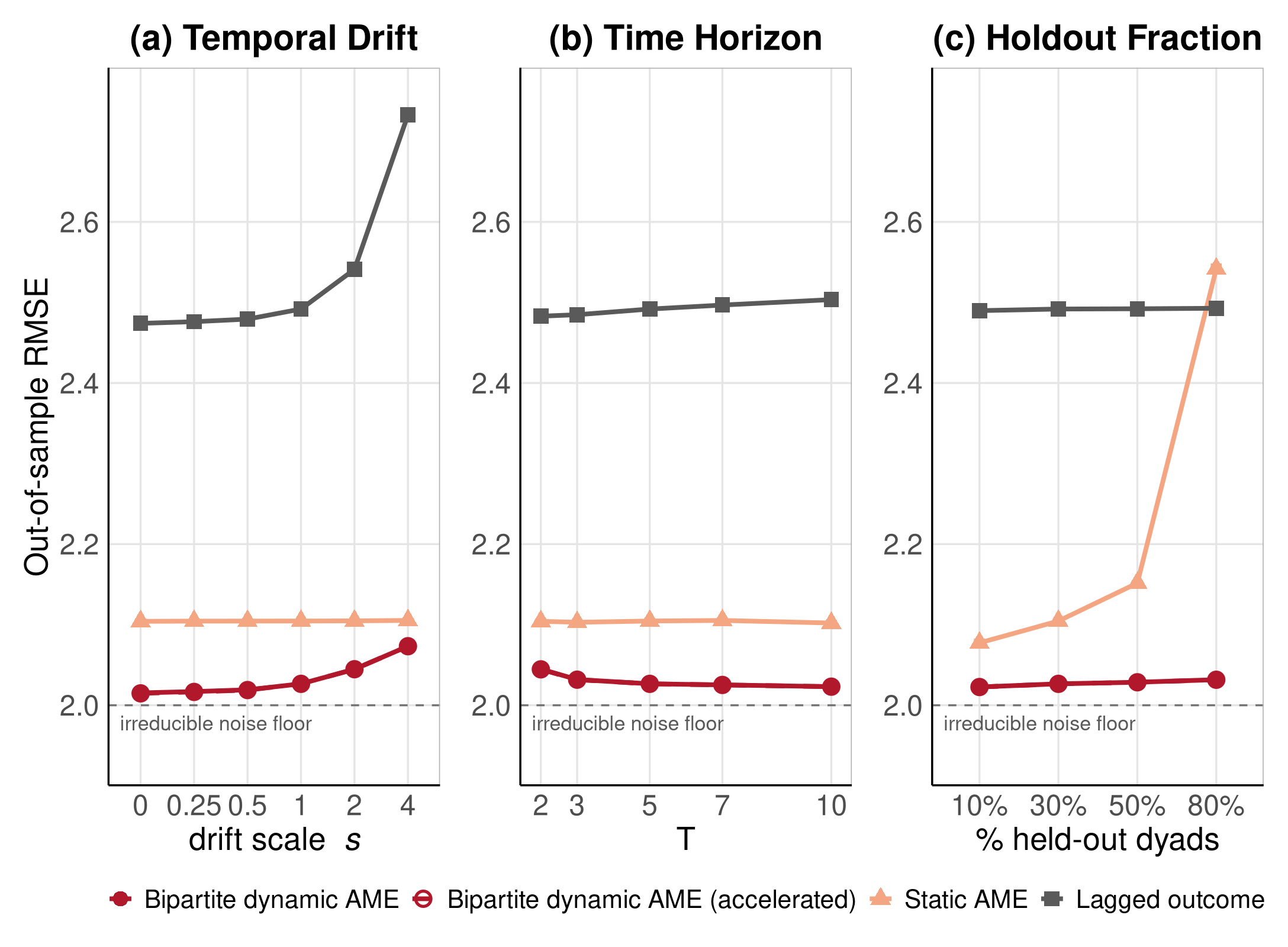}
     \label{fig:simulation_tempo}
\end{figure}

Figure~\ref{fig:simulation_tempo} presents results from 1,000 Monte Carlo replications.\footnote{The error bars report 95\% Monte Carlo intervals for the mean RMSE but are difficult to distinguish at the scale of the y-axis. Appendix Figure~\ref{fig:simulation_tempo2} provides a version in which they are more visible.} The static bipartite AME and lagged-outcome models provide benchmarks. The dashed horizontal line marks $\sigma_\epsilon=2$, reflecting the irreducible RMSE from the idiosyncratic error. Figure~\ref{fig:simulation_tempo} (a) shows that the bipartite dynamic AME estimator remains close to the irreducible noise floor across all levels of temporal drift, although RMSE increases as adjacent periods become less informative. The static AME model consistently produces higher RMSE, while the lagged-outcome model performs substantially worse and deteriorates sharply at the highest drift levels. Figure~\ref{fig:simulation_tempo} (b) shows that the RMSE of the bipartite dynamic AME estimator declines as the number of observed periods increases. In contrast, the static AME model shows little improvement as \(T\) increases. The lagged-outcome model remains substantially above the noise floor throughout. Because increasing \(T\) increases the amount of historical information available for estimation, this pattern suggests that the bipartite dynamic AME estimator benefits from additional temporal information. Figure~\ref{fig:simulation_tempo} (c) shows that the RMSE of the bipartite dynamic AME estimator remains nearly unchanged as the final-period holdout fraction increases from $10\%$ to $80\%$. By contrast, the static AME model deteriorates as fewer final-period dyads remain observed, particularly at the 80\% holdout level. The lagged-outcome model remains substantially above the noise floor throughout. Across all three panels, the ordinary and accelerated bipartite dynamic AME estimators produce nearly identical predictions. Overall, the second simulation shows that the bipartite dynamic AME estimator improves out-of-sample prediction by leveraging information across periods, especially when the final-period network is only partially observed.\footnote{Additional performance statistics for this simulation are reported in Appendix Table~\ref{tab:sim2_additional_stats}.}

The third simulation examines recovery of the multiplicative latent structure. Because the latent factors are not separately identified, I evaluate recovery using the interaction matrix $\mathbf{M}_t=\mathbf{U}_t\mathbf{V}_t^\top.$ I measure recovery using the relative Frobenius error:

\[
\frac{\left\lVert \widehat{\mathbf{M}}_t-\mathbf{M}_t\right\rVert_F}{\left\lVert \mathbf{M}_t\right\rVert_F},
\]

\noindent lower values indicate more accurate recovery of the multiplicative component. I also report the largest principal angle between the estimated and true latent subspaces, where smaller values indicate more accurate recovery of the leading latent dimensions.

I conduct two sets of experiments. First, I vary the latent variance, $\ell = K \sigma_U^2\sigma_V^2 \in \{0.5,1,2,4\},$ which corresponds to the variance of the multiplicative term $\mathbf{u}_{it}^{\top}\mathbf{v}_{jt}$. Because the observation error variance is fixed at $\sigma_\epsilon^2=4$, larger values of $\ell$ increase the variance of the multiplicative component relative to the idiosyncratic error. The number of latent dimensions is fixed at $K=2$. For each value of $\ell$, I set $\sigma_U^2 = \sigma_V^2 = \sqrt{\frac{\ell}{K}}$ and scale the latent factor innovation variances proportionally so that $\frac{\tau_U^2}{\sigma_U^2} = \frac{\tau_V^2}{\sigma_V^2} = 0.07^2$. This experiment therefore evaluates recovery as the variance of the multiplicative component increases relative to the idiosyncratic error. Second, I vary the number of latent dimensions included in the fitted model, \(K\in\{1,2,3,4\}\), while holding the true rank at $K_{\mathrm{true}}=2$ and the latent variance at \(\ell=2\). This experiment
assesses the robustness of latent-structure recovery to the choice of \(K\).

\begin{figure}[h]
    \centering
    \caption{Latent Structure Recovery}
    \includegraphics[width=16cm]{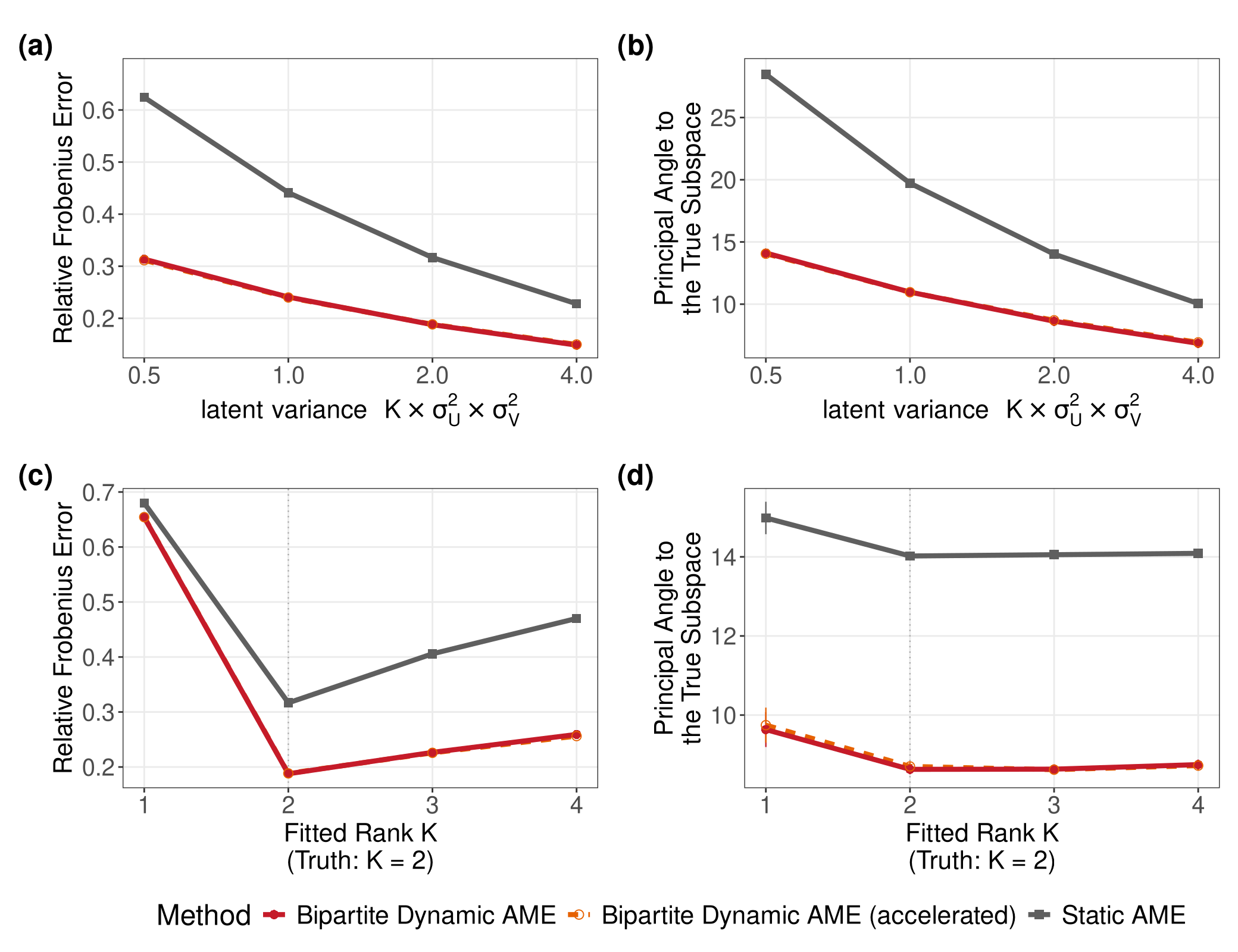}
    \label{fig:sim_latent}
\end{figure}

Figure~\ref{fig:sim_latent} presents the results from 1,000 Monte Carlo replications. I compare the bipartite dynamic AME estimator with a static bipartite AME model fitted separately to each period. Panels (a) and (b) report results across different levels of latent variance, while panels (c) and (d) report results across different numbers of latent dimensions.  Across all panels, the ordinary and accelerated bipartite dynamic AME estimators produce nearly identical results. Figure~\ref{fig:sim_latent} (a) shows that relative Frobenius error declines as latent variance increases, with the bipartite dynamic AME estimator consistently outperforming the static AME model. Figure~\ref{fig:sim_latent} (b) shows a similar pattern for latent-subspace recovery. The principal angle decreases as the latent variance increases, and the dynamic estimator produces smaller angles than the static model. These results indicate that both estimators recover the latent structure more accurately when the variance of the multiplicative component increases relative to the idiosyncratic error, while the dynamic estimator benefits from borrowing information across adjacent periods.

Figure~\ref{fig:sim_latent} (c) shows that fitting too few latent dimensions substantively limits recovery of the multiplicative component. When $K=1$, neither estimator can fully represent the multiplicative structure generated with $K_{\mathrm{true}}=2$ latent dimensions. The error declines sharply when $K$ equals the true rank. Increasing the fitted rank produces a gradual increase in Frobenius error. The dynamic estimator produces lower error than the static model across all fitted ranks. Figure~\ref{fig:sim_latent} (d) shows that the principal angle decreases between $K = 1$ and $K = 2$ and remains nearly unchanged for $K\geq2$. Thus, additional latent dimensions increase overall Frobenius error but have limited effect on recovery of the leading latent subspace. Overall, the third simulation shows that the bipartite dynamic AME estimator more accurately recovers both the multiplicative interaction matrix and its leading latent subspace than a static AME model.\footnote{Additional performance statistics for this simulation are reported in Appendix Table~\ref{tab:sim3_additional_stats}.} These results also suggest a practical approach to choosing \(K\) in empirical applications: fit the model across several candidate dimensions and compare their goodness of fit using network-level discrepancies, such as the Frobenius distance between observed and model-replicated networks. A parsimonious choice is the smallest \(K\) that provides adequate fit across these diagnostics.

\subsection{Empirical Application}
I use international trade data at the country-industry level to construct a bipartite global production network and apply the bipartite dynamic AME model. I show that the model (1) provides interpretable estimates of exogenous covariate effects, (2) uncovers substantively meaningful patterns of cross-country heterogeneity in global production, and (3) reveals evolving patterns of countries' engagement within the network. These patterns provide insights into countries’ participation in global production that are not captured by standard regression models or static AME specifications.

\subsubsection{Global Production Network}

I use international trade data to construct a longitudinal bipartite global production network linking countries to industries. The network is constructed by first identifying industry-level production chains for each final product from input–output tables and then mapping country–industry trade flows onto these chains.\footnote{Details of the network construction are provided in Appendix~\ref{sec:production_network}.}

I focus on production chains anchored in shipbuilding, a strategically important industry in which major economies have invested heavily in production capacity \citep{kalouptsidi2018detection, lane2025manufacturing}. The resulting network covers 190 countries and 143 industries observed annually from 2012 to 2024. Weighted edges represent export values from countries to industries. The model includes log GDP, log population, regime type, and log natural resources as observed covariates, together with additive and multiplicative effects.

\subsubsection{Empirical Findings and Goodness-of-Fit Diagnostics}

\begin{figure}[!htbp]
    \centering
    \caption{Estimated Covariate Effects}
    \includegraphics[width=14cm]{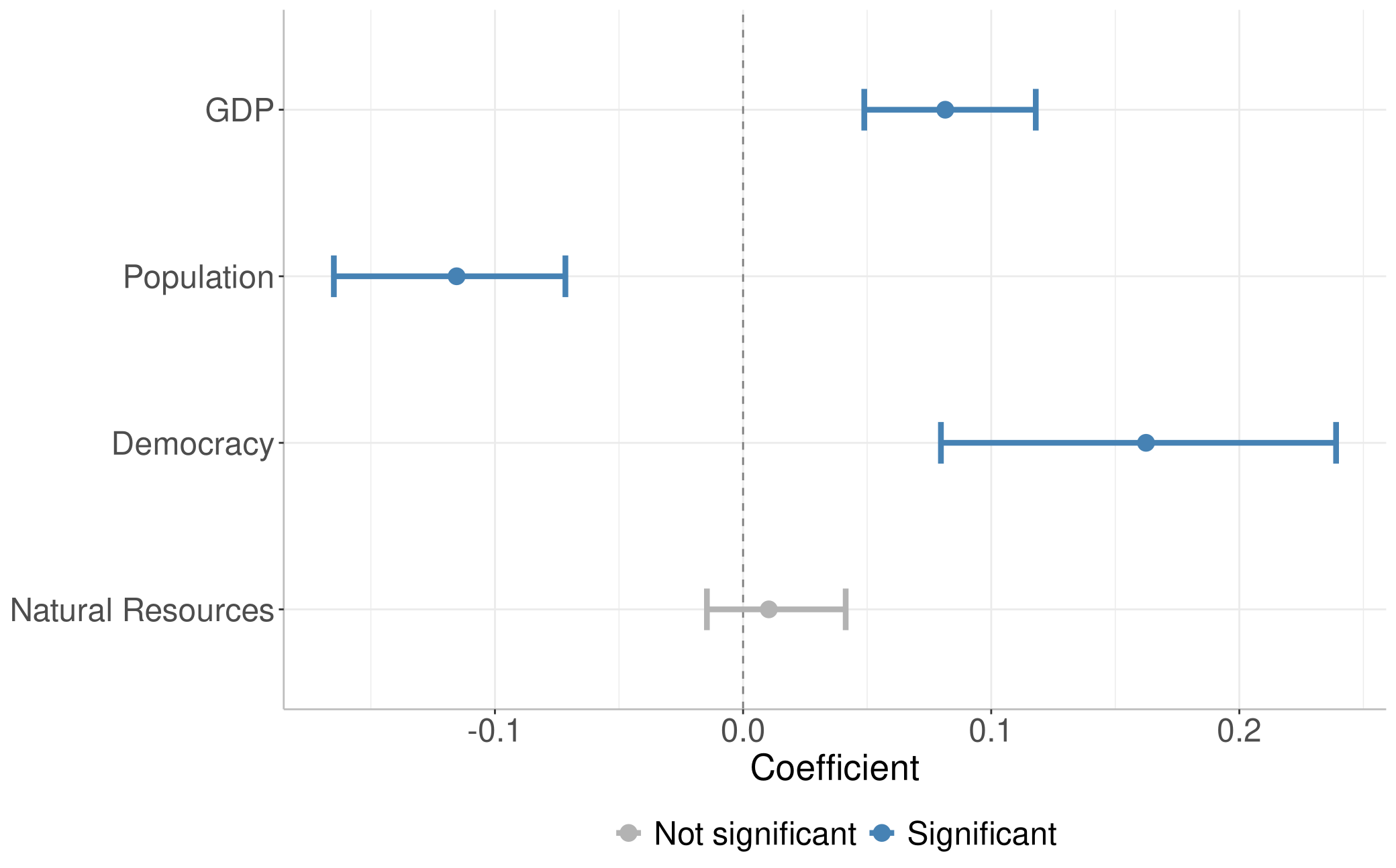}
    \label{fig:ship_covariates}
\end{figure}

Figure~\ref{fig:ship_covariates} presents the estimated covariate effects. The coefficients capture conditional associations between the observed covariates and engagement in the global production network, after accounting for additive actor-level heterogeneity and multiplicative network dependence. The results indicate that GDP, democracy, and larger population size are associated with greater engagement in the global production network, whereas natural resources have no significant effect, conditional on the other covariates.

\begin{figure}[!htbp]
    \centering
    \caption{Estimated Country Heterogeneity in Shipbuilding Production Networks}
    
    \begin{subfigure}[t]{0.49\textwidth}
        \centering
        \includegraphics[width=\linewidth]{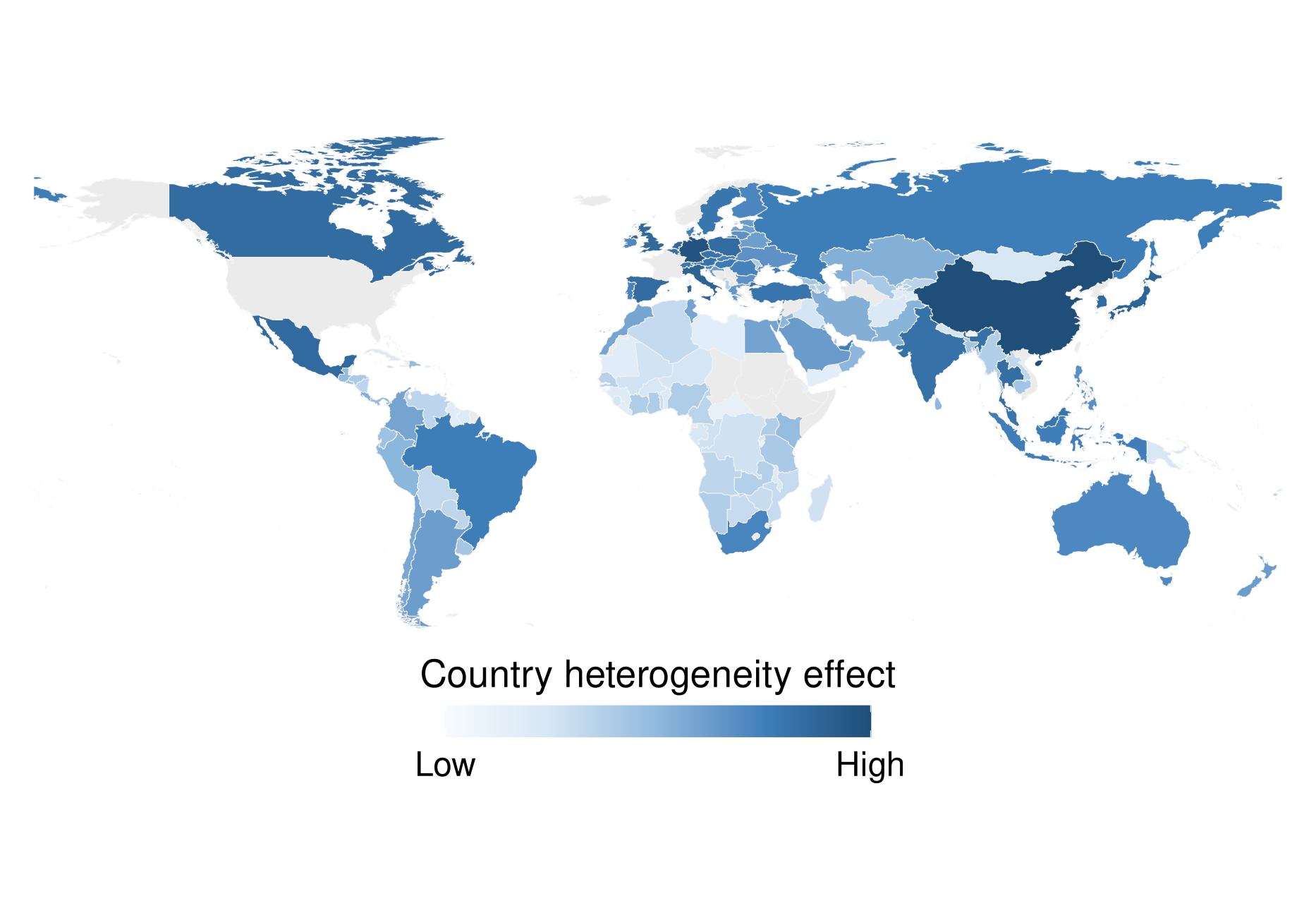}
        \caption{Global Distribution of Country Effects}
    \end{subfigure}
    \hfill
    \begin{subfigure}[t]{0.48\textwidth}
        \centering
         \includegraphics[width=\linewidth]{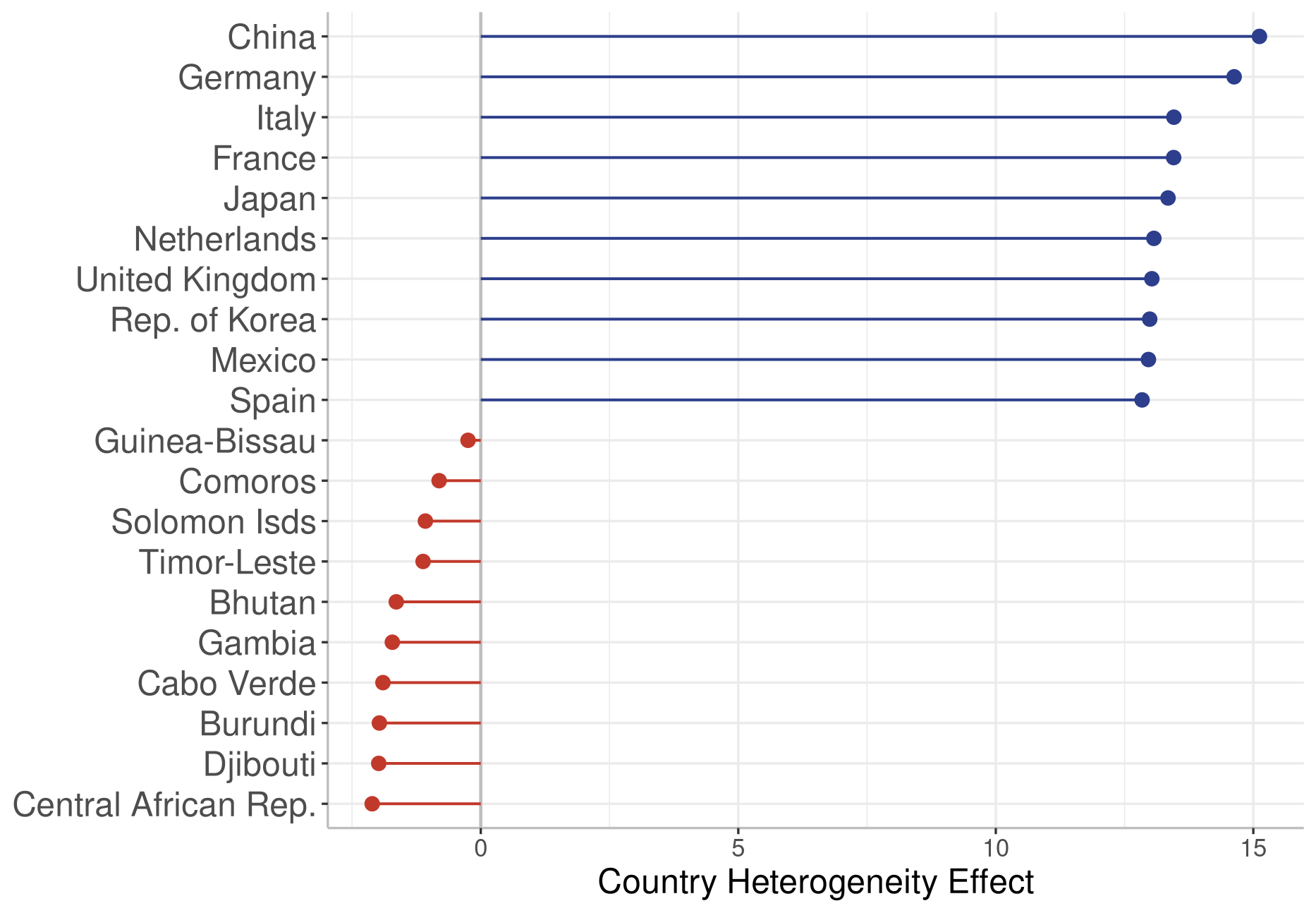} 
        \caption{Top and Bottom Countries by Estimated Effects}
    \end{subfigure}
    \label{fig:ship_alpha}
\end{figure}

\begin{figure}[!htbp]
    \centering
    \caption{Temporal Evolution of Country Positions in Latent Space}
    \includegraphics[width=13cm]{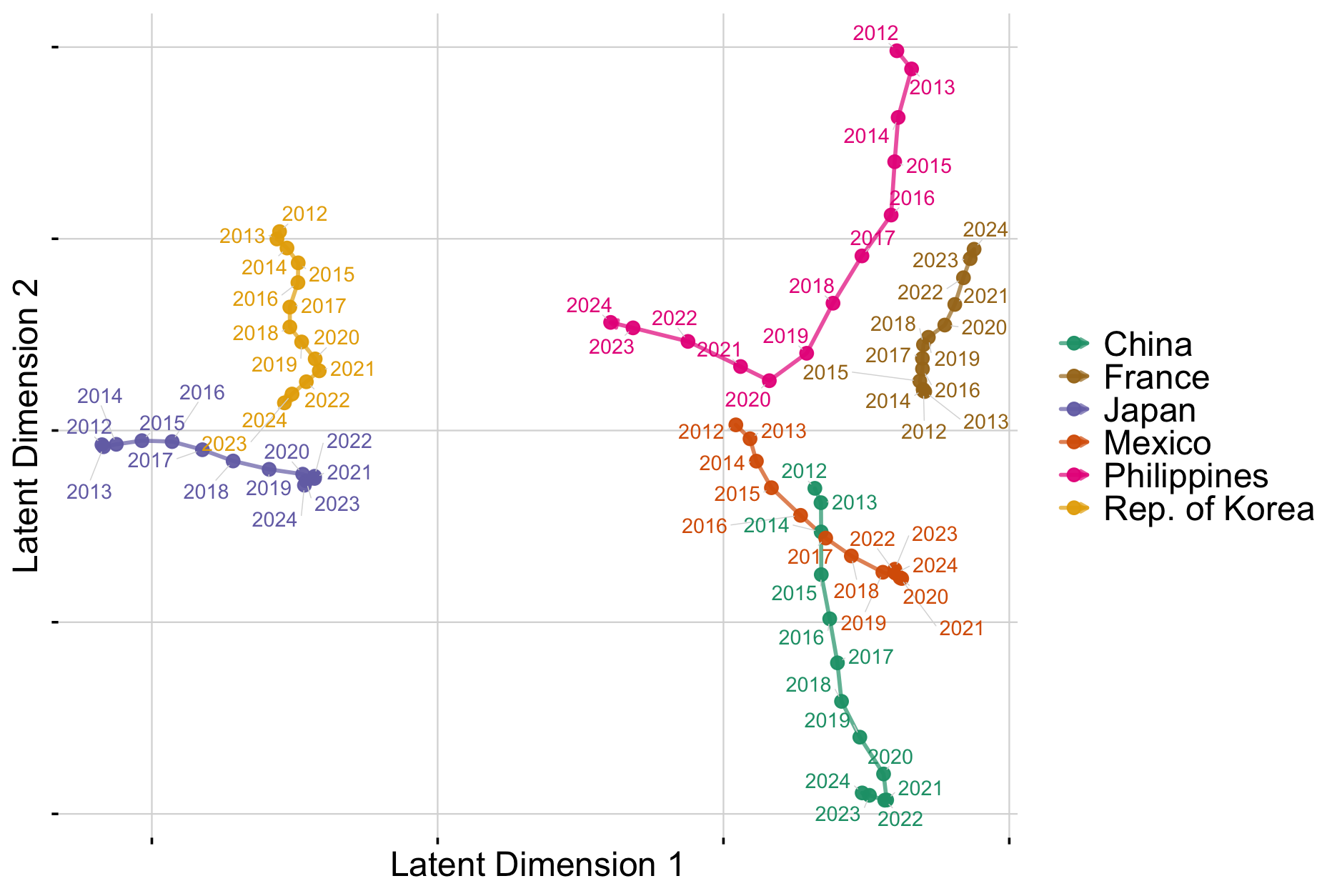}
    \label{fig:ship_latent}
\end{figure}

The additive effects capture residual heterogeneity across countries and industries after accounting for the observed covariates. Figure~\ref{fig:ship_alpha} focuses on country effects $a_i$. Figure~\ref{fig:ship_alpha} (a) shows the global distribution of $a_i$, while Figure~\ref{fig:ship_alpha} (b) highlights the gap between countries with the largest and smallest estimated effects. Both panels reveal substantial cross-country heterogeneity. Countries such as China, Germany, and Italy exhibit large positive $a_i$ estimates, indicating systematically higher levels of production participation. By contrast, several small and less economically integrated countries exhibit negative country effects, indicating lower levels of participation conditional on the observed covariates.

Turning to the multiplicative component, Figure~\ref{fig:ship_latent} traces changes in countries’ positions in the latent space over time. The two latent dimensions themselves have no substantive interpretation. Because the latent factors are aligned across periods, countries’ relative positions can be compared over time. I therefore focus on changes in countries’ relative positions rather than their locations along either dimension. Countries located closer exhibit more similar patterns of engagement in the shipbuilding production network.

Several patterns are visible. France and China begin relatively close in 2012 and then diverge, indicating that the two countries initially exhibit similar participation patterns but become increasingly distinct over time. China and Mexico are also relatively close during the middle of the period before moving apart, suggesting growing differences in their production engagement. The Philippines moves closer to France through 2019 before changing direction. Japan gradually moves closer to the Republic of Korea, indicating increasing similarity in their participation patterns. Together, these trajectories show that countries’ participation patterns evolve substantially over time. Such dynamics are difficult to capture with static latent structures or models based solely on observed covariates.

\begin{figure}[!htbp]
    \centering
    \caption{Goodness-of-Fit Diagnostics for the Shipbuilding Production Network}
    \includegraphics[width=15cm]{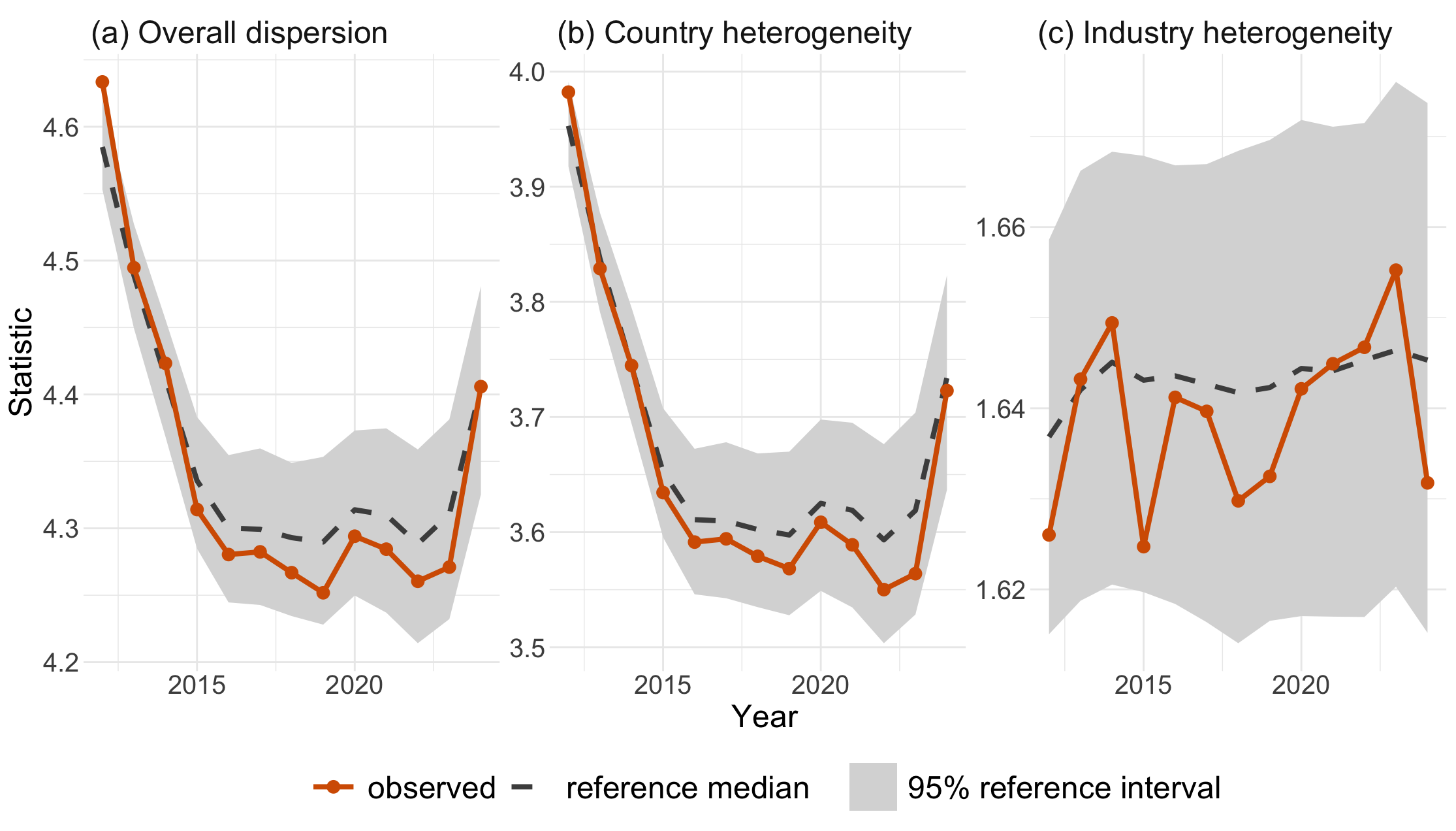}
    \label{fig:GOF}
\end{figure}

Figure~\ref{fig:GOF} evaluates goodness of fit by comparing statistics from networks simulated from the fitted model with their observed values.\footnote{Additional GOF diagnostics are reported in Appendix~\ref{sec:GOF_sup}. Other bipartite-specific diagnostics include four-cycle counts and bipartite degree distributions, which are typically defined for binary networks. Because the application uses continuous weighted edges, I instead evaluate fit using statistics that are directly defined for the weighted outcome. Appendix~\ref{sec:GOF_sup} provides further discussion of these alternative bipartite-specific diagnostics.} The dashed lines indicate the median statistics across simulated networks, and the shaded regions represent 95\% reference intervals. Figure~\ref{fig:GOF} (a) reports the standard deviation of dyad values in each period, capturing overall dispersion in the network. Figure~\ref{fig:GOF} (b) reports the standard deviation of country means, measuring cross-country variation in average engagement across industries. Figure~\ref{fig:GOF} (c) reports the corresponding variation across industry means. Across all three statistics, the observed values closely track the simulated medians and generally remain within the 95\% reference intervals over time. Together, these diagnostics indicate that the fitted model reproduces both the overall dispersion of the outcome and the observed heterogeneity across countries and industries.

\section{Conclusion}


This article develops an AME framework for longitudinal bipartite relational data. Bipartite networks differ from one-mode networks in their dependence structure. The proposed model is tailored to this structure while allowing covariate effects, additive heterogeneity, and multiplicative latent structure to evolve over time. I develop an empirical Bayes MAP estimator based on BCD, a conditional parametric bootstrap for uncertainty quantification, and a SQUAREM accelerator for computational efficiency.

The simulations demonstrate the value of modeling both bipartite network dependence and temporal structure. The bipartite dynamic AME estimator recovers covariate effects and provides well-calibrated confidence intervals when covariates are correlated with unobserved network components. Linking networks across periods also improves out-of-sample prediction, especially when the final-period network is only partially observed. The dynamic specification also more accurately recovers the multiplicative interaction structure and its leading latent subspace than separate static AME models.

The application to global production networks illustrates the substantive value of the framework. The model separates observed covariate effects from country- and industry-level heterogeneity and higher-order relational structure. In the shipbuilding production network, the additive effects reveal substantial cross-country heterogeneity, while the latent trajectories show how countries’ production participation evolves relative to one another over time. These dynamics are not captured by models based only on observed covariates or by static latent representations.

Looking ahead, the bipartite dynamic AME model opens promising avenues for both empirical and methodological research. Its flexibility allows applications across a wide range of political and social science contexts. For example, it can be used to study legislators’ engagement with bills, capturing observed covariate effects, heterogeneity across legislators and legislation, and the evolution of legislators’ engagement in Congress over time. It can also be applied to country-firm interactions, incorporating firm- and country-level covariates, capturing heterogeneity across actors, and tracing the evolution of countries’ participation in competition for foreign capital.

Methodologically, the proposed framework has several limitations. First, the estimator is developed for continuous outcomes. The framework could be extended to binary or count data through generalized linear models with corresponding modifications to the BCD updates. I leave these extensions to future work, including implementation in the \texttt{RAMEN} package. Second, the model uses a Gaussian random walk to allow the coefficients, additive effects, and latent factors to evolve gradually without imposing mean reversion toward a fixed long-run level. This is appropriate when parameters are expected to change incrementally rather than return to a stable equilibrium. I nevertheless acknowledge that the random walk is nonstationary, so shocks can accumulate and later states may drift farther from the initial state. An alternative is a stationary AR(1) specification, which imposes mean reversion and may be more appropriate when parameters are expected to fluctuate around a stable long-run level. Third, the conditional parametric bootstrap provides reliable but relatively conservative coverage in the simulations. Because it conditions on the fitted latent structure when generating bootstrap samples, uncertainty in the multiplicative component is only partially propagated. More generally, inference for the multiplicative component is conditional on both the fitted latent structure and the specified latent dimension \(K\). Future work could develop inference procedures that more fully account for uncertainty in the latent structure and its dimension while preserving scale and rotation invariance.


\section*{Acknowledgments}

The author used AI tools (ChatGPT and Claude) for language editing, proofreading, and coding tasks. These tools were accessed in 2026 and used without modification. The author reviewed and verified all outputs. All errors are the author’s own.

\bibliographystyle{chicago-apsr}
\bibliography{sample}

@article{boucher2019tariff,
  title={“I am a tariff man”: The power of populist foreign policy rhetoric under President Trump},
  author={Boucher, Jean-Christophe and Thies, Cameron G},
  journal={The Journal of Politics},
  volume={81},
  number={2},
  pages={712--722},
  year={2019},
  publisher={The University of Chicago Press Chicago, IL}
}

@article{wang2009exponential,
  title = {Exponential random graph (p$^*$) models for affiliation networks},
  author={Wang, Peng and Sharpe, Ken and Robins, Garry L and Pattison, Philippa E},
  journal={Social Networks},
  volume={31},
  number={1},
  pages={12--25},
  year={2009},
  publisher={Elsevier}
}

@article{stivala2025improving,
  title={Improving exponential-family random graph models for bipartite networks},
  author={Stivala, Alex and Wang, Peng and Lomi, Alessandro},
  journal={Journal of Complex Networks},
  volume={13},
  number={4},
  pages={cnaf017},
  year={2025},
  publisher={Oxford University Press}
}

@article{schoeneman2024bipartite,
  title={A bipartite network analysis of international trade volatility},
  author={Schoeneman, John and Brienen, Marten and Lambert, Lixia and Lambert, Dayton},
  journal={Applied Network Science},
  volume={9},
  number={1},
  pages={74},
  year={2024},
  publisher={Springer}
}

@article{lo2026statistical,
  title={A Statistical Model of Bipartite Networks: Application to Cosponsorship in the United States Senate},
  author={Lo, Adeline and Olivella, Santiago and Imai, Kosuke},
  journal={Political Analysis},
  volume={34},
  number={3},
  pages={451--470},
  year={2026},
  publisher={Cambridge University Press}
}

@article{battaglini2020effectiveness,
  title={Effectiveness of connected legislators},
  author={Battaglini, Marco and Sciabolazza, Valerio Leone and Patacchini, Eleonora},
  journal={American journal of political science},
  volume={64},
  number={4},
  pages={739--756},
  year={2020},
  publisher={Wiley Online Library}
}

@article{zelizer2019position,
  title={Is position-taking contagious? evidence of cue-taking from two field experiments in a state legislature},
  author={Zelizer, Adam},
  journal={American Political Science Review},
  volume={113},
  number={2},
  pages={340--352},
  year={2019},
  publisher={Cambridge University Press}
}

@article{aaroe2020cognitive,
  title={Cognitive biases and communication strength in social networks: The case of episodic frames},
  author={Aar{\o}e, Lene and Petersen, Michael Bang},
  journal={British Journal of Political Science},
  volume={50},
  number={4},
  pages={1561--1581},
  year={2020},
  publisher={Cambridge University Press}
}

@article{kinne2020guns,
  title={Guns or money? Defense co-operation and bilateral lending as coevolving networks},
  author={Kinne, Brandon J and Bunte, Jonas B},
  journal={British Journal of Political Science},
  volume={50},
  number={3},
  pages={1067--1088},
  year={2020},
  publisher={Cambridge University Press}
}

@article{du2020squarem,
  title={SQUAREM: An R package for off-the-shelf acceleration of EM, MM and other EM-like monotone algorithms},
  author={Du, Yu and Varadhan, Ravi},
  journal={Journal of Statistical Software},
  volume={92},
  pages={1--41},
  year={2020}
}

@article{cranmer2016critique,
  title={A critique of dyadic design},
  author={Cranmer, Skyler J and Desmarais, Bruce A},
  journal={International studies quarterly},
  volume={60},
  number={2},
  pages={355--362},
  year={2016},
  publisher={Oxford University Press}
}

@article{desmarais2015fates,
  title={The fates of challengers in US house elections: The role of extended party networks in supporting candidates and shaping electoral outcomes},
  author={Desmarais, Bruce A and La Raja, Raymond J and Kowal, Michael S},
  journal={American Journal of Political Science},
  volume={59},
  number={1},
  pages={194--211},
  year={2015},
  publisher={Wiley Online Library}
}

@article{kim2021mapping,
  title={Mapping political communities: A statistical analysis of lobbying networks in legislative politics},
  author={Kim, In Song and Kunisky, Dmitriy},
  journal={Political Analysis},
  volume={29},
  number={3},
  pages={317--336},
  year={2021},
  publisher={Cambridge University Press}
}

@article{cranmer2011inferential,
  title={Inferential network analysis with exponential random graph models},
  author={Cranmer, Skyler J and Desmarais, Bruce A},
  journal={Political analysis},
  volume={19},
  number={1},
  pages={66--86},
  year={2011},
  publisher={Cambridge University Press}
}

@article{acemoglu2012network,
  title={The network origins of aggregate fluctuations},
  author={Acemoglu, Daron and Carvalho, Vasco M and Ozdaglar, Asuman and Tahbaz-Salehi, Alireza},
  journal={Econometrica},
  volume={80},
  number={5},
  pages={1977--2016},
  year={2012},
  publisher={Wiley Online Library}
}

@article{hoff2021additive,
  title={Additive and multiplicative effects network models},
  author={Hoff, Peter},
  journal={Statistical Science},
  volume={36},
  number={1},
  pages={34--50},
  year={2021},
  publisher={JSTOR}
}

@article{kinne2024network,
  title={Network context and the effectiveness of international agreements},
  author={Kinne, Brandon J},
  journal={International Studies Quarterly},
  volume={68},
  number={3},
  pages={sqae099},
  year={2024},
  publisher={Oxford University Press}
}

@article{leifeld2017discourse,
  title={Discourse network analysis},
  author={Leifeld, Philip},
  journal={The Oxford handbook of political networks},
  pages={301--326},
  year={2017},
  publisher={Oxford University Press Oxford}
}

@article{wang2022resilience,
  title={Resilience to stress in bipartite networks: application to the Islamic State recruitment network},
  author={Wang, Selena and Edgerton, Jared},
  journal={Journal of Complex Networks},
  volume={10},
  number={4},
  pages={cnac017},
  year={2022},
  publisher={Oxford University Press}
}

@article{dorff2020networks,
  title={Networks of violence: Predicting conflict in Nigeria},
  author={Dorff, Cassy and Gallop, Max and Minhas, Shahryar},
  journal={The Journal of Politics},
  volume={82},
  number={2},
  pages={476--493},
  year={2020},
  publisher={The University of Chicago Press Chicago, IL}
}

@article{kim2023dynamic,
  title={A dynamic additive and multiplicative effects network model with application to the United Nations voting behaviors},
  author={Kim, Bomin and Niu, Xiaoyue and Hunter, David and Cao, Xun},
  journal={The annals of applied statistics},
  volume={17},
  number={4},
  pages={3283},
  year={2023}
}

@article{kalouptsidi2018detection,
  title={Detection and impact of industrial subsidies: The case of Chinese shipbuilding},
  author={Kalouptsidi, Myrto},
  journal={The Review of Economic Studies},
  volume={85},
  number={2},
  pages={1111--1158},
  year={2018},
  publisher={Oxford University Press}
}

@article{lane2025manufacturing,
  title={Manufacturing revolutions: Industrial policy and industrialization in South Korea},
  author={Lane, Nathan},
  journal={The Quarterly Journal of Economics},
  volume={140},
  number={3},
  pages={1683--1741},
  year={2025},
  publisher={Oxford University Press}
}

@article{minhas2024decomposing,
  title={Decomposing network influence: Social influence regression},
  author={Minhas, Shahryar and Hoff, Peter D},
  journal={Political Analysis},
  pages={1--20},
  year={2024},
  publisher={Cambridge University Press}
}

@article{minhas2019inferential,
  title={Inferential approaches for network analysis: AMEN for latent factor models},
  author={Minhas, Shahryar and Hoff, Peter D and Ward, Michael D},
  journal={Political Analysis},
  volume={27},
  number={2},
  pages={208--222},
  year={2019},
  publisher={Cambridge University Press}
}

\end{document}


\maketitle
\doublespacing
\tableofcontents

\thispagestyle{empty}

\newpage

\setcounter{page}{1}

\appendix
\addcontentsline{toc}{section}{Appendices}

\renewcommand{\thefigure}{\thesection\arabic{figure}}
\setcounter{figure}{0}

\section{Derivation of the MAP Objective Function}
\label{sec:Objective}

Given the dynamic bipartite AME model,

\[
Y_{ijt} = \mathbf{x}_{ijt}^{\top}\boldsymbol{\beta}_t +a_{it} +b_{jt} +\mathbf{u}_{it}^{\top}\mathbf{v}_{jt} +\epsilon_{ijt},
\qquad
\epsilon_{ijt}\overset{\mathrm{iid}}{\sim} \mathcal{N}(0,\sigma_\epsilon^2),
\]

the conditional distribution of the observed interaction is
\[
Y_{ijt}\mid\Theta \sim \mathcal{N}(\mu_{ijt},\sigma_\epsilon^2),
\]

where
\[
\mu_{ijt} = \mathbf{x}_{ijt}^{\top}\boldsymbol{\beta}_t +a_{it} +b_{jt} +\mathbf{u}_{it}^{\top}\mathbf{v}_{jt}.
\]

Ignoring constants that do not depend on the model parameters, the negative log-likelihood is
\[
\frac{1}{2\sigma_\epsilon^2} \sum_{t=1}^{T} \sum_{i=1}^{N} \sum_{j=1}^{M} \left( Y_{ijt} -\mathbf{x}_{ijt}^{\top}\boldsymbol{\beta}_t -a_{it} -b_{jt} -\mathbf{u}_{it}^{\top}\mathbf{v}_{jt} \right)^2.
\]

Let
\[
\mathbf{a}_t=(a_{1t},\ldots,a_{Nt})^\top,
\qquad
\mathbf{b}_t=(b_{1t},\ldots,b_{Mt})^\top,
\]

and let $\mathbf{U}_t$ and $\mathbf{V}_t$ denote the matrices whose rows are $\mathbf{u}_{it}^{\top}$ and $\mathbf{v}_{jt}^{\top}$, respectively.

The initial states are assigned Gaussian priors:
\begin{align*}
\boldsymbol{\beta}_1
&\sim
\mathcal{N}(\mathbf{0},\sigma_\beta^2\mathbf{I}_P),\\
a_{i1}
&\overset{\mathrm{iid}}{\sim}
\mathcal{N}(0,\sigma_a^2),\\
b_{j1}
&\overset{\mathrm{iid}}{\sim}
\mathcal{N}(0,\sigma_b^2),\\
\mathbf{u}_{i1}
&\overset{\mathrm{iid}}{\sim}
\mathcal{N}(\mathbf{0},\sigma_U^2\mathbf{I}_K),\\
\mathbf{v}_{j1}
&\overset{\mathrm{iid}}{\sim}
\mathcal{N}(\mathbf{0},\sigma_V^2\mathbf{I}_K).
\end{align*}

These priors contribute
\[
\frac{1}{2\sigma_\beta^2}
\|\boldsymbol{\beta}_1\|_2^2
+
\frac{1}{2\sigma_a^2}
\|\mathbf{a}_1\|_2^2
+
\frac{1}{2\sigma_b^2}
\|\mathbf{b}_1\|_2^2
+
\frac{1}{2\sigma_U^2}
\|\mathbf{U}_1\|_F^2
+
\frac{1}{2\sigma_V^2}
\|\mathbf{V}_1\|_F^2
\]
to the negative log posterior.

For $t=2,\ldots,T$, the time-varying components follow Gaussian random walks:
\begin{align*}
\boldsymbol{\beta}_t \mid \boldsymbol{\beta}_{t-1} &\sim \mathcal{N} \left( \boldsymbol{\beta}_{t-1}, \tau_\beta^2\mathbf{I}_P \right),\\
a_{it}\mid a_{i,t-1} &\sim \mathcal{N}\left( a_{i,t-1}, \tau_a^2 \right),\\
b_{jt}\mid b_{j,t-1} &\sim \mathcal{N} \left( b_{j,t-1}, \tau_b^2 \right),\\
\mathbf{u}_{it}\mid\mathbf{u}_{i,t-1} &\sim \mathcal{N} \left( \mathbf{u}_{i,t-1}, \tau_{UV}^2\mathbf{I}_K \right),\\
\mathbf{v}_{jt}\mid\mathbf{v}_{j,t-1} &\sim \mathcal{N} \left( \mathbf{v}_{j,t-1}, \tau_{UV}^2\mathbf{I}_K \right).
\end{align*}

The corresponding contributions to the negative log posterior are
\begin{align*}
&\frac{1}{2\tau_\beta^2} \sum_{t=2}^{T} \|\boldsymbol{\beta}_t-\boldsymbol{\beta}_{t-1}\|_2^2\\
&\quad+ \frac{1}{2\tau_a^2} \sum_{t=2}^{T} \|\mathbf{a}_t-\mathbf{a}_{t-1}\|_2^2\\
&\quad+ \frac{1}{2\tau_b^2} \sum_{t=2}^{T} \|\mathbf{b}_t-\mathbf{b}_{t-1}\|_2^2\\
&\quad+ \frac{1}{2\tau_{UV}^2} \sum_{t=2}^{T} \left[ \|\mathbf{U}_t-\mathbf{U}_{t-1}\|_F^2 + \|\mathbf{V}_t-\mathbf{V}_{t-1}\|_F^2
\right].
\end{align*}

The MAP estimator minimizes the negative log posterior, which is the sum of the negative log-likelihood and the negative log-prior contributions. Multiplying this expression by $2\sigma_\epsilon^2$ does not change its minimizer. Define
\[
\lambda_\beta = \frac{\sigma_\epsilon^2}{\sigma_\beta^2},
\qquad
\lambda_a = \frac{\sigma_\epsilon^2}{\sigma_a^2},
\qquad
\lambda_b = \frac{\sigma_\epsilon^2}{\sigma_b^2},
\qquad
\lambda_U = \frac{\sigma_\epsilon^2}{\sigma_U^2},
\qquad
\lambda_V = \frac{\sigma_\epsilon^2}{\sigma_V^2},
\]
and
\[
\gamma_\beta = \frac{\sigma_\epsilon^2}{\tau_\beta^2},
\qquad
\gamma_a = \frac{\sigma_\epsilon^2}{\tau_a^2},
\qquad
\gamma_b = \frac{\sigma_\epsilon^2}{\tau_b^2},
\qquad
\gamma_{UV} = \frac{\sigma_\epsilon^2}{\tau_{UV}^2}.
\]

The resulting penalized least-squares objective is
\begin{align*}
Q(\Theta) &= \sum_{t=1}^{T} \sum_{i=1}^{N} \sum_{j=1}^{M} \left( Y_{ijt} -a_{it} -b_{jt} -\mathbf{x}_{ijt}^{\top}\boldsymbol{\beta}_t -\mathbf{u}_{it}^{\top}\mathbf{v}_{jt} \right)^2\\
&\quad+ 
\lambda_\beta \|\boldsymbol{\beta}_1\|_2^2 + \lambda_a \|\mathbf{a}_1\|_2^2 + \lambda_b \|\mathbf{b}_1\|_2^2\\
&\quad+
\lambda_U \|\mathbf{U}_1\|_F^2 + \lambda_V \|\mathbf{V}_1\|_F^2\\
&\quad+
\gamma_\beta \sum_{t=2}^{T} \|\boldsymbol{\beta}_t-\boldsymbol{\beta}_{t-1}\|_2^2\\
&\quad+
\gamma_a \sum_{t=2}^{T} \|\mathbf{a}_t-\mathbf{a}_{t-1}\|_2^2\\
&\quad+
\gamma_b \sum_{t=2}^{T} \|\mathbf{b}_t-\mathbf{b}_{t-1}\|_2^2\\
&\quad+
\gamma_{UV} \sum_{t=2}^{T} \left[ \|\mathbf{U}_t-\mathbf{U}_{t-1}\|_F^2 + \|\mathbf{V}_t-\mathbf{V}_{t-1}\|_F^2
\right].
\end{align*}

\section{Estimation Details}\label{sec:Estimation}

The model is estimated using a nested procedure that combines trajectory-level block coordinate descent (BCD) with empirical Bayes updates of the variance components. Conditional on the current variance components, the inner BCD loop sequentially updates the complete coefficient trajectory, each Type 1 and Type 2 additive-effect trajectory, and each Type 1 and Type 2 latent-factor trajectory. Holding the remaining parameter blocks fixed, each update is a convex quadratic problem. The Gaussian random-walk penalties link adjacent periods, producing a tridiagonal or block-tridiagonal system that is solved directly.

The inner BCD loop continues until the penalized MAP objective stabilizes at the current penalty values. After the inner loop converges, the outer loop scale-normalizes the latent trajectories and aligns them to a common orientation before updating the variance components. The observation variance is updated from the residual sum of squares, while the initial-state and innovation variances are updated using blockwise conditional posterior moments. These variance estimates determine the penalties used in the next inner BCD loop. The outer loop continues until both the variance-component vector and the fitted values stabilize.


Let \(N\) and \(M\) denote the numbers of Type 1 and Type 2 actors, respectively; \(T\) the number of periods; \(P\) the number of covariates; and \(K\) the dimension of the multiplicative latent factors. Define the parameter trajectories as
\[ \Theta = \left\{ \boldsymbol{\beta}_{1:T}, \mathbf{a}_{1:T}, \mathbf{b}_{1:T}, \mathbf{U}_{1:T}, \mathbf{V}_{1:T} \right\}, \]
where
\[ \boldsymbol{\beta}_t\in\mathbb{R}^{P},
\qquad
\mathbf{a}_t\in\mathbb{R}^{N},
\qquad
\mathbf{b}_t\in\mathbb{R}^{M},
\qquad
\mathbf{U}_t\in\mathbb{R}^{N\times K},
\qquad
\mathbf{V}_t\in\mathbb{R}^{M\times K}. \]

\subsection*{Penalized MAP objective}

For fixed variance components, estimation is based on the following penalized least-squares objective:

\begin{align*}
Q(\Theta)
&= \sum_{i=1}^{N}\sum_{j=1}^{M}\sum_{t=1}^{T} \left( Y_{ijt}-a_{it}-b_{jt} -\mathbf{x}_{ijt}^{\top}\boldsymbol{\beta}_t -\mathbf{u}_{it}^{\top}\mathbf{v}_{jt} \right)^2 \\
&\quad
+\lambda_{\beta}\lVert\boldsymbol{\beta}_1\rVert_2^2
+\lambda_a\lVert\mathbf{a}_1\rVert_2^2
+\lambda_b\lVert\mathbf{b}_1\rVert_2^2
+\lambda_U\lVert\mathbf{U}_1\rVert_F^2
+\lambda_V\lVert\mathbf{V}_1\rVert_F^2 \\
&\quad
+\gamma_{\beta}\sum_{t=2}^{T}
\lVert\boldsymbol{\beta}_t-\boldsymbol{\beta}_{t-1}\rVert_2^2 \\
&\quad
+\gamma_a\sum_{t=2}^{T}
\lVert\mathbf{a}_t-\mathbf{a}_{t-1}\rVert_2^2 \\
&\quad
+\gamma_b\sum_{t=2}^{T}
\lVert\mathbf{b}_t-\mathbf{b}_{t-1}\rVert_2^2 \\
&\quad
+\gamma_U\sum_{t=2}^{T}
\lVert\mathbf{U}_t-\mathbf{U}_{t-1}\rVert_F^2 \\ 
&\quad
+\gamma_V\sum_{t=2}^{T}
\lVert\mathbf{V}_t-\mathbf{V}_{t-1}\rVert_F^2.
\end{align*}

The first line is the residual sum of squares over observed dyad-period observations. The parameters \(\lambda_{\beta},\lambda_a,\lambda_b,\lambda_U,\lambda_V\) are the initial-state penalty parameters induced by the Gaussian distributions assigned to the initial states. The parameters \(\gamma_{\beta},\gamma_a,\gamma_b,\gamma_U,\gamma_V\) are the temporal penalty parameters induced by the Gaussian random-walk innovations. Larger values of \(\lambda\) impose stronger shrinkage of the initial states toward zero, whereas larger values of \(\gamma\) impose stronger smoothing across adjacent periods. These penalty parameters are determined by the corresponding variance components:

\begin{align*}
\lambda_{\beta} &= \frac{\sigma_{\epsilon}^{2}}{\sigma_{\beta}^{2}},
&
\lambda_a &= \frac{\sigma_{\epsilon}^{2}}{\sigma_a^{2}},
&
\lambda_b &= \frac{\sigma_{\epsilon}^{2}}{\sigma_b^{2}}, \nonumber\\
\lambda_U &=\frac{\sigma_{\epsilon}^{2}}{\sigma_U^{2}},
&
\lambda_V &= \frac{\sigma_{\epsilon}^{2}}{\sigma_V^{2}}.
\end{align*}

Similarly,
\begin{align*}
\gamma_{\beta} &= \frac{\sigma_{\epsilon}^{2}}{\tau_{\beta}^{2}},
&
\gamma_a &= \frac{\sigma_{\epsilon}^{2}}{\tau_a^{2}},
&
\gamma_b &= \frac{\sigma_{\epsilon}^{2}}{\tau_b^{2}}, \nonumber\\
\gamma_U &= \frac{\sigma_{\epsilon}^{2}}{\tau_U^{2}},
& 
\gamma_V &= \frac{\sigma_{\epsilon}^{2}}{\tau_V^{2}}.
\end{align*}

I use a common innovation variance \(\tau_{UV}^2\) for the temporal evolution of the two latent-factor blocks. This restriction addresses a scale-related identification issue discussed below. Because the temporal penalties are determined by the innovation variances, it also implies a common temporal penalty \(\gamma_{UV}\):
\[
\tau_U^2=\tau_V^2=\tau_{UV}^2,
\qquad
\gamma_U=\gamma_V=\gamma_{UV}.
\]

Conditional on the variance components, minimizing \(Q(\Theta)\) yields the MAP estimates of the parameter trajectories.

\subsection*{Step 1: Initialization and multiple starting values}

The algorithm begins by initializing the model parameters and variance components. Because the objective function is non-convex due to the multiplicative latent component, different starting values may lead to different local optima. A multi-start strategy is therefore used to reduce sensitivity to initialization.

The additive effects are initialized using the row and column means of the outcome matrix:

\begin{align*}
a_{it}^{(0)} &= \bar{Y}_{i\cdot t},  & 
b_{jt}^{(0)} &= \bar{Y}_{\cdot jt} - \bar{Y}_{\cdot\cdot t}.
\end{align*}

where $b_{jt}^{(0)}$ is centered to have mean zero. The observation equation alone is invariant to location shifts, since, for any scalar \(c_t\),

\[ a_{it}^{'}+b_{jt}^{'} = (a_{it}+c)+(b_{jt}-c) = a_{it}+b_{jt}. \]

Centering \(b_{jt}^{(0)}\) provides a convenient and stable initial representation of the additive effects.

The regression coefficients are initialized at zero:

\[ \boldsymbol{\beta}_t^{(0)}=0, \qquad t=1,\ldots,T. \]

The latent factors are initialized from a balanced rank-$K$ singular-value decomposition. For each period, I construct the residualized outcome matrix:

\[ R_t^{(0)} = Y_t-A_t^{(0)}-B_t^{(0)}, \]

\noindent where $A_t^{(0)}$ and $B_t^{(0)}$ denote the matrices of sender- and receiver-specific additive effects implied by $a_{it}^{(0)}$ and $b_{jt}^{(0)}$, respectively. 

Let the rank-\(K\) truncated singular-value decomposition of \(R_t^{(0)}\) be 

\[ R_t^{(0)} \approx P_{t,K}D_{t,K}Q_{t,K}^{\top}, \]
where \(P_{t,K}\) and \(Q_{t,K}\) contain the leading \(K\) left and right singular vectors, and \(D_{t,K}\) is the diagonal matrix containing the corresponding singular values. I use the balanced factorization

\begin{align*}
U_t^{(0)} &= P_{t,K}D_{t,K}^{1/2}, \\
V_t^{(0)} &= Q_{t,K}D_{t,K}^{1/2},
\end{align*}

so that
\[ U_t^{(0)}V_t^{(0)\top} = P_{t,K}D_{t,K}Q_{t,K}^{\top}. \]

Because separate singular-value decompositions can produce arbitrary sign changes, dimension permutations, or rotations across periods, the baseline latent factors are sequentially aligned to construct a temporally coherent initial trajectory. Let 

\[ \mathbf{U}_1^{(0)*} = \mathbf{U}_1^{(0)},
\qquad
\mathbf{V}_1^{(0)*} = \mathbf{V}_1^{(0)}. \]

For \(t=2,\ldots,T\), define 

\[
\mathbf{O}_t^{(0)} =
\arg\min_{\mathbf{O}:\mathbf{O}^{\top}\mathbf{O}=\mathbf{I}_K} \left\{ \left\| \mathbf{U}_t^{(0)}\mathbf{O} -\mathbf{U}_{t-1}^{(0)*} \right\|_F^2 + \left\| \mathbf{V}_t^{(0)}\mathbf{O} - \mathbf{V}_{t-1}^{(0)*} \right\|_F^2 \right\}.
\]

The aligned factors are
\begin{align*}
\mathbf{U}_t^{(0)*} &= \mathbf{U}_t^{(0)}\mathbf{O}_t^{(0)},
&
\mathbf{V}_t^{(0)*} &= \mathbf{V}_t^{(0)}\mathbf{O}_t^{(0)}.
\end{align*}

Thus, the factors for each period are aligned with the already aligned factors from the preceding period. Applying the same orthogonal transformation to \(\mathbf{U}_t^{(0)}\) and \(\mathbf{V}_t^{(0)}\) leaves their product unchanged. This sequential alignment is used only to construct temporally coherent starting values. The resulting unperturbed trajectories are also retained as the reference orientation used later to identify the estimated latent trajectories.

To reduce sensitivity to the baseline initialization, additional candidate starting values are generated by adding independent Gaussian perturbations to the aligned latent factors. For candidate \(s=1,\ldots,S\),

\begin{align*}
U_t^{(0,s)} &= U_t^{(0)} + \Delta_{U,t}^{(s)}, \\
V_t^{(0,s)} &= V_t^{(0)} + \Delta_{V,t}^{(s)},
\end{align*}

where the elements of \(\Delta_{U,t}^{(s)}\) and
\(\Delta_{V,t}^{(s)}\) are independently drawn from \[ N(0,\eta^2). \] The perturbation scale \(\eta\) controls the magnitude of deviations from the baseline initialization. The initialization yielding the lowest penalized MAP objective is retained for the subsequent empirical Bayes updating procedure.

During the multi-start procedure, the penalty parameters $\lambda_\beta, \lambda_a, \lambda_b, \lambda_U, \lambda_V, \gamma_\beta, \gamma_a, \gamma_b, \gamma_U, \gamma_V$ are held fixed at the neutral starting values: 
\[  \lambda_\beta^{(0)} =  \lambda_a^{(0)} =  \lambda_b^{(0)} =\lambda_U^{(0)} = \lambda_V^{(0)} = \gamma_\beta^{(0)} = \gamma_a^{(0)} = \gamma_b^{(0)} = \gamma_U^{(0)} = \gamma_V^{(0)} = 1. \]

These starting values provide an initial regularization for both the initial-state priors and temporal evolution processes during the first BCD iteration. The penalty parameters are subsequently updated through the empirical Bayes variance-component updating procedure.


\subsection*{Step 2: Update \(\{\boldsymbol{\beta}_t\}_{t=1}^{T}\)}

The regression coefficients are updated as a complete trajectory block. Conditional on the remaining parameters and current variance components, the coefficient trajectory $\{\boldsymbol{\beta}_t\}_{t=1}^{T}$ is obtained by minimizing the conditional MAP objective function:

\[
\begin{aligned}
Q(\boldsymbol{\beta}) =& \sum_{t=1}^{T}\sum_{ij} \left( Y_{ijt} -  a_{it} -b_{jt} -\mathbf{x}_{ijt}^{\top}\boldsymbol{\beta}_t -\mathbf{u}_{it}^{\top}\mathbf{v}_{jt} \right)^2 \\
&+ \lambda_{\beta}\|\boldsymbol{\beta}_1\|^2 + \gamma_{\beta} \sum_{t=2}^{T} \|\boldsymbol{\beta}_t-\boldsymbol{\beta}_{t-1}\|^2 . \end{aligned} \]

Because the coefficient trajectory is updated jointly, I stack all time-specific coefficients into a single vector: 

\[ \boldsymbol{\beta} = \left(\boldsymbol{\beta}_1^{\top}, \boldsymbol{\beta}_2^{\top}, \ldots, \boldsymbol{\beta}_T^{\top} \right)^{\top}. \]

For the initial state (\(t=1\)), the first-order condition is:

\[
\begin{aligned}
\frac{\partial Q(\boldsymbol{\beta})}
{\partial\boldsymbol{\beta}_1}
={}&
-2\sum_{i=1}^{N}\sum_{j=1}^{M} \mathbf{x}_{ij1} \left( Y_{ij1} - a_{i1} - b_{j1} - \mathbf{u}_{i1}^{\top}\mathbf{v}_{j1} - \mathbf{x}_{ij1}^{\top}\boldsymbol{\beta}_1 \right) \\
&+ 2\lambda_{\beta}\boldsymbol{\beta}_1 + 2\gamma_{\beta} \left( \boldsymbol{\beta}_1 -\boldsymbol{\beta}_2 \right) = \mathbf{0}.
\end{aligned}
\]

Reorganize it gives: 

\[ \begin{aligned}
&
\left[
\sum_{i=1}^{N}\sum_{j=1}^{M} \mathbf{x}_{ij1}\mathbf{x}_{ij1}^{\top} + (\lambda_{\beta}+\gamma_{\beta})\mathbf{I}_P \right]\boldsymbol{\beta}_1 - \gamma_{\beta}\boldsymbol{\beta}_2 \\
&\qquad=
\sum_{i=1}^{N}\sum_{j=1}^{M} \mathbf{x}_{ij1} \left( Y_{ij1} - a_{i1} - b_{j1} - \mathbf{u}_{i1}^{\top}\mathbf{v}_{j1} \right).
\end{aligned} \]

For interior states (\(t=2,\ldots,T-1\)):

\[ \begin{aligned}
\frac{\partial Q(\boldsymbol{\beta})}{\partial\boldsymbol{\beta}_t} ={}&
-2\sum_{i=1}^{N}\sum_{j=1}^{M} \mathbf{x}_{ijt} \left( Y_{ijt} - a_{it} - b_{jt} - \mathbf{u}_{it}^{\top}\mathbf{v}_{jt} - \mathbf{x}_{ijt}^{\top}\boldsymbol{\beta}_t \right) \\
&+ 2\gamma_{\beta} \left( \boldsymbol{\beta}_t-\boldsymbol{\beta}_{t-1} \right) + 2\gamma_{\beta} \left( \boldsymbol{\beta}_t-\boldsymbol{\beta}_{t+1} \right) = \mathbf{0}.
\end{aligned} \]

Reorganize it gives: 

\[\begin{aligned}
& -\gamma_{\beta}\boldsymbol{\beta}_{t-1} + \left[ \sum_{i=1}^{N}\sum_{j=1}^{M} \mathbf{x}_{ijt}\mathbf{x}_{ijt}^{\top} + 2\gamma_{\beta}\mathbf{I}_P \right]\boldsymbol{\beta}_t - \gamma_{\beta}\boldsymbol{\beta}_{t+1} \\
&\qquad= \sum_{i=1}^{N}\sum_{j=1}^{M} \mathbf{x}_{ijt} \left( Y_{ijt} - a_{it} - b_{jt} - \mathbf{u}_{it}^{\top}\mathbf{v}_{jt} \right).
\end{aligned} \]

For the final state (\(t=T\)):

\[\begin{aligned}
\frac{\partial Q(\boldsymbol{\beta})}{\partial\boldsymbol{\beta}_T}
={}& -2\sum_{i=1}^{N}\sum_{j=1}^{M} \mathbf{x}_{ijT} \left( Y_{ijT} - a_{iT} - b_{jT} - \mathbf{u}_{iT}^{\top}\mathbf{v}_{jT} - \mathbf{x}_{ijT}^{\top}\boldsymbol{\beta}_T \right) \\
&+ 2\gamma_{\beta} \left( \boldsymbol{\beta}_T-\boldsymbol{\beta}_{T-1} \right) = \mathbf{0}.
\end{aligned} \]

Reorganize it gives: 

\[\begin{aligned}
& -\gamma_{\beta}\boldsymbol{\beta}_{T-1} + \left[ \sum_{i=1}^{N}\sum_{j=1}^{M} \mathbf{x}_{ijT}\mathbf{x}_{ijT}^{\top} + \gamma_{\beta}\mathbf{I}_P \right]\boldsymbol{\beta}_T \\
&\qquad= \sum_{i=1}^{N}\sum_{j=1}^{M} \mathbf{x}_{ijT} \left( Y_{ijT} - a_{iT} - b_{jT} - \mathbf{u}_{iT}^{\top}\mathbf{v}_{jT} \right).
\end{aligned} \]

Collecting these first-order conditions gives the block tridiagonal linear system:

\[
H_{\beta}
\begin{bmatrix}
\boldsymbol{\beta}_1\\
\boldsymbol{\beta}_2\\
\vdots\\
\boldsymbol{\beta}_T
\end{bmatrix} = r_{\beta}. \]

The diagonal blocks of \(H_{\beta}\) contain the observation information and temporal penalties, while the off-diagonal blocks are generated by the Gaussian random-walk prior and equal \(-\gamma_{\beta}I\). Solving this system jointly yields the updated coefficient trajectory: $\{\boldsymbol{\beta}_t\}_{t=1}^{T}.$

Because the Gaussian random-walk penalty links each state only to its immediate neighbors, the resulting linear system has a sparse block tridiagonal structure. I exploit this structure using a block tridiagonal solver rather than directly inverting the full \(TP\times TP\) coefficient matrix.

\subsection*{Step 3: Update the Type 1 Actor Additive-Effect Trajectories}

The Type 1 actor additive effects are updated as complete trajectory blocks. Conditional on the remaining parameters and current variance components, the trajectory for Type 1 actor \(i\), $\mathbf{a}_i =\left(a_{i1}, a_{i2}, \ldots, a_{iT} \right)^{\top}$, is obtained by minimizing the conditional MAP objective function:

\[\begin{aligned}
Q(\mathbf{a}_i) =& \sum_{t=1}^{T}\sum_j\left(Y_{ijt} -a_{it} -b_{jt} -\mathbf{x}_{ijt}^{\top}\boldsymbol{\beta}_t -\mathbf{u}_{it}^{\top}\mathbf{v}_{jt} \right)^2 \\
&+ \lambda_a a_{i1}^{2} + \gamma_a \sum_{t=2}^{T} \left(a_{it}-a_{i,t-1} \right)^2 . \end{aligned} \]

For the initial state (\(t=1\)), the first-order condition is

\[ \begin{aligned}
\frac{\partial Q(a_{i1})}{\partial a_{i1}} =& -2\sum_j \left( Y_{ij1} -a_{i1} -b_{j1} -\mathbf{x}_{ij1}^{\top}\boldsymbol{\beta}_1 -\mathbf{u}_{i1}^{\top}\mathbf{v}_{j1} \right) \\
&+ 2\lambda_a a_{i1} + 2\gamma_a \left( a_{i1}-a_{i2} \right)=0.\end{aligned} \]

Reorganizing terms gives:

\[\begin{aligned}
& \sum_j a_{i1} + \lambda_a a_{i1} + \gamma_a \left( a_{i1}-a_{i2} \right) \\
&= \sum_j \left( Y_{ij1}-b_{j1} -\mathbf{x}_{ij1}^{\top}\boldsymbol{\beta}_1 -\mathbf{u}_{i1}^{\top}\mathbf{v}_{j1} \right).
\end{aligned} \]

For interior states (\(t=2,\ldots,T-1\)), the first-order condition is

\[\begin{aligned}
\frac{\partial Q(a_{it})}{\partial a_{it}} =& -2\sum_j \left( Y_{ijt} -a_{it} -b_{jt} -\mathbf{x}_{ijt}^{\top}\boldsymbol{\beta}_t -\mathbf{u}_{it}^{\top}\mathbf{v}_{jt} \right)\\
&+ 2\gamma_a \left( a_{it}-a_{i,t-1} \right) + 2\gamma_a \left( a_{it}-a_{i,t+1} \right)=0.\end{aligned}\]

Reorganizing terms gives:

\[\begin{aligned}
& \sum_j a_{it} + \gamma_a \left( a_{it}-a_{i,t-1} \right) + \gamma_a \left( a_{it}-a_{i,t+1} \right) \\
&= \sum_j \left(Y_{ijt} -b_{jt} -\mathbf{x}_{ijt}^{\top}\boldsymbol{\beta}_t -\mathbf{u}_{it}^{\top}\mathbf{v}_{jt} \right). \end{aligned} \]

Equivalently,

\[
\begin{aligned}
& -\gamma_a a_{i,t-1} + \left( \sum_j 1+2\gamma_a \right)a_{it} -\gamma_a a_{i,t+1} \\
&= \sum_j \left( Y_{ijt} -b_{jt} -\mathbf{x}_{ijt}^{\top}\boldsymbol{\beta}_t -\mathbf{u}_{it}^{\top}\mathbf{v}_{jt} \right). \end{aligned} \]

For the final state (\(t=T\)), the first-order condition is

\[\begin{aligned}
\frac{\partial Q(a_{iT})}{\partial a_{iT}} 
=& -2\sum_j \left( Y_{ijT} -a_{iT} -b_{jT} -\mathbf{x}_{ijT}^{\top}\boldsymbol{\beta}_T -\mathbf{u}_{iT}^{\top}\mathbf{v}_{jT} \right) \\
&+ 2\gamma_a \left( a_{iT}-a_{i,T-1} \right) =0.\end{aligned} \]

Reorganizing terms gives:

\[
\begin{aligned}
& \sum_j a_{iT} + \gamma_a \left( a_{iT}-a_{i,T-1} \right) \\
&= \sum_j \left( Y_{ijT} -b_{jT} -\mathbf{x}_{ijT}^{\top}\boldsymbol{\beta}_T -\mathbf{u}_{iT}^{\top}\mathbf{v}_{jT} \right).
\end{aligned} \]

Equivalently,

\[\begin{aligned}
& -\gamma_a a_{i,T-1} + \left( \sum_j 1+\gamma_a \right)a_{iT} \\
&= \sum_j \left( Y_{ijT} -b_{jT} -\mathbf{x}_{ijT}^{\top}\boldsymbol{\beta}_T -\mathbf{u}_{iT}^{\top}\mathbf{v}_{jT} \right).
\end{aligned} \]

Collecting these first-order conditions gives the tridiagonal linear system

\[
\mathbf{H}_{a_i}
\begin{bmatrix}
a_{i1}\\
a_{i2}\\
\vdots\\
a_{iT}
\end{bmatrix}
=
\mathbf{r}_{a_i}.
\]

The diagonal elements of \(H_{a_i}\) contain the observation information and temporal penalties, while the off-diagonal elements are generated by the Gaussian random-walk prior and equal \(-\gamma_a\). Solving this system jointly yields the updated Type 1 actor additive-effect trajectory $\mathbf{a}_i = \left(a_{i1}, a_{i2}, \ldots, a_{iT} \right)^{\top}.$ This update is repeated for \(i=1,\ldots,N\).


\subsection*{Step 4: Update the Type 2 Actor Additive-Effect Trajectories}

The Type 2 actor additive effects are updated as complete trajectory blocks. Conditional on the remaining parameters and current variance components, the trajectory for Type 2 actor \(j\), $\mathbf{b}_j = \left(b_{j1}, b_{j2}, \ldots, b_{jT} \right)^{\top},$ is obtained by minimizing the conditional MAP objective function:

\[
\begin{aligned}
Q(\mathbf{b}_j) =& \sum_{t=1}^{T}\sum_i \left(Y_{ijt} -a_{it} -b_{jt} -\mathbf{x}_{ijt}^{\top}\boldsymbol{\beta}_t -\mathbf{u}_{it}^{\top}\mathbf{v}_{jt} \right)^2 \\
&+ \lambda_b b_{j1}^{2} + \gamma_b \sum_{t=2}^{T} \left( b_{jt}-b_{j,t-1} \right)^2 .
\end{aligned} \]

For the initial state (\(t=1\)), the first-order condition is:

\[\begin{aligned}
\frac{\partial Q(b_{j1})}{\partial b_{j1}}
=& -2\sum_i \left(Y_{ij1} -a_{i1} -b_{j1}-\mathbf{x}_{ij1}^{\top}\boldsymbol{\beta}_1 -\mathbf{u}_{i1}^{\top}\mathbf{v}_{j1} \right) \\
&+ 2\lambda_b b_{j1} + 2\gamma_b \left( b_{j1}-b_{j2} \right) =0.
\end{aligned} \]

Reorganizing terms gives:

\[ \begin{aligned}
& \left( \sum_i 1 +\lambda_b +\gamma_b \right)b_{j1} -\gamma_b b_{j2} \\
&= \sum_i \left( Y_{ij1} -a_{i1} -\mathbf{x}_{ij1}^{\top}\boldsymbol{\beta}_1 -\mathbf{u}_{i1}^{\top}\mathbf{v}_{j1} \right).
\end{aligned} \]

For interior states (\(t=2,\ldots,T-1\)), the first-order condition is:

\[\begin{aligned}
\frac{\partial Q(b_{jt})}{\partial b_{jt}}
=& -2\sum_i\left( Y_{ijt} -a_{it} -b_{jt} -\mathbf{x}_{ijt}^{\top}\boldsymbol{\beta}_t -\mathbf{u}_{it}^{\top}\mathbf{v}_{jt} \right) \\
&+ 2\gamma_b \left( b_{jt}-b_{j,t-1} \right) + 2\gamma_b \left( b_{jt}-b_{j,t+1} \right) =0.
\end{aligned} \]

Reorganizing terms gives:

\[\begin{aligned} 
& -\gamma_b b_{j,t-1} + \left( \sum_i 1 + 2\gamma_b \right)b_{jt} -\gamma_b b_{j,t+1} \\
&= \sum_i \left( Y_{ijt} -a_{it} -\mathbf{x}_{ijt}^{\top}\boldsymbol{\beta}_t -\mathbf{u}_{it}^{\top}\mathbf{v}_{jt} \right).
\end{aligned} \]

For the final state (\(t=T\)), the first-order condition is:

\[ \begin{aligned}
\frac{\partial Q(b_{jT})}{\partial b_{jT}}
=& -2\sum_i \left( Y_{ijT} -a_{iT} -b_{jT} -\mathbf{x}_{ijT}^{\top}\boldsymbol{\beta}_T -\mathbf{u}_{iT}^{\top}\mathbf{v}_{jT} \right) \\
&+ 2\gamma_b \left( b_{jT}-b_{j,T-1} \right) =0.
\end{aligned} \]

Reorganizing terms gives:

\[ \begin{aligned}
& -\gamma_b b_{j,T-1} + \left(\sum_i 1 + \gamma_b \right)b_{jT} \\
&= \sum_i \left( Y_{ijT} -a_{iT} -\mathbf{x}_{ijT}^{\top}\boldsymbol{\beta}_T -\mathbf{u}_{iT}^{\top}\mathbf{v}_{jT} \right).
\end{aligned} \]

Collecting these first-order conditions gives the tridiagonal linear system:

\[
\mathbf{H}_{b_j}
\begin{bmatrix}
b_{j1}\\
b_{j2}\\
\vdots\\
b_{jT}
\end{bmatrix}
=
\mathbf{r}_{b_j}.
\]

The diagonal elements of \(H_{b_j}\) contain the observation information and temporal penalties, whereas the off-diagonal elements are generated by the Gaussian random-walk prior and equal \(-\gamma_b\). Solving this system jointly yields the updated Type 2 additive-effect trajectory: $\mathbf{b}_j = \left( b_{j1}, b_{j2}, \ldots, b_{jT} \right)^{\top}.$ This update is repeated for \(j=1,\ldots,M\).

\subsection*{Step 5: Update the Type 1 Latent-Factor Trajectories}

The Type 1 latent factors are updated as complete trajectory blocks. Conditional on the remaining parameters and current variance components, the latent trajectory for Type 1 actor \(i\), $\mathbf{u}_i = \left(\mathbf{u}_{i1}^{\top}, \mathbf{u}_{i2}^{\top}, \ldots, \mathbf{u}_{iT}^{\top} \right)^{\top}$, is obtained by minimizing the conditional MAP objective function:

\[ \begin{aligned}
Q(\mathbf{u}_i) =& \sum_{t=1}^{T}\sum_j \left(Y_{ijt} -a_{it} -b_{jt} -\mathbf{x}_{ijt}^{\top}\boldsymbol{\beta}_t -\mathbf{u}_{it}^{\top}\mathbf{v}_{jt} \right)^2 \\
&+ \lambda_U \left\| \mathbf{u}_{i1} \right\|^2 + \gamma_U \sum_{t=2}^{T} \left\| \mathbf{u}_{it} - \mathbf{u}_{i,t-1} \right\|^2 .
\end{aligned} \]

For the initial state (\(t=1\)), the first-order condition is:

\[\begin{aligned}
\frac{\partial Q(\mathbf{u}_{i1})} {\partial \mathbf{u}_{i1}} =&
-2\sum_j \mathbf{v}_{j1} \left(Y_{ij1} -a_{i1} -b_{j1} -\mathbf{x}_{ij1}^{\top}\boldsymbol{\beta}_1 -\mathbf{u}_{i1}^{\top}\mathbf{v}_{j1} \right) \\
&+ 2\lambda_U\mathbf{u}_{i1} + 2\gamma_U \left( \mathbf{u}_{i1} - \mathbf{u}_{i2} \right) =0.
\end{aligned} \]

Reorganizing terms gives:

\[\begin{aligned}
& \left( \sum_j \mathbf{v}_{j1}\mathbf{v}_{j1}^{\top} + (\lambda_U+\gamma_U)I_K \right) \mathbf{u}_{i1} - \gamma_U\mathbf{u}_{i2} \\ 
&= \sum_j \mathbf{v}_{j1} \left( Y_{ij1} -a_{i1} -b_{j1} -\mathbf{x}_{ij1}^{\top}\boldsymbol{\beta}_1 \right).
\end{aligned} \]

For interior states (\(t=2,\ldots,T-1\)), the first-order condition is:

\[\begin{aligned}
\frac{\partial Q(\mathbf{u}_{it})} {\partial \mathbf{u}_{it}} 
=& -2\sum_j \mathbf{v}_{jt} \left( Y_{ijt} -a_{it} -b_{jt} -\mathbf{x}_{ijt}^{\top}\boldsymbol{\beta}_t -\mathbf{u}_{it}^{\top}\mathbf{v}_{jt} \right) \\
&+ 2\gamma_U \left( \mathbf{u}_{it} - \mathbf{u}_{i,t-1} \right) + 2\gamma_U \left( \mathbf{u}_{it} - \mathbf{u}_{i,t+1} \right) =0.
\end{aligned} \]

Reorganizing terms gives:

\[ \begin{aligned}
& -\gamma_U\mathbf{u}_{i,t-1} + \left(\sum_j \mathbf{v}_{jt}\mathbf{v}_{jt}^{\top} + 2\gamma_U I_K \right) \mathbf{u}_{it} - \gamma_U\mathbf{u}_{i,t+1} \\
&= \sum_j \mathbf{v}_{jt} \left(Y_{ijt} -a_{it} -b_{jt} -\mathbf{x}_{ijt}^{\top}\boldsymbol{\beta}_t \right).
\end{aligned} \]

For the final state (\(t=T\)), the first-order condition is:

\[ \begin{aligned}
\frac{\partial Q(\mathbf{u}_{iT})}{\partial \mathbf{u}_{iT}}=&
-2\sum_j \mathbf{v}_{jT} \left(Y_{ijT} -a_{iT} -b_{jT} -\mathbf{x}_{ijT}^{\top}\boldsymbol{\beta}_T -\mathbf{u}_{iT}^{\top}\mathbf{v}_{jT} \right) \\
&+ 2\gamma_U \left( \mathbf{u}_{iT} - \mathbf{u}_{i,T-1} \right) =0.
\end{aligned} \]

Reorganizing terms gives:

\[\begin{aligned}
& -\gamma_U\mathbf{u}_{i,T-1} + \left( \sum_j \mathbf{v}_{jT}\mathbf{v}_{jT}^{\top} + \gamma_U I_K \right) \mathbf{u}_{iT} \\
&= \sum_j \mathbf{v}_{jT} \left( Y_{ijT} -a_{iT} -b_{jT} -\mathbf{x}_{ijT}^{\top}\boldsymbol{\beta}_T \right).
\end{aligned} \]

Collecting these first-order conditions gives the block tridiagonal linear system:

\[
\mathbf{H}_{U_i}
\begin{bmatrix}
\mathbf{u}_{i1}\\
\mathbf{u}_{i2}\\
\vdots\\
\mathbf{u}_{iT}
\end{bmatrix}
=
\mathbf{r}_{U_i}.
\]

The diagonal blocks of \(H_{U_i}\) contain the observation information and temporal penalties, whereas the off-diagonal blocks are generated by the Gaussian random-walk prior and equal \(-\gamma_U I_K\). Solving this block tridiagonal system jointly yields the updated Type 1 latent-position trajectory: $\mathbf{u}_i = \left( \mathbf{u}_{i1}^{\top}, \mathbf{u}_{i2}^{\top}, \ldots, \mathbf{u}_{iT}^{\top} \right)^{\top}.$ This update is repeated for \(i=1,\ldots,N\).

\subsection*{Step 6: Update the Type 2 Latent-Position Trajectories}

The Type 2 latent positions are updated as complete trajectory blocks. Conditional on the remaining parameters and current variance components, the latent trajectory for Type 2 actor \(j\), $\mathbf{v}_j = \left( \mathbf{v}_{j1}^{\top}, \mathbf{v}_{j2}^{\top}, \ldots, \mathbf{v}_{jT}^{\top} \right)^{\top},$ is obtained by minimizing the conditional MAP objective function:

\[ \begin{aligned}
Q(\mathbf{v}_j) =& \sum_{t=1}^{T}\sum_i \left(Y_{ijt} -a_{it} -b_{jt} -\mathbf{x}_{ijt}^{\top}\boldsymbol{\beta}_t -\mathbf{u}_{it}^{\top}\mathbf{v}_{jt} \right)^2 \\
&+ \lambda_V \left\| \mathbf{v}_{j1} \right\|^2 + \gamma_V \sum_{t=2}^{T} \left\| \mathbf{v}_{jt} - \mathbf{v}_{j,t-1} \right\|^2 .
\end{aligned} \]

For the initial state (\(t=1\)), the first-order condition is:

\[
\begin{aligned}
\frac{\partial Q(\mathbf{v}_{j1})} {\partial \mathbf{v}_{j1}} =&
-2\sum_i \mathbf{u}_{i1} \left(Y_{ij1} -a_{i1} -b_{j1} -\mathbf{x}_{ij1}^{\top}\boldsymbol{\beta}_1 -\mathbf{u}_{i1}^{\top}\mathbf{v}_{j1} \right) \\
&+ 2\lambda_V\mathbf{v}_{j1} + 2\gamma_V \left(\mathbf{v}_{j1} -\mathbf{v}_{j2} \right) =0.
\end{aligned} \]

Reorganizing terms gives:

\[ \begin{aligned}
& \left( \sum_i \mathbf{u}_{i1}\mathbf{u}_{i1}^{\top} +(\lambda_V+\gamma_V)I_K \right) \mathbf{v}_{j1} - \gamma_V\mathbf{v}_{j2} \\
&= \sum_i \mathbf{u}_{i1} \left( Y_{ij1} -a_{i1} -b_{j1} -\mathbf{x}_{ij1}^{\top}\boldsymbol{\beta}_1 \right).
\end{aligned} \]

For interior states (\(t=2,\ldots,T-1\)), the first-order condition is:

\[ \begin{aligned}
\frac{\partial Q(\mathbf{v}_{jt})}{\partial \mathbf{v}_{jt}}
=& -2\sum_i \mathbf{u}_{it} \left( Y_{ijt} -a_{it} -b_{jt} -\mathbf{x}_{ijt}^{\top}\boldsymbol{\beta}_t -\mathbf{u}_{it}^{\top}\mathbf{v}_{jt} \right) \\
&+ 2\gamma_V \left( \mathbf{v}_{jt} - \mathbf{v}_{j,t-1} \right) + 2\gamma_V \left( \mathbf{v}_{jt} - \mathbf{v}_{j,t+1} \right) =0. 
\end{aligned} \]

Reorganizing terms gives:

\[ \begin{aligned} 
& -\gamma_V\mathbf{v}_{j,t-1} + \left( \sum_i \mathbf{u}_{it}\mathbf{u}_{it}^{\top} + 2\gamma_V I_K \right) \mathbf{v}_{jt} - \gamma_V\mathbf{v}_{j,t+1} \\
&= \sum_i \mathbf{u}_{it} \left( Y_{ijt} -a_{it} -b_{jt} -\mathbf{x}_{ijt}^{\top}\boldsymbol{\beta}_t \right).
\end{aligned} \]

For the final state (\(t=T\)), the first-order condition is:

\[\begin{aligned}
\frac{\partial Q(\mathbf{v}_{jT})}{\partial \mathbf{v}_{jT}}
=& -2\sum_i \mathbf{u}_{iT} \left( Y_{ijT} -a_{iT} -b_{jT} -\mathbf{x}_{ijT}^{\top}\boldsymbol{\beta}_T -\mathbf{u}_{iT}^{\top}\mathbf{v}_{jT} \right) \\
&+ 2\gamma_V \left( \mathbf{v}_{jT} - \mathbf{v}_{j,T-1} \right) =0.
\end{aligned} \]

Reorganizing terms gives:

\[ \begin{aligned}
& -\gamma_V\mathbf{v}_{j,T-1} + \left( \sum_i \mathbf{u}_{iT}\mathbf{u}_{iT}^{\top} + \gamma_V I_K \right) \mathbf{v}_{jT} \\
&= \sum_i \mathbf{u}_{iT} \left( Y_{ijT} -a_{iT} -b_{jT} -\mathbf{x}_{ijT}^{\top}\boldsymbol{\beta}_T \right).
\end{aligned} \]

Collecting these first-order conditions gives the block tridiagonal linear system:

\[
\mathbf{H}_{V_j}
\begin{bmatrix}
\mathbf{v}_{j1}\\
\mathbf{v}_{j2}\\
\vdots\\
\mathbf{v}_{jT}
\end{bmatrix}
=
\mathbf{r}_{V_j}.
\]

The diagonal blocks of \(H_{V_j}\) contain the observation information and temporal penalties, whereas the off-diagonal blocks are generated by the Gaussian random-walk prior and equal \(-\gamma_V I_K\). Solving this block tridiagonal system jointly yields the updated Type 2 latent-position trajectory: $\mathbf{v}_j = \left( \mathbf{v}_{j1}^{\top}, \mathbf{v}_{j2}^{\top}, \ldots, \mathbf{v}_{jT}^{\top} \right)^{\top}. $ This update is repeated for \(j=1,\ldots,M\).

\subsection*{Step 7: Whole-Trajectory Reciprocal Scale Normalization}\label{sec:indetermine}

The multiplicative component retains a reciprocal scale indeterminacy. Specifically, for any scalar \(c>0\),

\[
U_tV_t^{\top} = (cU_t) \left(\frac{1}{c}V_t\right)^{\top},
\qquad t=1,\ldots,T.
\]

The observation model identifies the interaction matrices \(U_tV_t^{\top}\), but it does not determine how the magnitude of the multiplicative component should be divided between the Type 1 and Type 2 latent factors. This reciprocal scale indeterminacy is consequential for the plug-in empirical Bayes variance-component updates. Under the transformation

\[
U_t^{*}=cU_t,
\qquad V_t^{*}=\frac{1}{c}V_t,
\]

the corresponding latent-factor variance components transform as

\[
\sigma_U^{2*}=c^2\sigma_U^2,
\qquad \tau_U^{2*}=c^2\tau_U^2,
\]

and

\[
\sigma_V^{2*}=c^{-2}\sigma_V^2,
\qquad \tau_V^{2*}=c^{-2}\tau_V^2.
\]

Thus, observationally equivalent latent-factor representations can produce arbitrarily different variance-component estimates. Without an additional scale convention, one set of latent factors may become progressively smaller while the other becomes progressively larger across outer iterations, even though the fitted multiplicative components remain unchanged.

To fix this remaining scale freedom, I normalize the complete latent trajectories before updating the latent-factor variance components. Let 

\[
S_U = \frac{1}{NKT}\sum_{t=1}^{T}\|U_t\|_F^2, 
\qquad
S_V = \frac{1}{MKT}\sum_{t=1}^{T} \|V_t\|_F^2
\]

denote the average squared coordinate magnitudes of the Type 1 and Type 2 latent trajectories, respectively. I define the reciprocal scaling factor

\[
c = \left( \frac{S_V}{S_U} \right)^{1/4},
\]

and normalize the complete latent trajectories as

\[
U_t^{\dagger} = cU_t,
\qquad V_t^{\dagger} = \frac{1}{c}V_t,
\qquad t=1,\ldots,T.
\]

The normalized latent trajectories satisfy the scale convention

\[
\frac{1}{NKT} \sum_{t=1}^{T} \|U_t^{\dagger}\|_F^2 = \frac{1}{MKT} \sum_{t=1}^{T} \|V_t^{\dagger}\|_F^2.
\]

To verify this result, note that

\[
\frac{1}{NKT} \sum_{t=1}^{T} \|U_t^{\dagger}\|_F^2 = c^2S_U,
\qquad 
\frac{1}{MKT} \sum_{t=1}^{T} \|V_t^{\dagger}\|_F^2 = c^{-2}S_V.
\]

Because

\[
c^4=\frac{S_V}{S_U},
\]

it follows that

\[
c^2S_U=c^{-2}S_V.
\]

The reciprocal normalization preserves the fitted multiplicative component at every period:

\[
U_t^{\dagger}V_t^{\dagger\top} = (cU_t) \left(\frac{1}{c}V_t \right)^{\top} = U_tV_t^{\top}.
\]

It therefore leaves the fitted values and estimated relational structure unchanged. The normalization only selects a common scale from the class of reciprocally rescaled latent-factor representations.

Importantly, this scale convention does not impose equality between the Type 1 and Type 2 variance components. The parameters

\[
\sigma_U^2,\qquad
\sigma_V^2,\qquad
\tau_U^2,\qquad
\tau_V^2
\]

remain distinct and are estimated separately from the normalized latent trajectories. The normalization fixes the arbitrary relative unit of measurement between \(U_t\) and \(V_t\), while preserving the ability of the bipartite model to represent different initial dispersion and temporal variation for the two node sets.

The normalized trajectories \(\{U_t^{\dagger},V_t^{\dagger}\}_{t=1}^{T}\) are subsequently used in the plug-in empirical Bayes variance-component updates. For notational simplicity, I suppress the superscript \(\dagger\) in the remaining steps and denote the normalized latent trajectories by \(U_t\) and
\(V_t\).

\subsection*{Step 8: Latent-Factor Identification by Common Orthogonal Alignment}

After scale normalization, I use a common orthogonal rotation to select a stable and reproducible representation of the latent space. Because the multiplicative component enters the model only through the product \(U_tV_t^{\top}\), the individual latent factors are not uniquely identified from the observation model. For example, for any invertible \(K\times K\) matrix \(A\),

\[
U_tV_t^{\top} = (U_tA)(V_tA^{-T})^{\top}.
\]

The initial-state and temporal penalties regularize the relative scale and geometry of the sender and receiver latent spaces. In particular, the penalties

\[
\lambda_U\|U_1\|_F^2 + \lambda_V\|V_1\|_F^2
\]

and

\[
\gamma_U \sum_{t=2}^{T} \|U_t-U_{t-1}\|_F^2 + \gamma_V \sum_{t=2}^{T} \|V_t-V_{t-1}\|_F^2
\]

are generally not invariant to arbitrary invertible transformations. These penalties therefore regularize the relative scale, geometry, and temporal orientation of the Type 1 and Type 2 latent-factor spaces. A common orthogonal rotation applied to all periods nevertheless remains unidentified. 
For any orthogonal matrix \(R\) satisfying

\[
R^{\top}R=I_K,
\]

the transformation

\[
U_t^{*}=U_tR,
\qquad V_t^{*}=V_tR,
\qquad t=1,\ldots,T,
\]

preserves the fitted interaction component because

\[
U_t^{*}V_t^{*\top} = U_tRR^{\top}V_t^{\top} = U_tV_t^{\top}.
\]

Applying the same orthogonal transformation to every period also preserves the initial-state penalties:

\[
\|U_1^{*}\|_F^2 = \|U_1R\|_F^2 = \|U_1\|_F^2,
\]

and

\[
\|V_1^{*}\|_F^2 = \|V_1R\|_F^2 = \|V_1\|_F^2.
\]

Similarly, the temporal penalties are unchanged because

\[
\begin{aligned}
\|U_t^{*}-U_{t-1}^{*}\|_F^2 &= \|(U_t-U_{t-1})R\|_F^2 \\
&= \|U_t-U_{t-1}\|_F^2,
\end{aligned}
\]

and

\[
\begin{aligned}
\|V_t^{*}-V_{t-1}^{*}\|_F^2 &= \|(V_t-V_{t-1})R\|_F^2 \\
&= \|V_t-V_{t-1}\|_F^2.
\end{aligned}
\]

Therefore, a common orthogonal rotation preserves both the fitted values and the complete penalized MAP objective.

To select a reproducible orientation, let \(\{\widetilde U_t,\widetilde V_t\}_{t=1}^{T}\) denote the latent trajectories obtained from Steps 5 and 6, and let \(\{U_t^{\mathrm{ref}},V_t^{\mathrm{ref}}\}_{t=1}^{T}\) denote the temporally aligned retained initialization from Step 1. I obtain a single common rotation matrix by solving

\[
R^{*} = \arg\min_{R:R^{\top}R=I_K} \sum_{t=1}^{T} \left\{ \|\widetilde U_tR-U_t^{\mathrm{ref}}\|_F^2 + \|\widetilde V_tR-V_t^{\mathrm{ref}}\|_F^2 \right\}.
\]

The identified latent trajectories are then defined as

\[
U_t^{*} = \widetilde U_tR^{*}, 
\qquad V_t^{*} = \widetilde V_tR^{*},
\qquad t=1,\ldots,T.
\]

Because the same rotation is applied to all periods,

\[
U_t^{*}V_t^{*\top} = \widetilde U_t\widetilde V_t^{\top},
\]

and the initial-state and temporal penalties remain unchanged. The alignment therefore selects a common coordinate convention without altering the estimated relational structure or the value of the MAP objective.

This procedure does not identify a substantively unique absolute orientation of the latent coordinate system. Rather, it selects one representative from a class of globally rotated but observationally equivalent solutions. Substantive interpretation therefore focuses on rotation-invariant quantities, particularly the fitted multiplicative component \(U_tV_t^{\top}\). Individual latent coordinates are interpreted only relative to the selected reference orientation.

\subsection*{Step 9: Update Variance Components}

After the inner trajectory-level BCD loop converges, the variance components are updated using an empirical Bayes procedure  based on blockwise conditional posterior moments. Let $\Omega$ denote the collection of variance components, including the observation variance, initial-state variances, and temporal innovation variances. 

\[
\Omega = \left( \sigma_\epsilon^{2}, \sigma_\beta^{2}, \sigma_a^{2}, \sigma_b^{2}, \sigma_U^{2}, \sigma_V^{2}, \tau_\beta^{2}, \tau_a^{2}, \tau_b^{2}, \tau_U^{2}, \tau_V^{2} \right)
\]

Under the default shared-latent-innovation specification, however, \(\tau_U^2=\tau_V^2=\tau_{UV}^2\).

\paragraph{Updating the observation variance.}

The observation variance is updated using the residual variance of the fitted model:

\[
\widehat{\sigma}_\epsilon^2 = \frac{1}{N_{\mathrm{obs}}} \sum_{i=1}^{N} \sum_{j=1}^{M} \sum_{t=1}^{T} \left(Y_{ijt} - a_{it} - b_{jt} -\mathbf{x}_{ijt}^{\top}\boldsymbol{\beta}_t - \mathbf{u}_{it}^{\top}\mathbf{v}_{jt} \right)^2,
\]

where $N_{\mathrm{obs}}$ is the number of observed dyad-time observations.

\paragraph{Blockwise conditional posterior moments.}

Conditional on the variance components, the complete-data distribution can be factored as

\[
p(Y,\Theta\mid\Omega) = p(Y\mid\Theta,\sigma_\epsilon^2) p(\Theta_1\mid\boldsymbol{\Sigma}_0) p(\Theta_{2:T}\mid\Theta_{1:T-1},\mathcal{T}),
\]

where \(\boldsymbol{\Sigma}_0\) collects the initial-state variance components and \(\mathcal{T}\) collects the temporal innovation variances. In an exact Expectation–Maximization formulation, the variance components would be updated by maximizing the expected complete-data log-likelihood
\[
\mathcal{Q}_{\mathrm{EB}} \left( \Omega\mid\Omega^{(r)} \right)
= E_{\Theta\mid Y,\Omega^{(r)}} \left[ \log p(Y,\Theta\mid\Omega) \right],
\]
where \(\Omega^{(r)}\) denotes the variance components from the previous outer iteration.

Because the multiplicative component prevents the joint posterior of all parameter trajectories from having a Gaussian form, I do not evaluate this joint expectation directly. Instead, the initial-state and innovation variance updates use quadratic moments calculated from the blockwise conditional posterior distributions.

Updating the remaining variance components requires accounting for uncertainty in the estimated parameter trajectories. This uncertainty is calculated after updating \(\sigma_\epsilon^2\). Holding the remaining parameter blocks fixed, the model is linear in the trajectory block being updated. Combined with the Gaussian initial-state and random-walk distributions, this gives a Gaussian conditional posterior for each trajectory block. For a generic block \(\boldsymbol{\theta}\),
\[
\boldsymbol{\theta} \mid Y,\widehat{\Theta}_{-\theta},\Omega^{(r)} \sim \mathcal{N} \left( \widehat{\boldsymbol{\theta}}, \boldsymbol{\Sigma}_{\theta} \right),
\]

where \(\widehat{\boldsymbol{\theta}}\) denotes its estimated trajectory and \(\widehat{\Theta}_{-\theta}\) denotes the current estimates of the remaining parameter blocks.

Let \(\mathbf{H}_{\theta}\) denote the tridiagonal or block-tridiagonal system matrix used in the BCD update for \(\boldsymbol{\theta}\). The corresponding conditional covariance matrix is
\[
\boldsymbol{\Sigma}_{\theta}
=
\widehat{\sigma}_\epsilon^2
\mathbf{H}_{\theta}^{-1}.
\]

The variance updates require only the diagonal and first off-diagonal blocks of \(\mathbf{H}_{\theta}^{-1}\). These blocks are computed using a selected-inverse procedure for tridiagonal or block-tridiagonal systems. The selected inversion has the same computational order as the corresponding trajectory solve and avoids forming the full inverse.

For a generic initial state \(\boldsymbol{\theta}_1\), the required conditional posterior moment is
\[
E \left[ \left\| \boldsymbol{\theta}_1 \right\|_2^2 \mid Y \right] = \left\| \widehat{\boldsymbol{\theta}}_1 \right\|_2^2 + \operatorname{tr} \left( \boldsymbol{\Sigma}_{\theta,11} \right).
\]

For a temporal innovation, the corresponding moment is
\[
\begin{aligned}
E\left[ \left\| \boldsymbol{\theta}_t-\boldsymbol{\theta}_{t-1} \right\|_2^2 \mid Y \right]
={}& \left\| \widehat{\boldsymbol{\theta}}_t - \widehat{\boldsymbol{\theta}}_{t-1} \right\|_2^2 \\
&+ \operatorname{tr} \left( \boldsymbol{\Sigma}_{\theta,tt} + \boldsymbol{\Sigma}_{\theta,t-1,t-1} - 2\boldsymbol{\Sigma}_{\theta,t,t-1} \right).
\end{aligned}
\]

The first term measures variation in the estimated trajectory, while the trace term accounts for uncertainty in its blockwise conditional posterior distribution.

\paragraph{Updating the initial-state variances.}

The initial-state variances are updated using the expected quadratic forms of their Gaussian initial-state distributions. For the coefficient trajectory,

\[
\widehat{\sigma}_\beta^2 = \frac{ \|\widehat{\boldsymbol{\beta}}_1\|_2^2 + \operatorname{tr}(\Sigma_{\beta,11}) }{P},
\]

\[
\widehat{\sigma}_a^2 = \frac{1}{N} \sum_{i=1}^{N} \left( \hat a_{i1}^{2} + \Sigma_{a,i1,i1} \right),
\]

\[
\widehat{\sigma}_b^2 = \frac{1}{M} \sum_{j=1}^{M} \left( \hat b_{j1}^{2} + \Sigma_{b,j1,j1} \right),
\]

\[
\widehat{\sigma}_U^2 = \frac{1}{NK} \sum_{i=1}^{N} \left[ \|\hat{\mathbf{u}}_{i1}\|_2^2 + \operatorname{tr}(\Sigma_{U,i1,i1})
\right], 
\]

and
\[
\widehat{\sigma}_V^2 = \frac{1}{MK} \sum_{j=1}^{M} \left[ \|\hat{\mathbf{v}}_{j1}\|_2^2 + \operatorname{tr}(\Sigma_{V,j1,j1}) \right].
\]

\paragraph{Updating temporal innovation variances.}

For each dynamic component, the temporal innovation variance is updated using the expected squared innovation under the conditional posterior distribution.

For the time-varying regression coefficients,

\[ \boldsymbol{\beta}_t|\boldsymbol{\beta}_{t-1} \sim N(\boldsymbol{\beta}_{t-1},\tau_\beta^2 I_P), \]

Equivalently, the innovation term

\[ \boldsymbol{\Delta}_{\beta,t} = \boldsymbol{\beta}_t-\boldsymbol{\beta}_{t-1} \]

follows

\[ \boldsymbol{\Delta}_{\beta,t} \sim N(0,\tau_\beta^2 I_P). \]

If the coefficient trajectory were observed, the likelihood of the innovations would be

\[ p(\boldsymbol{\beta}_{2:T}|\boldsymbol{\beta}_{1:T-1},\tau_\beta^2) = \prod_{t=2}^{T} (2\pi\tau_\beta^2)^{-P/2} \exp \left(-\frac{ \|\boldsymbol{\beta}_t-\boldsymbol{\beta}_{t-1}\|_2^2 }{2\tau_\beta^2} \right). \]

The corresponding log-likelihood is

\[
\ell(\tau_\beta^2) = -\frac{P(T-1)}{2}\log(\tau_\beta^2) - \frac{1}{2\tau_\beta^2} \sum_{t=2}^{T} \| \boldsymbol{\beta}_t-\boldsymbol{\beta}_{t-1} \|_2^2 + C. \]

Taking the derivative with respect to $\tau_\beta^2$ and setting the derivative equal to zero gives

\[ \tau_\beta^2 = \frac{1}{P(T-1)} \sum_{t=2}^{T} \| \boldsymbol{\beta}_t-\boldsymbol{\beta}_{t-1} \|_2^2 . \]

In the blockwise empirical Bayes update, however, the latent trajectories are not directly observed. Therefore, the complete-data quadratic term is replaced by its conditional expectation under the posterior distribution:

\[ \tau_\beta^2 = \frac{1}{P(T-1)} \sum_{t=2}^{T} E\left[ \| \boldsymbol{\beta}_t-\boldsymbol{\beta}_{t-1} \|_2^2 \mid Y \right]. \]

Combining the posterior mean and posterior uncertainty components gives

\[
\begin{aligned}
E\left[\| \boldsymbol{\beta}_t-\boldsymbol{\beta}_{t-1} \|_2^2 \mid Y \right] = & \| \widehat{\boldsymbol{\beta}}_t - \widehat{\boldsymbol{\beta}}_{t-1} \|_2^2 \\
& + \operatorname{tr} \left( \Sigma_{\beta,tt} + \Sigma_{\beta,t-1,t-1} - 2\Sigma_{\beta,t,t-1} \right).
\end{aligned} \]

Therefore, the blockwise empirical Bayes update for the temporal innovation variance is

\[ \begin{aligned}
\tau_\beta^2 = \frac{1}{P(T-1)} \sum_{t=2}^{T} \Big[
& \| \widehat{\boldsymbol{\beta}}_t - \widehat{\boldsymbol{\beta}}_{t-1} \|_2^2 \\
& + \operatorname{tr} (\Sigma_{\beta,tt} + \Sigma_{\beta,t-1,t-1} - 2\Sigma_{\beta,t,t-1} ) \Big].
\end{aligned} \]

For Type 1 additive effects,

\[ a_{it}|a_{i,t-1} \sim N(a_{i,t-1},\tau_a^2), \]

the update is

\[\begin{aligned}
\tau_a^2 = \frac{1}{N(T-1)} \sum_{i=1}^{N} \sum_{t=2}^{T} \Big[ & (\hat a_{it}-\hat a_{i,t-1})^2 \\
& + \Sigma_{a,it,it} + \Sigma_{a,i,t-1,i,t-1} - 2\Sigma_{a,it,i,t-1} \Big].
\end{aligned} \]

For Type 2 additive effects,

\[ b_{jt}|b_{j,t-1} \sim N(b_{j,t-1},\tau_b^2), \]

the update is

\[ \begin{aligned}
\tau_b^2 = \frac{1}{M(T-1)} \sum_{j=1}^{M} \sum_{t=2}^{T} \Big[ & (\hat b_{jt}-\hat b_{j,t-1})^2 \\
& + \Sigma_{b,jt,jt} + \Sigma_{b,j,t-1,j,t-1} - 2\Sigma_{b,jt,j,t-1} \Big].
\end{aligned} \]

For Type 1 latent positions,

\[ \mathbf{u}_{it}|\mathbf{u}_{i,t-1} \sim N(\mathbf{u}_{i,t-1},\tau_U^2 I_K), \]

the update is

\[ \begin{aligned}
\widetilde{\tau}_U^2 = \frac{1}{NK(T-1)} \sum_{i=1}^{N} \sum_{t=2}^{T} \Big[ & \| \hat{\mathbf{u}}_{it} - \hat{\mathbf{u}}_{i,t-1} \|_2^2 \\
& + \operatorname{tr} (\Sigma_{U,it,it} + \Sigma_{U,i,t-1,i,t-1} - 2\Sigma_{U,it,i,t-1} ) \Big].
\end{aligned} \]

Similarly, for Type 2 latent positions,

\[ \mathbf{v}_{jt}|\mathbf{v}_{j,t-1} \sim N(\mathbf{v}_{j,t-1},\tau_V^2 I_K), \]

the update is

\[ \begin{aligned}
\widetilde{\tau}_V^2 = \frac{1}{MK(T-1)} \sum_{j=1}^{M} \sum_{t=2}^{T} \Big[
& \| \hat{\mathbf{v}}_{jt} - \hat{\mathbf{v}}_{j,t-1} \|_2^2 \\
& + \operatorname{tr} ( \Sigma_{V,jt,jt} + \Sigma_{V,j,t-1,j,t-1} - 2\Sigma_{V,jt,j,t-1} ) \Big].
\end{aligned} \]

Under the default shared-latent-innovation specification, the two provisional latent-factor innovation variances are combined using their geometric mean:
\[
\widehat{\tau}_{UV}^2 = \sqrt{ \widetilde{\tau}_U^2 \widetilde{\tau}_V^2 }.
\]

The common estimate is then assigned to both latent-factor blocks:
\[
\widehat{\tau}_U^2 = \widehat{\tau}_V^2 = \widehat{\tau}_{UV}^2.
\]

This restriction prevents the empirical Bayes updates from allocating temporal variation arbitrarily between the two latent-factor blocks.

\paragraph{Updating the penalty parameters.}

The updated variance components are transformed into the penalty parameters used in the MAP objective:

\[
\lambda_\beta = \frac{\sigma_\epsilon^2}{\sigma_\beta^2},
\qquad \lambda_a = \frac{\sigma_\epsilon^2}{\sigma_a^2},
\qquad \lambda_b = \frac{\sigma_\epsilon^2}{\sigma_b^2}, \]

\[
\lambda_U = \frac{\sigma_\epsilon^2}{\sigma_U^2},
\qquad \lambda_V = \frac{\sigma_\epsilon^2}{\sigma_V^2}, \]

and

\[
\gamma_\beta = \frac{\sigma_\epsilon^2}{\tau_\beta^2},
\qquad \gamma_a = \frac{\sigma_\epsilon^2}{\tau_a^2},
\qquad \gamma_b = \frac{\sigma_\epsilon^2}{\tau_b^2},
\]

and, under the default shared-latent-innovation specification,
\[
\gamma_U = \gamma_V = \gamma_{UV} = \frac{\widehat{\sigma}_\epsilon^2}{\widehat{\tau}_{UV}^2}.
\]

The updated penalty parameters are then held fixed during the subsequent inner trajectory-level BCD loop.

\subsection*{Step 10: Convergence}

The inner trajectory-level BCD loop and the outer empirical Bayes variance-component updates use separate convergence criteria because the penalty parameters remain fixed within the inner loop but change across outer iterations. Within outer iteration $m$, let $k$ index the inner BCD iterations. The penalty parameters are held fixed at their current values, $\boldsymbol{\lambda}^{(m)}$ and $\boldsymbol{\gamma}^{(m)}$, while the state trajectories are updated as
\[
\Theta^{(m,0)} \rightarrow \Theta^{(m,1)} \rightarrow \Theta^{(m,2)} \rightarrow \cdots .
\]

Because each inner iteration evaluates the same penalized MAP objective, inner-loop convergence is assessed using the relative change in the objective function:
\[
\frac{ \left| Q\!\left( \Theta^{(m,k)}; \boldsymbol{\lambda}^{(m)}, \boldsymbol{\gamma}^{(m)} \right) - Q\!\left( \Theta^{(m,k-1)}; \boldsymbol{\lambda}^{(m)}, \boldsymbol{\gamma}^{(m)}
\right)
\right|}{ \left| Q\!\left( \Theta^{(m,k-1)}; \boldsymbol{\lambda}^{(m)}, \boldsymbol{\gamma}^{(m)} \right) \right| +\delta } < \epsilon_Q,
\]

where $\delta>0$ is a small numerical constant used to prevent division by zero.

After the inner loop converges, the latent trajectories are scale-normalized and aligned as described in Steps 7 and 8. The resulting state-trajectory estimate is denoted by $\widehat{\Theta}^{(m)}$. The variance components are then updated using the empirical Bayes procedure described in Step 9. Let
\[
\Omega^{(m)} = \left( \sigma_\epsilon^{2(m)}, \sigma_\beta^{2(m)}, \sigma_a^{2(m)}, \sigma_b^{2(m)}, \sigma_U^{2(m)}, \sigma_V^{2(m)}, \tau_\beta^{2(m)}, \tau_a^{2(m)}, \tau_b^{2(m)}, \tau_{UV}^{2(m)} \right) \]
denote the vector of variance-component estimates at outer iteration $m$.

The penalized objective is not compared across outer iterations because the penalty parameters change after each variance-component update. Instead, outer-loop convergence is assessed using the stability of both the variance components and the fitted values.

First, the relative change in the variance-component vector must satisfy

\[
\frac{ \left\| \Omega^{(m)}-\Omega^{(m-1)} \right\|_2 }{ \left\| \Omega^{(m-1)} \right\|_2+\delta
} < \epsilon_\Omega. \]

Second, define the fitted value at outer iteration $m$ as
\[
\widehat{Y}_{ijt}^{(m)} = a_{it}^{(m)} + b_{jt}^{(m)} + \mathbf{x}_{ijt}^{\top}\boldsymbol{\beta}_t^{(m)} + \mathbf{u}_{it}^{(m)\top}\mathbf{v}_{jt}^{(m)}. 
\]

Let $\widehat{\mathbf{Y}}^{(m)}$ denote the collection of fitted values over all observed dyad-periods. The relative change in fitted values must satisfy
\[
\frac{ \left\| \widehat{\mathbf{Y}}^{(m)} - \widehat{\mathbf{Y}}^{(m-1)} \right\|_F }{\left\| \widehat{\mathbf{Y}}^{(m-1)} \right\|_F +\delta } < \epsilon_{\mathrm{fit}}.
\]

The outer procedure terminates when both the variance-component and fitted-value criteria are satisfied. Otherwise, the updated variance components are converted into the corresponding penalty parameters, and the next outer iteration begins. By default,
\[
\epsilon_Q=10^{-5},
\qquad
\epsilon_\Omega=\epsilon_{\mathrm{fit}}=10^{-4}.
\]

Upon convergence, the procedure returns the trajectory-level MAP estimates and the empirical Bayes estimates of the variance components.

\section{Inference}\label{sec:inference}

\subsection{Full-Model Parametric Bootstrap}

To evaluate estimator performance across the entire estimation procedure, I implement a full-model parametric bootstrap. Specifically, I treat the estimated model as the data-generating process and generate repeated realizations of bipartite dynamic networks from the estimated parameters and variance components. Unlike the conditional bootstrap, this design allows the regression coefficients, additive effects, and latent factors to vary across bootstrap realizations, thereby reflecting variation generated by the fitted model. The covariates $\mathbf{x}_{ijt}$ are held at their observed values in every replication, and dyad-periods that are unobserved in the data remain unobserved in each simulated panel. For each bootstrap realization, I re-estimate the model using the complete estimation procedure, including multistart initialization, latent-factor estimation, identification, scale normalization, and variance-component updates. No warm start from the original estimate is used because the generating parameters of a replicate differ from that estimate. Replications that fail to converge are discarded, and the number of successful replications is reported.

Because each bootstrap realization is generated from its own underlying parameter trajectories, the dispersion of the resulting estimates does not isolate estimation uncertainty because it combines process variation and estimation error. Instead, the full-model parametric bootstrap provides a direct evaluation of estimator recovery. For each realization, I compare the estimated parameters with the generating parameters used to produce that realization. Estimation accuracy is summarized using mean error, which captures systematic bias, and root mean squared error (RMSE), which captures overall estimation error. These summaries are computed for the dynamic regression coefficients $\boldsymbol{\beta}_t$ and the additive effects $\mathbf{a}_t$ and $\mathbf{b}_t$.

The same evaluation is applied to the variance components governing the dynamic processes. The latent initial-state variances require additional care. Because $\mathbf{U}_t$ and $\mathbf{V}_t$ enter the observation model only through their product, the reciprocal rescaling
\[
\mathbf{U}_t \rightarrow c\mathbf{U}_t,
\qquad
\mathbf{V}_t \rightarrow c^{-1}\mathbf{V}_t
\]
leaves the fitted interaction structure unchanged while transforming $\sigma_U^2$ to $c^2\sigma_U^2$ and $\sigma_V^2$ to $c^{-2}\sigma_V^2$. The generating latent trajectories need not satisfy the estimator's scale convention, whereas the estimated trajectories do. Comparing $\sigma_U^2$ and $\sigma_V^2$ separately would therefore depend on an arbitrary allocation of scale between the two latent-factor blocks. Their product, $\sigma_U^2\sigma_V^2$, is invariant to reciprocal rescaling and is reported instead. Under the shared-latent-innovation specification, $\tau_{UV}^2$ is estimated as a single variance component. Mean errors and RMSEs are accordingly reported for
\[
\Omega^{\mathrm{inv}} = \left\{ \sigma_{\epsilon}^{2},\sigma_{\beta}^{2}, \sigma_{a}^{2}, \sigma_{b}^{2}, \tau_{\beta}^{2}, \tau_{a}^{2}, \tau_{b}^{2}, \sigma_{U}^{2}\sigma_{V}^{2}, \tau_{UV}^{2} \right\}.
\]

For the latent multiplicative component, I evaluate estimator performance at the level of the implied interaction matrix, which represents the identified relational structure of the model. For each bootstrap realization $b$, let
\[
\mathbf{M}_{t}^{(b)} = \mathbf{U}_{t}^{(b)} \mathbf{V}_{t}^{(b)\top},
\]

denote the generating interaction matrix, and let
\[
\widehat{\mathbf{M}}_{t}^{(b)} = \widehat{\mathbf{U}}_{t}^{(b)} \widehat{\mathbf{V}}_{t}^{(b)\top},
\]

denote the estimated interaction matrix obtained from the simulated network. Because the interaction matrix is invariant to rotations and reciprocal rescaling of the latent factors, these two quantities can be compared directly.

I summarize the discrepancy between the estimated and generating latent relational structures using two matrix-level measures. First, I calculate the relative Frobenius error:
\[
D_{t}^{(b)} = \frac{ \left\| \widehat{\mathbf{M}}_{t}^{(b)} - \mathbf{M}_{t}^{(b)} \right\|_{F}}{ \left\| \mathbf{M}_{t}^{(b)} \right\|_{F}
}.
\]

This measure captures the magnitude of the discrepancy between the estimated and generating interaction matrices.

Second, I calculate the matrix correlation:
\[
r_{t}^{(b)} = \operatorname{cor} \left( \operatorname{vec} \left( \widehat{\mathbf{M}}_{t}^{(b)} \right), \operatorname{vec}\left( \mathbf{M}_{t}^{(b)}\right)\right), \]

which measures similarity in the pattern of dyadic interactions. Values of $D_{t}^{(b)}$ closer to zero and $r_{t}^{(b)}$ closer to one indicate more accurate recovery of the generating relational structure. Both measures are calculated separately for each time period, and their means and standard deviations across replications are reported, allowing recovery of the dynamic relational structure to be evaluated over the full trajectory rather than only in aggregate.

\section{Simulation Details}\label{sec:Simulation}

\subsection{Common Data-Generating Process}

All three simulation studies use a common bipartite dynamic data-generating process (DGP). Unless otherwise varied, each simulated panel contains $N=80$ Type 1 actors, $M=100$ Type 2 actors, $T=5$ periods, and $P=3$ covariates. The true multiplicative component has rank $K=2$. The outcome is generated from
\begin{equation}
Y_{ijt} = \mathbf{x}_{ijt}^{\top}\boldsymbol{\beta}_t + a_{it} + b_{jt} +
\mathbf{u}_{it}^{\top}\mathbf{v}_{jt} + \epsilon_{ijt},
\qquad \epsilon_{ijt} \overset{\mathrm{iid}}{\sim} \mathcal{N}(0,\sigma_\epsilon^2).
\end{equation}

The additive effects induce dependence among observations that share a Type 1 or Type 2 actor, while the multiplicative component generates higher-order dependence across multiple dyads. The coefficients, additive effects, and latent factors follow the Gaussian random walks specified in the bipartite dynamic AME model: 
\begin{align*}
\boldsymbol{\beta}_t\mid\boldsymbol{\beta}_{t-1} &\sim \mathcal{N}\left( \boldsymbol{\beta}_{t-1}, \tau_\beta^2\mathbf{I}_P \right),
\nonumber\\
a_{it}\mid a_{i,t-1} &\sim \mathcal{N}\left( a_{i,t-1}, \tau_a^2 \right), 
&
b_{jt}\mid b_{j,t-1} &\sim \mathcal{N}\left( b_{j,t-1}, \tau_b^2 \right),
\nonumber\\
\mathbf{u}_{it}\mid\mathbf{u}_{i,t-1} &\sim \mathcal{N}\left( \mathbf{u}_{i,t-1}, \tau_U^2\mathbf{I}_K \right),
&
\mathbf{v}_{jt}\mid\mathbf{v}_{j,t-1} &\sim \mathcal{N}\left( \mathbf{v}_{j,t-1}, \tau_V^2\mathbf{I}_K \right).
\end{align*}

The period-to-period changes are independent across periods and parameter blocks and, for the additive effects and latent factors, across actors.

The DGP includes three covariates, $x_{1ijt}$, $x_{2ijt}$, and $x_{3ijt}$. I construct the first two to be correlated with the unobserved multiplicative and additive components, respectively, while $x_{3ijt}$ is independent of both. To generate these correlations, each Type 1 actor is assigned a scalar attribute $g_i$ and a $K$-dimensional attribute $\mathbf{z}_i$, while each Type 2 actor is assigned corresponding attributes $h_j$ and $\mathbf{w}_j$. The scalar attributes $g_i$ and $h_j$ provide a common source of variation for the initial additive effects and $x_{2ijt}$, thereby inducing correlation between $x_{2ijt}$ and the unobserved additive component $a_{it}+b_{jt}$. Similarly, $\mathbf{z}_i$ and $\mathbf{w}_j$ provide a common source of variation for the initial latent factors and $x_{1ijt}$, thereby inducing correlation between $x_{1ijt}$ and the multiplicative component $\mathbf{u}_{it}^{\top}\mathbf{v}_{jt}$:

\[
g_i,h_j \overset{\mathrm{iid}}{\sim} \mathcal{N}(0,1),
\qquad
\mathbf{z}_i,\mathbf{w}_j \overset{\mathrm{iid}}{\sim}
\mathcal{N}(\mathbf{0},\mathbf{I}_K).
\]

The initial additive effects are generated as
\[
a_{i1} = \sigma_a \left( 0.8g_i+0.6\xi_{a,i} \right),
\qquad
b_{j1} = \sigma_b \left( 0.8h_j+0.6\xi_{b,j} \right),
\]
where $\xi_{a,i},\xi_{b,j} \overset{\mathrm{iid}}{\sim} \mathcal{N}(0,1)$; the weights $0.8$ and $0.6$ induce the intended correlation while preserving the specified initial-state variances. Because $0.8^2+0.6^2=1$, these constructions preserve the specified marginal initial-state variances while allowing the actor attributes to account for part of the variation in the additive effects and latent factors.

Similarly, the initial latent factors are generated as
\[
\mathbf{u}_{i1} = \sigma_U \left( 0.8\mathbf{z}_i+0.6\boldsymbol{\xi}_{U,i} \right),
\qquad
\mathbf{v}_{j1} = \sigma_V \left( 0.8\mathbf{w}_j+0.6\boldsymbol{\xi}_{V,j} \right),
\]
where $\boldsymbol{\xi}_{U,i}, \boldsymbol{\xi}_{V,j} \overset{\mathrm{iid}}{\sim} \mathcal{N}(\mathbf{0},\mathbf{I}_K).$

The covariates are constructed from the same actor attributes. Define
\[
L_{ij} = \operatorname{std}\left( \mathbf{z}_i^\top\mathbf{w}_j \right),
\qquad
A_{ij} = \operatorname{std}\left( g_i+h_j \right),
\]

where $\operatorname{std}(\cdot)$ denotes standardization across all $NM$ dyads. For each dyad and period, the three covariates are generated as
\begin{align}
x_{1ijt} &= \rho_x L_{ij} + \sqrt{1-\rho_x^2}\,\zeta_{1ijt},
\nonumber\\
x_{2ijt} &= \rho_x A_{ij} + \sqrt{1-\rho_x^2}\,\zeta_{2ijt},
\nonumber\\
x_{3ijt} &= \zeta_{3ijt},
\end{align}
where

\[
\zeta_{1ijt},\zeta_{2ijt},\zeta_{3ijt}
\overset{\mathrm{iid}}{\sim}
\mathcal{N}(0,1).
\]

The first covariate shares actor attributes with the multiplicative component, and the second shares actor attributes with the additive effects. The third covariate is generated independently of both forms of unobserved dependence. The baseline setting is $\rho_x=0.40$.

Unless otherwise varied, the initial-state variances are
\[
\sigma_a^2 = \sigma_b^2 = \sigma_U^2 = \sigma_V^2 = 1,
\]
and the observation-error variance is

\[
\sigma_\epsilon^2=4.
\]

The baseline innovation variances are
\[
\tau_\beta^2=0.02^2,
\qquad
\tau_a^2=\tau_b^2=0.10^2,
\qquad
\tau_U^2=\tau_V^2=0.07^2.
\]

Thus, the DGP uses the common latent innovation variance imposed by the default bipartite dynamic AME specification. All Gaussian draws are mutually independent except for the dependence deliberately introduced through the shared actor attributes. The initial coefficient values and the settings varied across the three simulation studies are described in the corresponding sections below. All simulations run with $B = 500$ bootstrap refits.

\subsection{Simulation 1}

Table~\ref{tab:sim1_additional_stats} reports additional performance statistics for the first simulation. The columns are defined as follows:

\begin{itemize}
\item \texttt{Covariate}: $x_1$ is correlated with the multiplicative component, $x_2$ with the additive component, and $x_3$ is independent of both.
\item $N_{\mathrm{est}}$: number of coefficient estimates contributing to the row, one per replication and period.
\item \texttt{Bias}: mean error, $\operatorname{mean}(\widehat{\beta} - \beta)$.
\item \texttt{RMSE}: root mean squared error, $\sqrt{\operatorname{mean}((\widehat{\beta} - \beta)^2)}$, across replications.
\item \texttt{MC SD}: Monte Carlo standard deviation of the estimation error, $\operatorname{sd}(\widehat{\beta} - \beta)$. This is the sampling variability that a standard error is meant to estimate, so the ratio of the mean reported standard error to this column indicates whether uncertainty is understated or conservative.
\item \texttt{Coverage}: share of nominal 95\% intervals containing the true coefficient.
\item \texttt{Type I Error}: one minus coverage, that is, the share of replications in which a 5\% test of the true coefficient value is rejected. Under correct calibration this equals 0.05.
\end{itemize}

\begin{table}[htbp]
\centering
\caption{Simulation 1: Additional Performance Statistics}
\label{tab:sim1_additional_stats}

\small
\setlength{\tabcolsep}{5pt}
\renewcommand{\arraystretch}{1.05}

\begin{tabular}{llrrrrrr}
\toprule
Method & Covariate & $N_{\mathrm{est}}$ & Bias & RMSE & MC SD & Coverage & Type I Error \\
\midrule

Lagged outcome & $x_1$ & 5000 & 0.1567 & 0.1593 & 0.0287 & 0.0000 & 1.0000 \\
Node-by-year FE & $x_1$ & 5000 & 0.3611 & 0.3643 & 0.0480 & 0.0000 & 1.0000 \\
Pooled OLS & $x_1$ & 5000 & 0.3617 & 0.3646 & 0.0459 & 0.0000 & 1.0000 \\
RAMEN dynamic AME & $x_1$ & 5000 & 0.0023 & 0.0172 & 0.0170 & 0.9818 & 0.0182 \\
RAMEN dynamic AME (accelerated) & $x_1$ & 5000 & 0.0024 & 0.0172 & 0.0171 & 0.9782 & 0.0218 \\
Row/col random effects & $x_1$ & 5000 & 0.3558 & 0.3586 & 0.0446 & 0.0000 & 1.0000 \\
Static AME & $x_1$ & 5000 & 0.0172 & 0.0306 & 0.0253 & 0.8710 & 0.1290 \\
Two-way FE & $x_1$ & 5000 & 0.3557 & 0.3585 & 0.0446 & 0.0000 & 1.0000 \\

\midrule

Lagged outcome & $x_2$ & 5000 & 0.3005 & 0.3019 & 0.0293 & 0.0000 & 1.0000 \\
Node-by-year FE & $x_2$ & 5000 & -0.0002 & 0.0301 & 0.0302 & 0.9514 & 0.0486 \\
Pooled OLS & $x_2$ & 5000 & 0.4478 & 0.4497 & 0.0408 & 0.0000 & 1.0000 \\
RAMEN dynamic AME & $x_2$ & 5000 & 0.0048 & 0.0181 & 0.0175 & 0.9698 & 0.0302 \\
RAMEN dynamic AME (accelerated) & $x_2$ & 5000 & 0.0048 & 0.0182 & 0.0175 & 0.9638 & 0.0362 \\
Row/col random effects & $x_2$ & 5000 & 0.0067 & 0.0239 & 0.0229 & 0.7262 & 0.2738 \\
Static AME & $x_2$ & 5000 & 0.0235 & 0.0348 & 0.0256 & 0.7392 & 0.2608 \\
Two-way FE & $x_2$ & 5000 & -0.0002 & 0.0229 & 0.0229 & 0.7512 & 0.2488 \\

\midrule

Lagged outcome & $x_3$ & 5000 & 0.0001 & 0.0228 & 0.0228 & 0.7760 & 0.2240 \\
Node-by-year FE & $x_3$ & 5000 & 0.0002 & 0.0272 & 0.0272 & 0.9518 & 0.0482 \\
Pooled OLS & $x_3$ & 5000 & 0.0000 & 0.0225 & 0.0225 & 0.7792 & 0.2208 \\
RAMEN dynamic AME & $x_3$ & 5000 & -0.0002 & 0.0164 & 0.0164 & 0.9872 & 0.0128 \\
RAMEN dynamic AME (accelerated) & $x_3$ & 5000 & -0.0002 & 0.0165 & 0.0165 & 0.9820 & 0.0180 \\
Row/col random effects & $x_3$ & 5000 & 0.0002 & 0.0213 & 0.0213 & 0.7388 & 0.2612 \\
Static AME & $x_3$ & 5000 & -0.0006 & 0.0230 & 0.0230 & 0.9882 & 0.0118 \\
Two-way FE & $x_3$ & 5000 & 0.0002 & 0.0213 & 0.0213 & 0.7384 & 0.2616 \\

\bottomrule
\end{tabular}

\end{table}

\newpage
\subsection{Simulation 2}

\begin{figure}[!htbp]
    \centering
    \caption{Out-of-Sample Prediction under Temporal Dependence with Error Bars}
    \includegraphics[width=14cm]{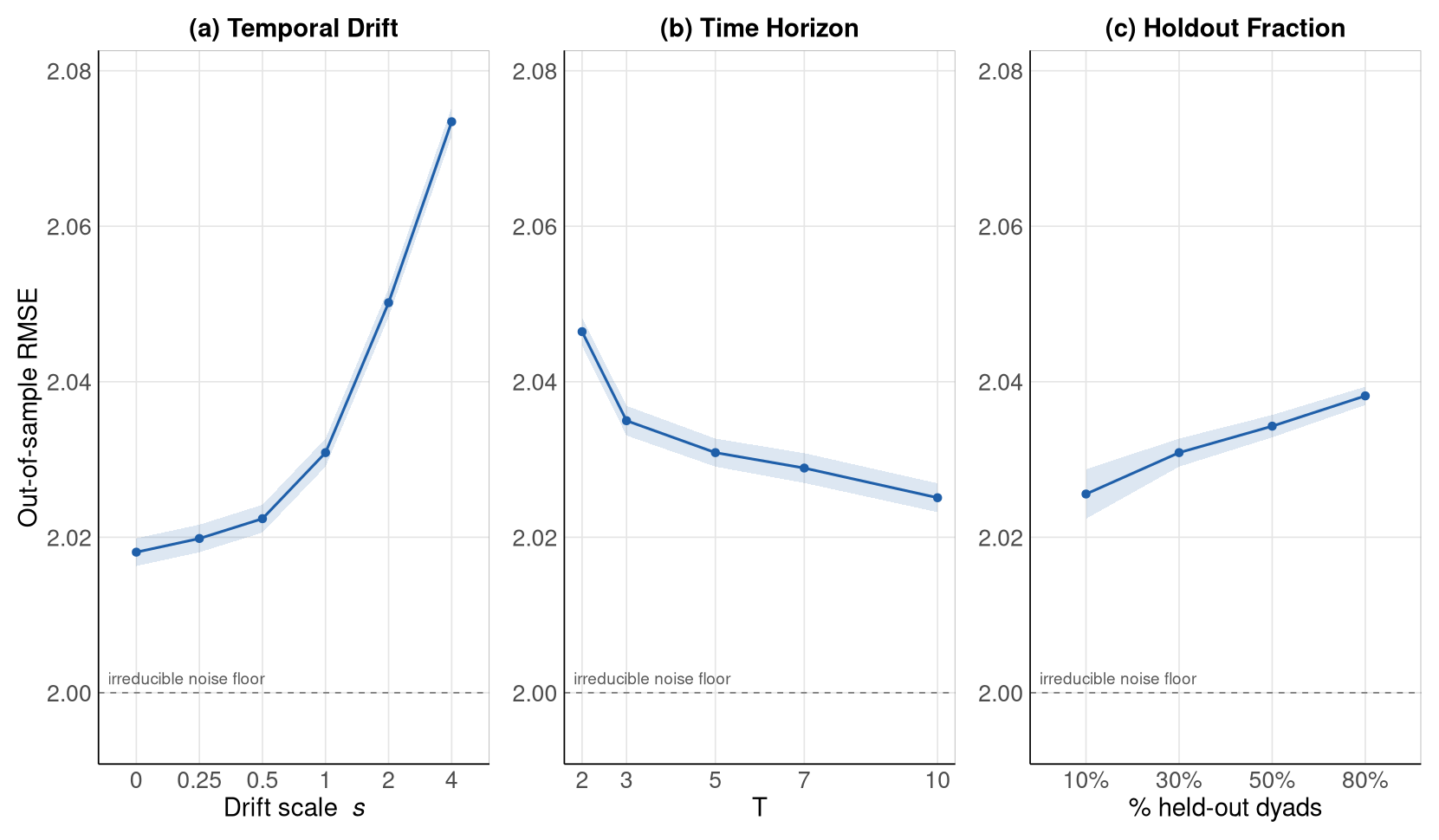}
     \label{fig:simulation_tempo2}
\end{figure}

Figure 4 in the manuscript includes error bars that are difficult to distinguish because of the scale of the y-axis. Figure~\ref{fig:simulation_tempo2} therefore reproduces the out-of-sample RMSE results with more visible error bars.

\newpage
Table~\ref{tab:sim2_additional_stats} reports additional performance statistics for the second simulation. The columns are defined as follows:

\begin{itemize}
\item \texttt{Method}: estimator, either the bipartite dynamic AME model, its accelerated variant, the static bipartite AME model fitted separately to each period, or the lagged-outcome model.
\item $\tau$ \texttt{Scale}: temporal drift scale $s$, which multiplies every innovation standard deviation, $\tau_\theta(s) = s\,\tau_\theta$. Thus $s = 0$ holds the trajectories constant and $s = 1$ is the baseline data-generating process.
\item $T$: number of periods in the simulated panel.
\item \texttt{Holdout}: share of final-period dyads held out as the test set.
\item $N_{\mathrm{rep}}$: number of Monte Carlo replications contributing to the row.
\item \texttt{RMSE}: out-of-sample root mean squared error against the observed outcome $Y_{ijt}$ in the held-out cells. It includes the idiosyncratic error and therefore cannot fall below $\sigma_\epsilon = 2$.
\item \texttt{Signal RMSE}: out-of-sample root mean squared error against the systematic component $\mathbf{x}_{ijt}^{\top}\boldsymbol{\beta}_t + a_{it} + b_{jt} + \mathbf{u}_{it}^{\top}\mathbf{v}_{jt}$, which excludes the idiosyncratic error and has no such floor. This is the sharper measure of how well the data-generating process is recovered.
\item \texttt{Signal Cor.}: correlation between the predicted values and that systematic component in the held-out cells.
\item \texttt{Convergence}: share of fits in which the estimator met its convergence criteria.
\end{itemize}

\scriptsize
\setlength{\tabcolsep}{4pt}
\renewcommand{\arraystretch}{1.05}

\begin{longtable}{lrrrrrrrr}
\caption{Simulation 2: Out-of-Sample Prediction}
\label{tab:sim2_additional_stats}\\

\toprule
Method &
$\tau$ Scale &
$T$ &
Holdout &
$N_{\mathrm{rep}}$ &
RMSE &
Signal RMSE &
Signal Cor. &
Convergence \\
\midrule
\endfirsthead

\multicolumn{9}{c}%
{{\tablename\ \thetable{} -- continued from previous page}}\\
\toprule
Method &
$\tau$ Scale &
$T$ &
Holdout &
$N_{\mathrm{rep}}$ &
RMSE &
Signal RMSE &
Signal Cor. &
Convergence \\
\midrule
\endhead

\midrule
\multicolumn{9}{r}{Continued on next page}\\
\endfoot

\bottomrule
\endlastfoot

RAMEN dynamic AME & 0 & 5 & 0.3 & 1000 & 2.0153 & 0.2478 & 0.9941 & 0.709 \\
RAMEN dynamic AME (accelerated) & 0 & 5 & 0.3 & 1000 & 2.0148 & 0.2432 & 0.9943 & 1.000 \\
Static AME & 0 & 5 & 0.3 & 1000 & 2.1042 & 0.6527 & 0.9584 & 1.000 \\
Lagged outcome & 0 & 5 & 0.3 & 1000 & 2.4741 & 1.4558 & 0.7676 & NA \\

\midrule

RAMEN dynamic AME & 0.25 & 5 & 0.3 & 1000 & 2.0170 & 0.2535 & 0.9938 & 0.735 \\
RAMEN dynamic AME (accelerated) & 0.25 & 5 & 0.3 & 1000 & 2.0165 & 0.2496 & 0.9940 & 1.000 \\
Static AME & 0.25 & 5 & 0.3 & 1000 & 2.1045 & 0.6523 & 0.9586 & 1.000 \\
Lagged outcome & 0.25 & 5 & 0.3 & 1000 & 2.4762 & 1.4574 & 0.7678 & NA \\

\midrule

RAMEN dynamic AME & 0.5 & 5 & 0.3 & 1000 & 2.0192 & 0.2703 & 0.9930 & 0.846 \\
RAMEN dynamic AME (accelerated) & 0.5 & 5 & 0.3 & 1000 & 2.0189 & 0.2679 & 0.9931 & 1.000 \\
Static AME & 0.5 & 5 & 0.3 & 1000 & 2.1045 & 0.6523 & 0.9588 & 1.000 \\
Lagged outcome & 0.5 & 5 & 0.3 & 1000 & 2.4794 & 1.4628 & 0.7675 & NA \\

\midrule

RAMEN dynamic AME & 1 & 2 & 0.3 & 1000 & 2.0448 & 0.4254 & 0.9826 & 1.000 \\
RAMEN dynamic AME (accelerated) & 1 & 2 & 0.3 & 1000 & 2.0446 & 0.4244 & 0.9827 & 1.000 \\
Static AME & 1 & 2 & 0.3 & 1000 & 2.1041 & 0.6531 & 0.9587 & 1.000 \\
Lagged outcome & 1 & 2 & 0.3 & 1000 & 2.4830 & 1.4711 & 0.7646 & NA \\

\midrule

RAMEN dynamic AME & 1 & 3 & 0.3 & 1000 & 2.0319 & 0.3605 & 0.9876 & 1.000 \\
RAMEN dynamic AME (accelerated) & 1 & 3 & 0.3 & 1000 & 2.0317 & 0.3596 & 0.9876 & 1.000 \\
Static AME & 1 & 3 & 0.3 & 1000 & 2.1030 & 0.6509 & 0.9593 & 1.000 \\
Lagged outcome & 1 & 3 & 0.3 & 1000 & 2.4848 & 1.4743 & 0.7652 & NA \\

\midrule

RAMEN dynamic AME & 1 & 5 & 0.1 & 1000 & 2.0227 & 0.3106 & 0.9909 & 0.997 \\
RAMEN dynamic AME (accelerated) & 1 & 5 & 0.1 & 1000 & 2.0226 & 0.3102 & 0.9910 & 0.999 \\
Static AME & 1 & 5 & 0.1 & 1000 & 2.0775 & 0.5678 & 0.9695 & 1.000 \\
Lagged outcome & 1 & 5 & 0.1 & 1000 & 2.4899 & 1.4829 & 0.7663 & NA \\

\midrule

RAMEN dynamic AME & 1 & 5 & 0.3 & 1000 & 2.0266 & 0.3216 & 0.9903 & 0.997 \\
RAMEN dynamic AME (accelerated) & 1 & 5 & 0.3 & 1000 & 2.0266 & 0.3212 & 0.9903 & 0.999 \\
Static AME & 1 & 5 & 0.3 & 1000 & 2.1046 & 0.6522 & 0.9598 & 1.000 \\
Lagged outcome & 1 & 5 & 0.3 & 1000 & 2.4919 & 1.4840 & 0.7660 & NA \\

\midrule

RAMEN dynamic AME & 1 & 5 & 0.5 & 1000 & 2.0286 & 0.3343 & 0.9895 & 0.996 \\
RAMEN dynamic AME (accelerated) & 1 & 5 & 0.5 & 1000 & 2.0285 & 0.3338 & 0.9895 & 1.000 \\
Static AME & 1 & 5 & 0.5 & 1000 & 2.1517 & 0.7918 & 0.9407 & 1.000 \\
Lagged outcome & 1 & 5 & 0.5 & 1000 & 2.4922 & 1.4846 & 0.7660 & NA \\

\midrule

RAMEN dynamic AME & 1 & 5 & 0.8 & 1000 & 2.0318 & 0.3558 & 0.9881 & 0.953 \\
RAMEN dynamic AME (accelerated) & 1 & 5 & 0.8 & 1000 & 2.0316 & 0.3552 & 0.9882 & 1.000 \\
Static AME & 1 & 5 & 0.8 & 1000 & 2.5417 & 1.5648 & 0.7811 & 0.986 \\
Lagged outcome & 1 & 5 & 0.8 & 1000 & 2.4928 & 1.4862 & 0.7662 & NA \\

\midrule

RAMEN dynamic AME & 1 & 7 & 0.3 & 1000 & 2.0253 & 0.3130 & 0.9909 & 0.985 \\
RAMEN dynamic AME (accelerated) & 1 & 7 & 0.3 & 1000 & 2.0252 & 0.3127 & 0.9910 & 0.995 \\
Static AME & 1 & 7 & 0.3 & 1000 & 2.1053 & 0.6522 & 0.9603 & 1.000 \\
Lagged outcome & 1 & 7 & 0.3 & 1000 & 2.4968 & 1.4935 & 0.7665 & NA \\

\midrule

RAMEN dynamic AME & 1 & 10 & 0.3 & 1000 & 2.0231 & 0.3104 & 0.9913 & 0.949 \\
RAMEN dynamic AME (accelerated) & 1 & 10 & 0.3 & 1000 & 2.0231 & 0.3102 & 0.9913 & 0.972 \\
Static AME & 1 & 10 & 0.3 & 1000 & 2.1021 & 0.6518 & 0.9614 & 0.999 \\
Lagged outcome & 1 & 10 & 0.3 & 1000 & 2.5035 & 1.5070 & 0.7682 & NA \\

\midrule

RAMEN dynamic AME & 2 & 5 & 0.3 & 1000 & 2.0447 & 0.4211 & 0.9847 & 0.995 \\
RAMEN dynamic AME (accelerated) & 2 & 5 & 0.3 & 1000 & 2.0447 & 0.4210 & 0.9847 & 0.998 \\
Static AME & 2 & 5 & 0.3 & 1000 & 2.1048 & 0.6524 & 0.9632 & 1.000 \\
Lagged outcome & 2 & 5 & 0.3 & 1000 & 2.5411 & 1.5654 & 0.7614 & NA \\

\midrule

RAMEN dynamic AME & 4 & 5 & 0.3 & 1000 & 2.0733 & 0.5423 & 0.9814 & 0.993 \\
RAMEN dynamic AME (accelerated) & 4 & 5 & 0.3 & 1000 & 2.0733 & 0.5423 & 0.9814 & 0.992 \\
Static AME & 4 & 5 & 0.3 & 1000 & 2.1053 & 0.6531 & 0.9730 & 1.000 \\
Lagged outcome & 4 & 5 & 0.3 & 1000 & 2.7328 & 1.8605 & 0.7530 & NA \\

\end{longtable}
\normalsize

\subsection{Simulation 3}

Table~\ref{tab:sim3_additional_stats} reports additional performance statistics for the third simulation. The columns are defined as follows:

\begin{itemize}
\item \texttt{cell}: design-cell label, identifying the combination of fitted rank and latent variance.
\item \texttt{K\_fit}: latent dimension supplied to the estimator. The true rank is $K = 2$.
\item \texttt{latvar}: latent variance $\ell = K\sigma_U^2\sigma_V^2$, the variance of the multiplicative term $\mathbf{u}_{it}^{\top}\mathbf{v}_{jt}$. The observation error variance is fixed at $\sigma_\epsilon^2 = 4$.
\item \texttt{method}: estimator, either the bipartite dynamic AME model or the static bipartite AME model fitted separately to each period.
\item \texttt{n\_rep}: number of Monte Carlo replications contributing to the row.
\item \texttt{bias\_cell}: mean cell-level error, $\operatorname{mean}(\widehat{\mathbf{M}}_t - \mathbf{M}_t)$, where $\mathbf{M}_t = \mathbf{U}_t\mathbf{V}_t^{\top}$. Measures systematic over- or under-estimation; positive and negative errors offset.
\item \texttt{rmse\_cell}: root mean squared cell-level error, in the units of the outcome.
\item \texttt{frob\_rel}: relative Frobenius error, $\|\widehat{\mathbf{M}}_t - \mathbf{M}_t\|_F / \|\mathbf{M}_t\|_F$. Charges the estimator for structure it cannot represent.
\item \texttt{angle\_deg}: largest principal angle, in degrees, between the estimated and generating latent subspaces. Ignores magnitude and measures only whether the structure is correctly oriented.
\item \texttt{corr}: correlation between the cells of $\widehat{\mathbf{M}}_t$ and $\mathbf{M}_t$. Insensitive to overall scale.
\end{itemize}

\begin{table}[htbp]
\centering
\caption{Simulation 3: Latent-Structure Recovery}
\label{tab:sim3_additional_stats}

\small
\setlength{\tabcolsep}{4pt}
\renewcommand{\arraystretch}{1.05}

\resizebox{\textwidth}{!}{%
\begin{tabular}{rrlrrrrrr}
\toprule
$K_{\mathrm{fit}}$ &
Latent Variance &
Method &
$N_{\mathrm{rep}}$ &
Bias &
Cell RMSE &
Relative Frobenius Error &
Principal Angle ($^\circ$) &
Correlation \\
\midrule

1 & 2.0 & RAMEN dynamic AME
& 1000 & -0.0002 & 0.9322 & 0.6538 & 9.6331 & 0.7553 \\
1 & 2.0 & RAMEN dynamic AME (accelerated)
& 1000 & -0.0002 & 0.9329 & 0.6543 & 9.7446 & 0.7549 \\
1 & 2.0 & Static AME
& 1000 & -0.0002 & 0.9690 & 0.6799 & 14.9837 & 0.7317 \\

\midrule

2 & 0.5 & RAMEN dynamic AME
& 1000 & -0.0001 & 0.2226 & 0.3135 & 14.1113 & 0.9504 \\
2 & 0.5 & RAMEN dynamic AME (accelerated)
& 1000 & -0.0001 & 0.2212 & 0.3115 & 14.0563 & 0.9509 \\
2 & 0.5 & Static AME
& 1000 & 0.0000 & 0.4427 & 0.6242 & 28.4758 & 0.8072 \\

\midrule

2 & 1.0 & RAMEN dynamic AME
& 1000 & -0.0001 & 0.2419 & 0.2408 & 10.9674 & 0.9706 \\
2 & 1.0 & RAMEN dynamic AME (accelerated)
& 1000 & -0.0001 & 0.2410 & 0.2399 & 10.9608 & 0.9708 \\
2 & 1.0 & Static AME
& 1000 & -0.0001 & 0.4428 & 0.4413 & 19.7229 & 0.9019 \\

\midrule

2 & 2.0 & RAMEN dynamic AME
& 1000 & -0.0001 & 0.2667 & 0.1877 & 8.6279 & 0.9821 \\
2 & 2.0 & RAMEN dynamic AME (accelerated)
& 1000 & -0.0002 & 0.2676 & 0.1884 & 8.7009 & 0.9819 \\
2 & 2.0 & Static AME
& 1000 & -0.0001 & 0.4494 & 0.3167 & 14.0222 & 0.9494 \\

\midrule

2 & 4.0 & RAMEN dynamic AME
& 1000 & -0.0003 & 0.2983 & 0.1485 & 6.8491 & 0.9888 \\
2 & 4.0 & RAMEN dynamic AME (accelerated)
& 1000 & -0.0004 & 0.3008 & 0.1497 & 6.9287 & 0.9886 \\
2 & 4.0 & Static AME
& 1000 & -0.0002 & 0.4573 & 0.2279 & 10.0521 & 0.9738 \\

\midrule

3 & 2.0 & RAMEN dynamic AME
& 1000 & -0.0001 & 0.3218 & 0.2266 & 8.6328 & 0.9741 \\
3 & 2.0 & RAMEN dynamic AME (accelerated)
& 1000 & -0.0001 & 0.3206 & 0.2258 & 8.6194 & 0.9743 \\
3 & 2.0 & Static AME
& 1000 & -0.0002 & 0.5760 & 0.4058 & 14.0548 & 0.9189 \\

\midrule

4 & 2.0 & RAMEN dynamic AME
& 1000 & -0.0002 & 0.3682 & 0.2592 & 8.7476 & 0.9663 \\
4 & 2.0 & RAMEN dynamic AME (accelerated)
& 1000 & -0.0002 & 0.3639 & 0.2563 & 8.7225 & 0.9670 \\
4 & 2.0 & Static AME
& 1000 & -0.0001 & 0.6672 & 0.4701 & 14.0887 & 0.8942 \\

\bottomrule
\end{tabular}%
}

\end{table}

\newpage
\section{Supplemental Simulation}

I compare the computation time of the standard estimation procedure with that of the SQUAREM-accelerated procedure. This simulation uses the same DGP as the three simulations reported in the main manuscript, with $N = 80$, $M = 100$, $K = 2$, $P = 3$, and $\rho_x = 0.4$. I vary $T \in \{2, 5, 10, 20\}$ over 10 replications, with $B = 500$ bootstrap refits.

\begin{figure}[ht]
    \centering
    \caption{Comparing Computation Time: Standard Estimation vs. SQUAREM Acceleration}
    \includegraphics[width=14cm]{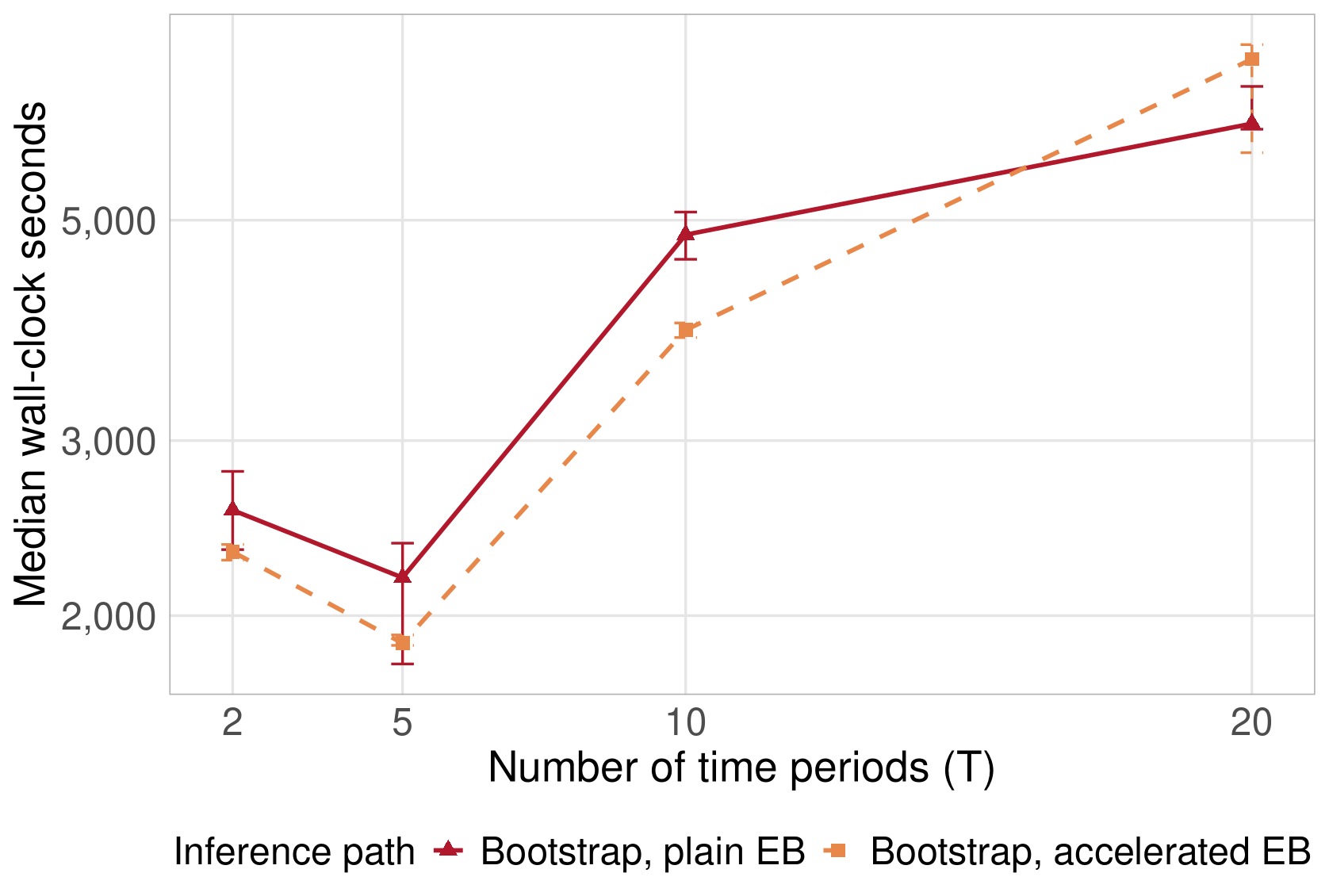}
     \label{fig:simulation_compute}
\end{figure}

Figure~\ref{fig:simulation_compute} reports the results. SQUAREM acceleration generally reduces computation time, with the largest gains at moderate values of $T$; however, this advantage diminishes as $T$ increases and disappears at $T = 20$. SQUAREM accelerates the outer loop by extrapolating in the direction of its current iterations, which is most beneficial when many iterations remain. Because each bootstrap replicate is warm-started at the original estimate and typically requires only a few outer-loop iterations, the fixed computational cost of extrapolation can outweigh its savings when $T = 20$, where bootstrap refits account for most of the total computation time.

\newpage
\section{Empirical Evaluation}

\subsection{Construct the global production network}\label{sec:production_network}

I use international trade data to study countries’ engagement in global production networks for strategically important products. The construction of the production network proceeds in two steps: (1) identifying benchmark production chains, and (2) assembling the corresponding global production network.

\subsection*{Step 1a: construct the production chain (benchmark)}

\begin{itemize}
    \item[1.] I use the input-output table from the U.S. Bureau of Economic Analysis (BEA) to construct the production chain.\footnote{https://www.bea.gov/industry/input-output-accounts-data, accessed 18th Jan, 2026.} Steps 2–4 follow the standard procedure for constructing the production network following \citet{acemoglu2012network}.

    \item[2.] The output table reports the value of commodity $k$ produced by industry $i$, capturing the linkage from industry $i$ to commodity $k$. I measure this linkage as: 
    \begin{align*}
        D_{i,k} = \frac{M_{i,k}}{\sum_i M_{i, k}},
    \end{align*}
    where $D_{i,k}$ denotes the share of total production of commodity $k$ accounted for by industry $i$, and the denominator is the total production of commodity $k$ across all industries.

    \item[3.] The input table reports the value of commodity $k$ used by industry $j$, capturing the linkage from commodity $k$ to industry $j$. I measure this linkage as:
    \begin{align*}
        B_{k, j} = \frac{U_{k, j}}{\sum_k U_{k, j}},
    \end{align*}
    where $B_{k,j}$ denotes the share of total inputs of industry $j$ accounted for by commodity $k$, and the denominator is the total value of inputs used by industry $j$ across all commodities.

    \item[4.] Merging the output and input tables by commodity $k$, I construct industry-to-industry linkages. The tie from industry $i$ to industry $j$ is defined as:
    \begin{align*}
        A_{i, j} = \sum_k D_{i, k} B_{k, j}
    \end{align*}
    where $A_{i,j}$ measures the extent to which industry $i$ contributes to the production of industry $j$ through intermediate inputs.

    \item[5.] To identify economically meaningful supply chains, I exclude service sectors to prevent chains from becoming excessively long or dominated by downstream service activities (e.g., pet food manufacturing $\rightarrow$ veterinary services).

    \item[6.] Using the remaining industries, I construct a directed production network with edge weights given by $A_{i,j}$.
\end{itemize}

\subsection*{Step 1b: identify the production chain (benchmark)}

\begin{figure}[!htbp]
    \centering
    \caption{Main Production Chain in Shipbuilding}
    \includegraphics[width=8cm]{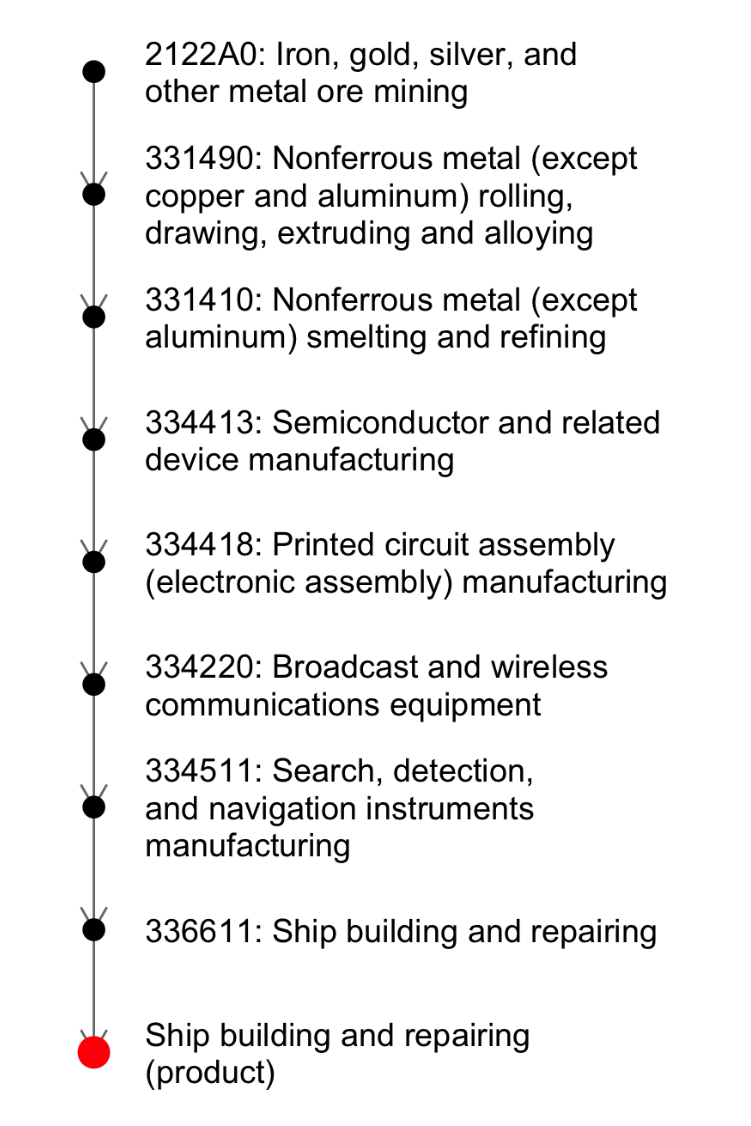}
    \caption*{\footnotesize Note: The figure illustrates only the identified main production stream. }
    \label{fig:ship_chain}
\end{figure}

\begin{figure}[!htbp]
    \centering
    \caption{The distribution of the longest shortest path and $A_{i, j}$ (top\_k = 1)}
    \label{fig:distribution1}
    \begin{minipage}{0.48\linewidth}
        \centering
        \includegraphics[width=\linewidth]{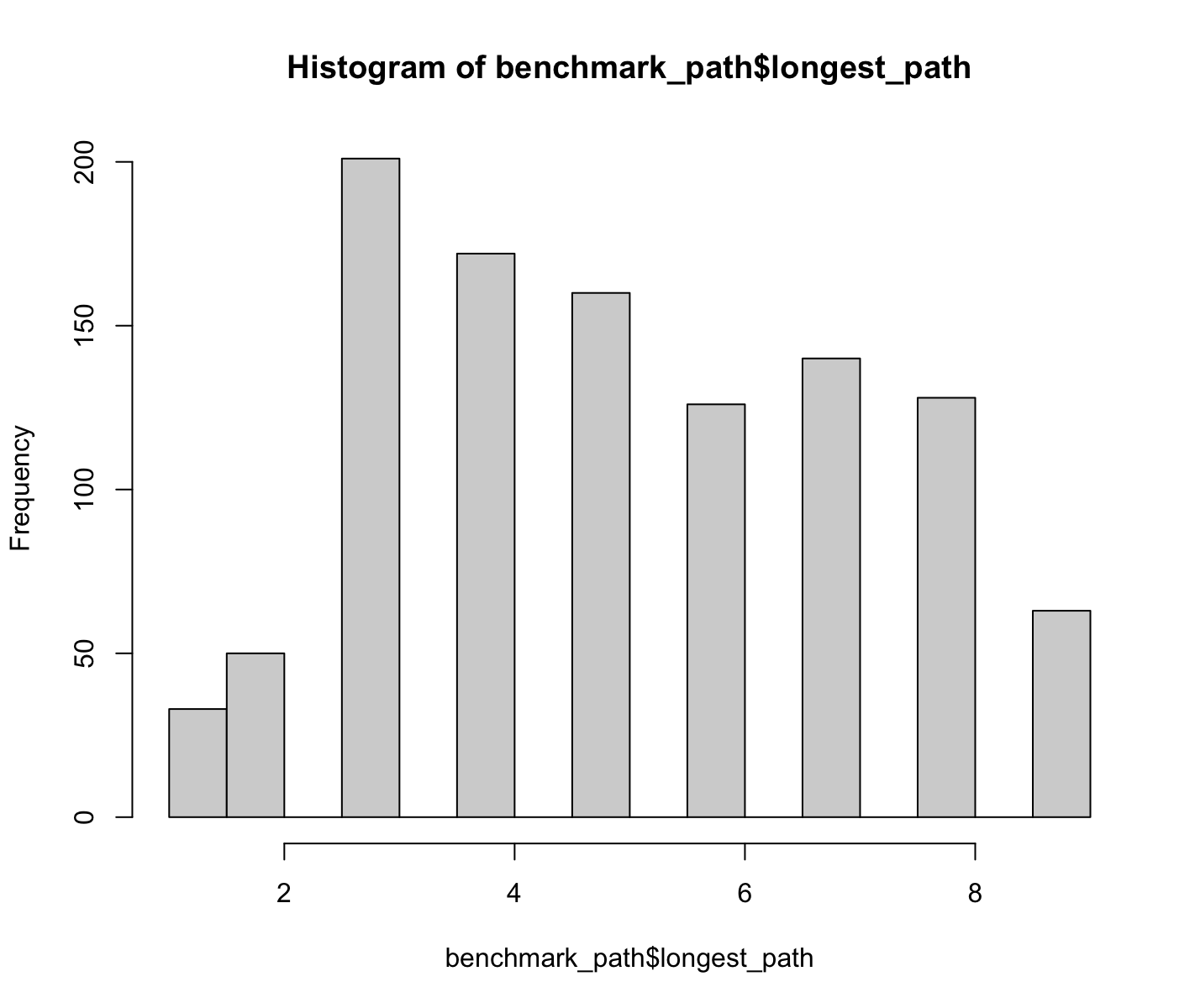}
        \subcaption{The distribution of the longest shortest path}
    \end{minipage}
    \hfill
    \begin{minipage}{0.48\linewidth}
        \centering
        \includegraphics[width=\linewidth]{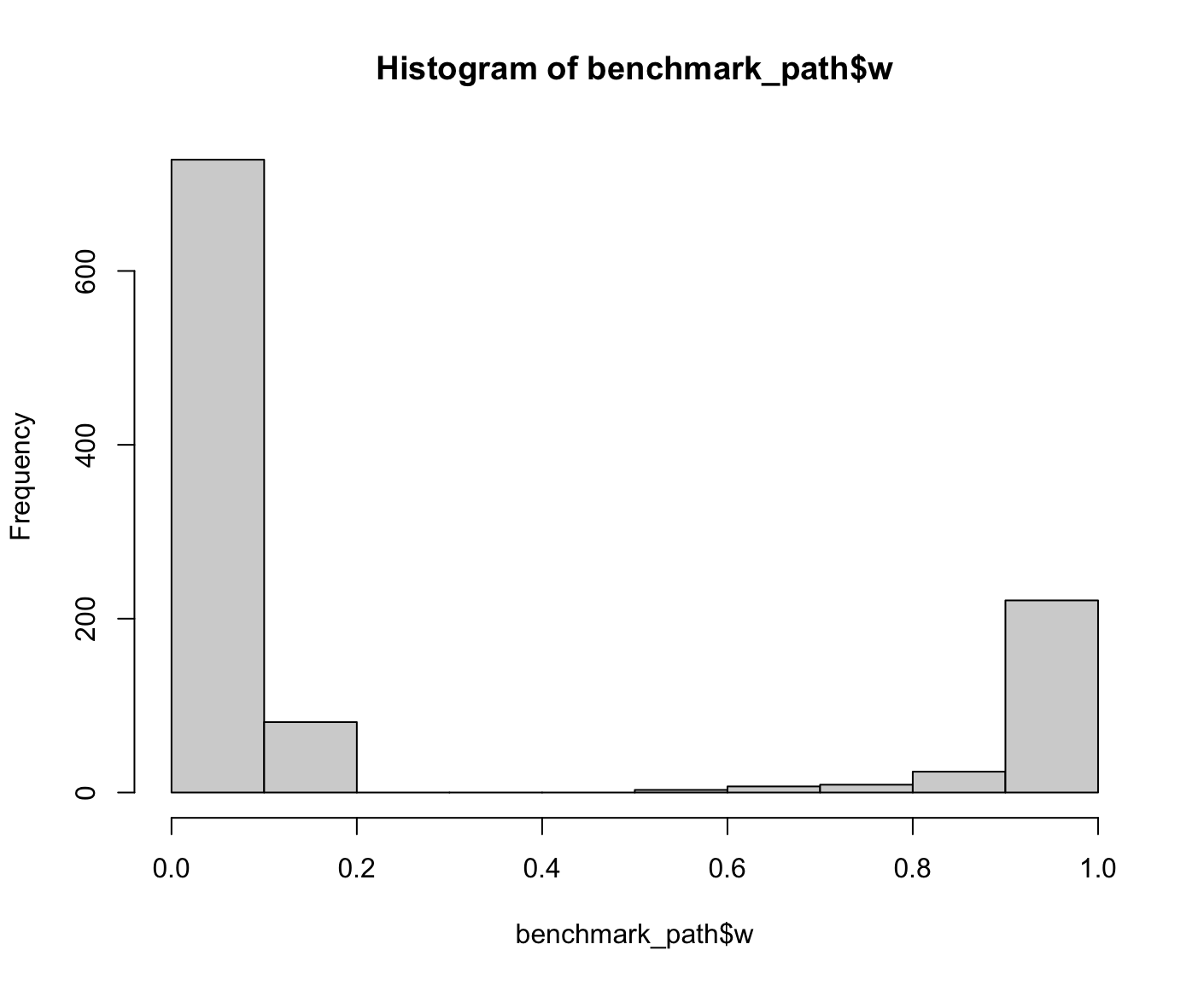}
        \subcaption{Distribution of Upstream Input Shares}
        
    \end{minipage}
\end{figure}

\begin{figure}[!htbp]
    \centering
    \caption{The distribution of the longest shortest path and $A_{i, j}$ (top\_k = 2)}
     \label{fig:distribution2}
    \begin{minipage}{0.48\linewidth}
        \centering
        \includegraphics[width=\linewidth]{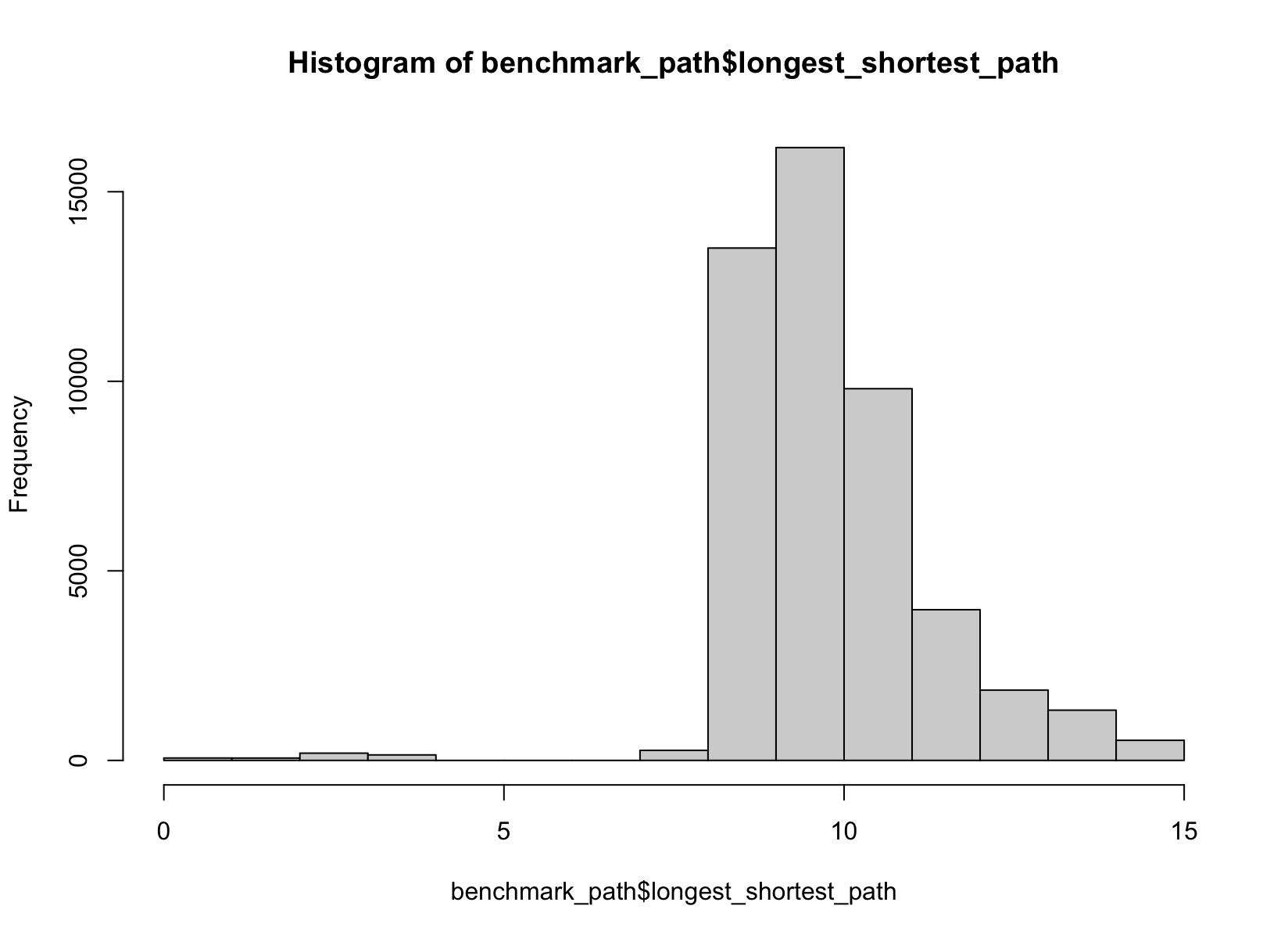}
        \subcaption{The distribution of the longest shortest path}
    \end{minipage}
    \hfill
    \begin{minipage}{0.48\linewidth}
        \centering
        \includegraphics[width=\linewidth]{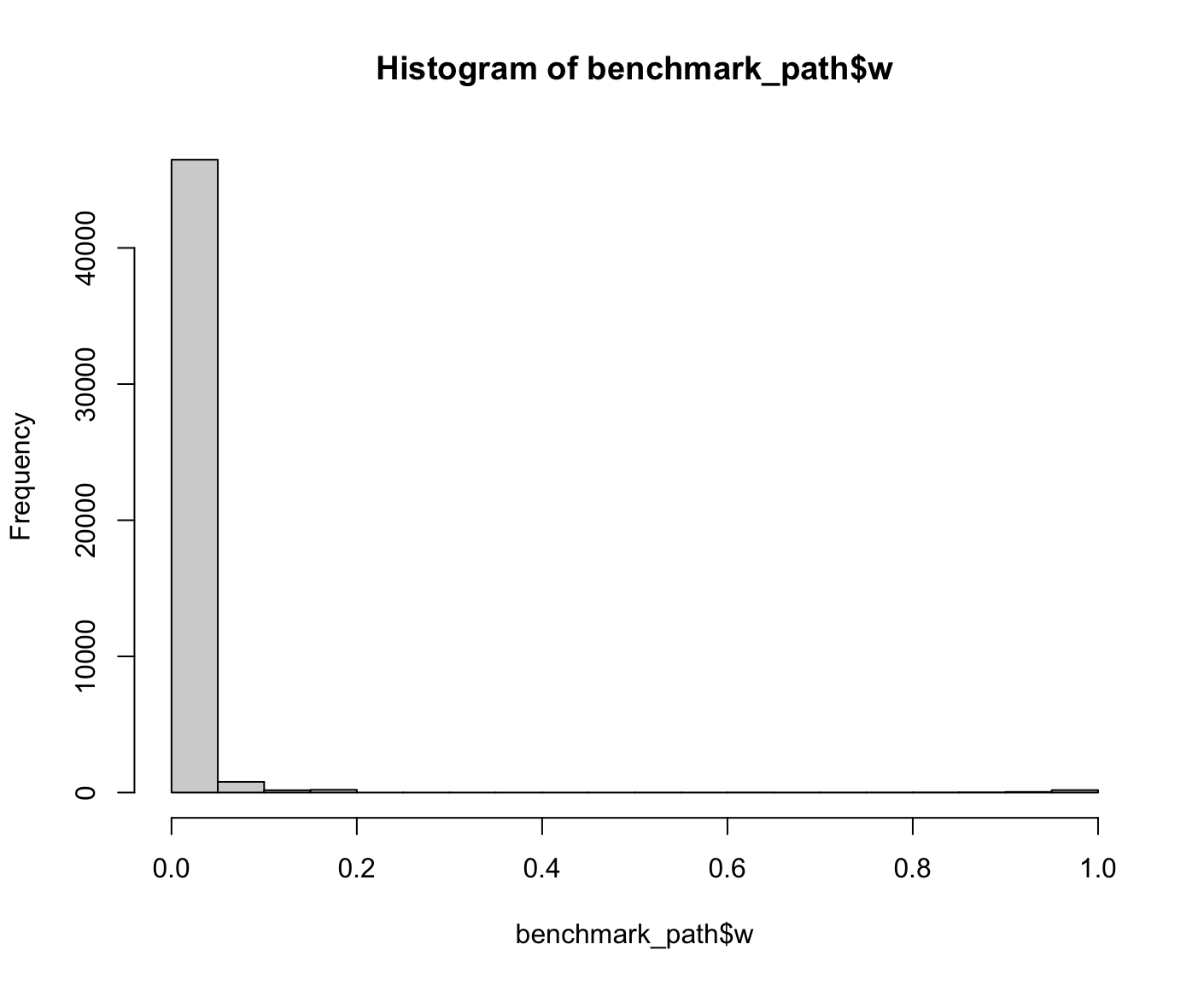}
        \subcaption{The distribution of upstream industry inputs to the downstream}

    \end{minipage}
\end{figure}

\begin{figure}[!htbp]
    \centering
    \caption{The distribution of the longest shortest path and $A_{i, j}$ (top\_k = 3)}
    \label{fig:distribution3}
    \begin{minipage}{0.48\linewidth}
        \centering
        \includegraphics[width=\linewidth]{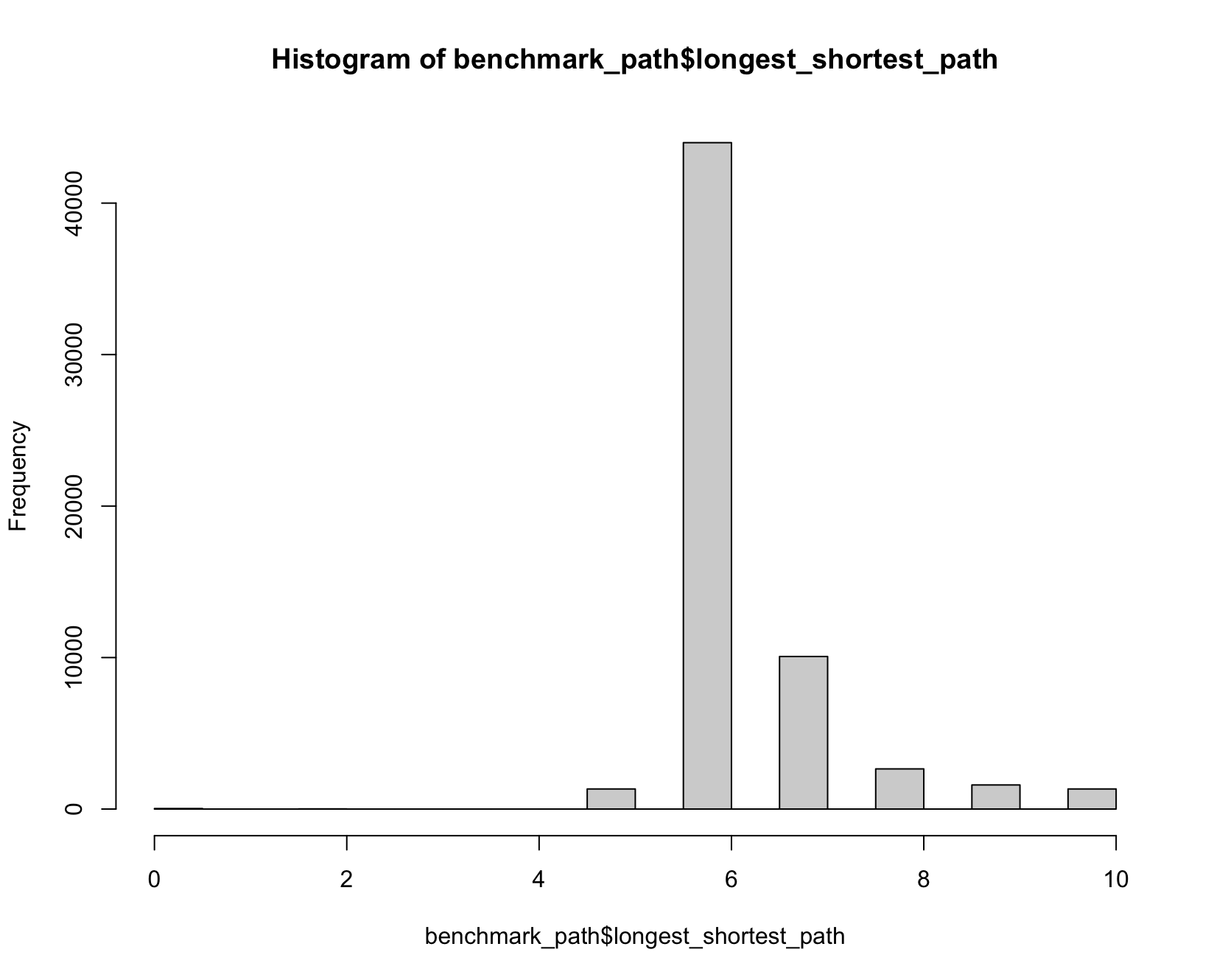}
        \subcaption{The distribution of the longest shortest path}
        
    \end{minipage}
    \hfill
    \begin{minipage}{0.48\linewidth}
        \centering
        \includegraphics[width=\linewidth]{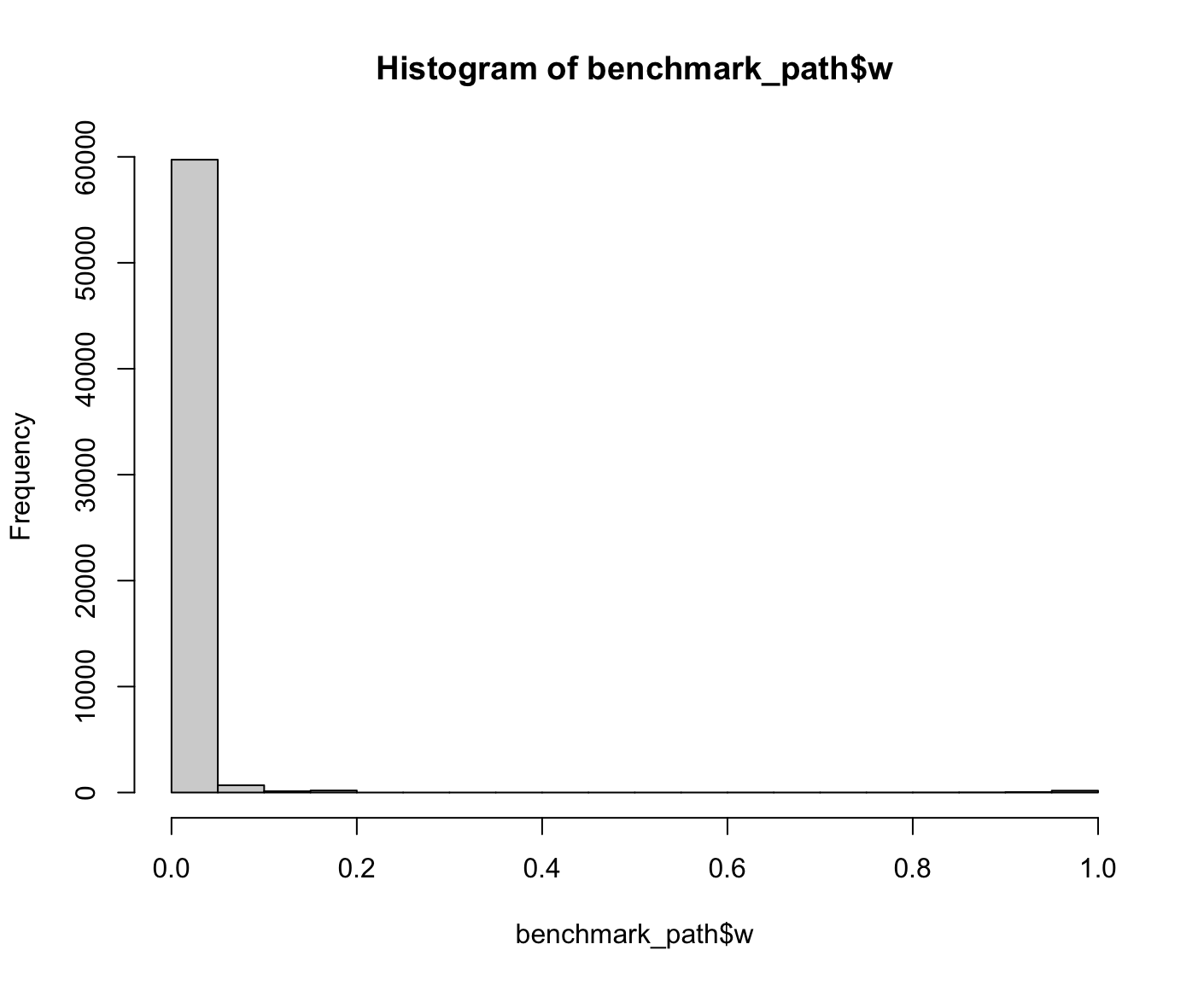}
        \subcaption{The distribution of upstream industry inputs to the downstream}
    \end{minipage}
\end{figure}

\begin{itemize}
    \item[1.] The production chain identification proceeds backward from final goods to trace upstream production stages. I first use the output table to identify final manufacturing products by excluding service products such as legal services and advertising.

    \item[2.] The upstream tracing procedure is based on the industry-to-industry linkage $A_{i,j}$. In the production network, each industry is connected to multiple upstream suppliers, reflecting that several industries jointly contribute to the next stage of production. At each step, I trace the top two upstream industries with the largest input shares (i.e., top\_k = 2 based on $A_{i,j}$) to construct the benchmark chain.

    \item[3.] Figure~\ref{fig:ship_chain} presents an example of the identified chains when tracing the single largest upstream input (top\_k = 1). Black nodes denote industries and the red node denotes the final product. Each chain traces the transformation from upstream industry to the final product through successive intermediate industries.

    \item[4.] To validate the choice of top\_k = 2, I construct benchmark supply chains and measure the length of each chain using the longest shortest path. Figures~\ref{fig:distribution1}, \ref{fig:distribution2} and \ref{fig:distribution3} present the resulting distributions of chain lengths and input intensities $A_{i,j}$ for top\_k = 1, 2, and 3, respectively. These figures indicate that top\_k = 2 provides an optimal parameter for tracing economically meaningful supply chains. When top\_k = 1, the resulting chains are typically short and omit substantively important upstream linkages, whereas top\_k = 3 introduces many industries with negligible input shares, generating excessively diffuse downstream chains.
\end{itemize}

\newpage
\subsection*{Step 2: Construct the global production network}


\begin{itemize}
    \item[1.] I focus on shipbuilding and construct the global production network. In this bipartite network, one set of nodes corresponds to countries and the other to industries. The weighted edges represent the value of exports (log) from each country to industries identified within the shipbuilding production network.\footnote{Data from \href{https://www.cepii.fr/DATA_DOWNLOAD/baci/doc/baci_webpage.html\#download-links}{CEPII BACI database}, accessed April 10, 2026.} 
\end{itemize}

The empirical application uses $B=1000$ bootstrap replications for analysis.


\subsection{Supplemental Goodness-of-fit Diagnostics}\label{sec:GOF_sup}

The goodness-of-fit diagnostics compare statistics from the observed panel with reference distributions obtained from 1,000 conditional replications of the fitted model. Each replication holds the estimated multiplicative component fixed, introduces the additional variation implied by the estimated variance components for the coefficient and additive-effect trajectories, and draws new idiosyncratic errors. The manuscript reports diagnostics for overall dispersion and heterogeneity across countries and industries. Here, I supplement these results with two additional diagnostics. The first supports the model specification by assessing the need for a multiplicative component, while the second evaluates the Gaussian error assumption using the fitted residuals.

\begin{figure}[!htbp]
    \centering
    \caption{Goodness-of-Fit Diagnostics for the Shipbuilding Production Network}
    \includegraphics[width=15cm]{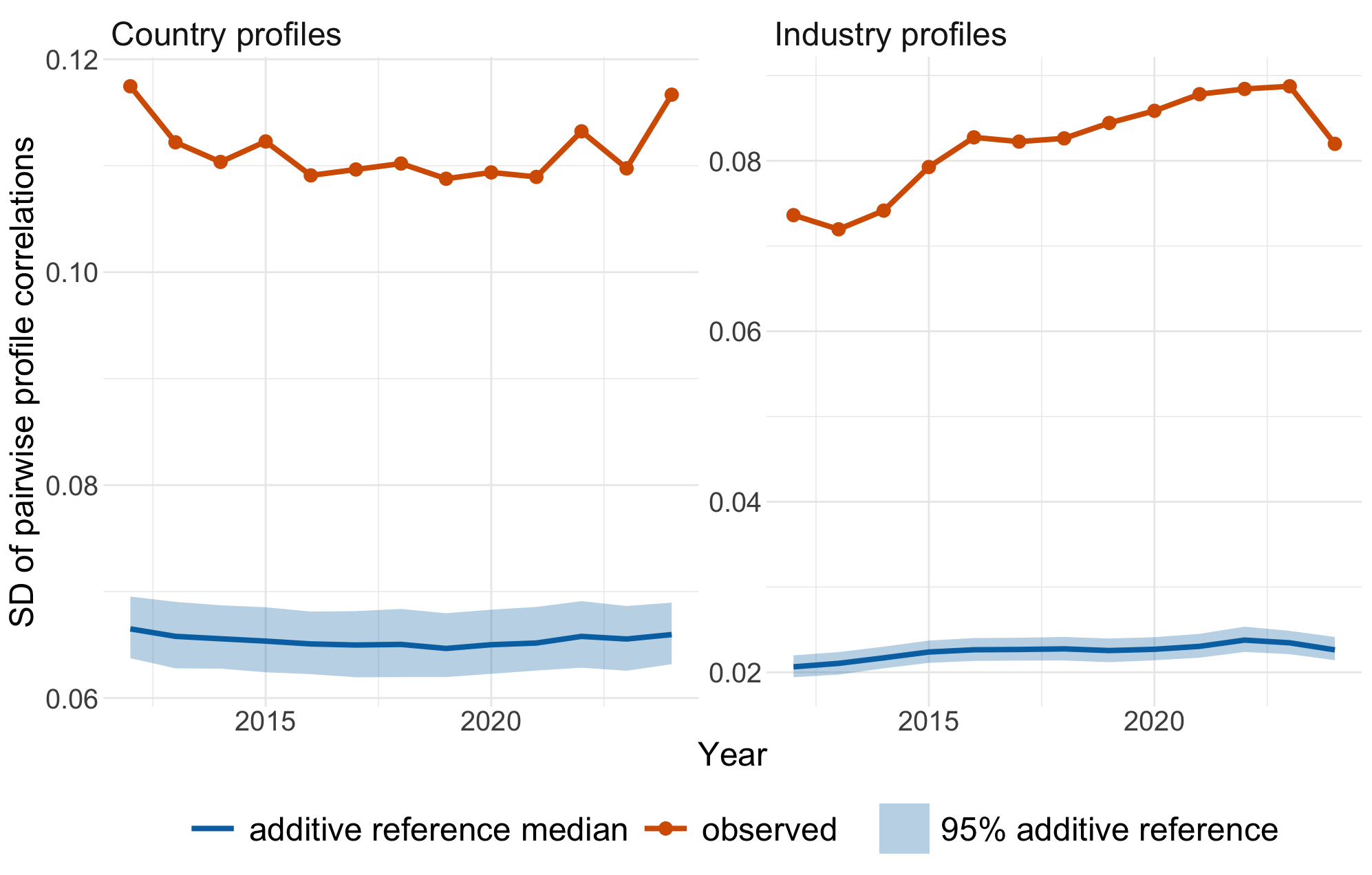}
    \label{fig:GOF2}
\end{figure}

Figure~\ref{fig:GOF2} examines whether the observed networks contain cross-dyad structure beyond the covariate and additive components by comparing statistics from the observed networks with an additive reference distribution. The additive reference retains the estimated covariate and additive components but replaces the remaining dyad-level variation with independent Gaussian noise having the same mean and variance. For the country profiles, I calculate the correlation between each pair of countries' engagement profiles across industries and report the standard deviation of these pairwise correlations. The industry-profile statistic is constructed similarly by comparing industries' engagement profiles across countries.

If the covariate and additive components adequately represented the dependence in the observed networks, the observed profile-correlation statistics would be close to their additive reference distributions. A larger difference indicates greater structured dyadic variation that cannot be reproduced by additive effects and independent noise, and therefore a greater role for the multiplicative component. In every year, the observed statistics lie well above the 95\% additive reference intervals for both country and industry profiles, with particularly large differences for the industry profiles. The observed networks therefore contain substantial cross-dyad structure beyond differences in countries' and industries' average levels of engagement. The persistence of this structure across periods supports the use of a bipartite dynamic AME model that includes a time-varying multiplicative component.

\begin{figure}[!htbp]
    \centering
    \caption{Goodness-of-Fit Diagnostics for the Shipbuilding Production Network}
    \includegraphics[width=13cm]{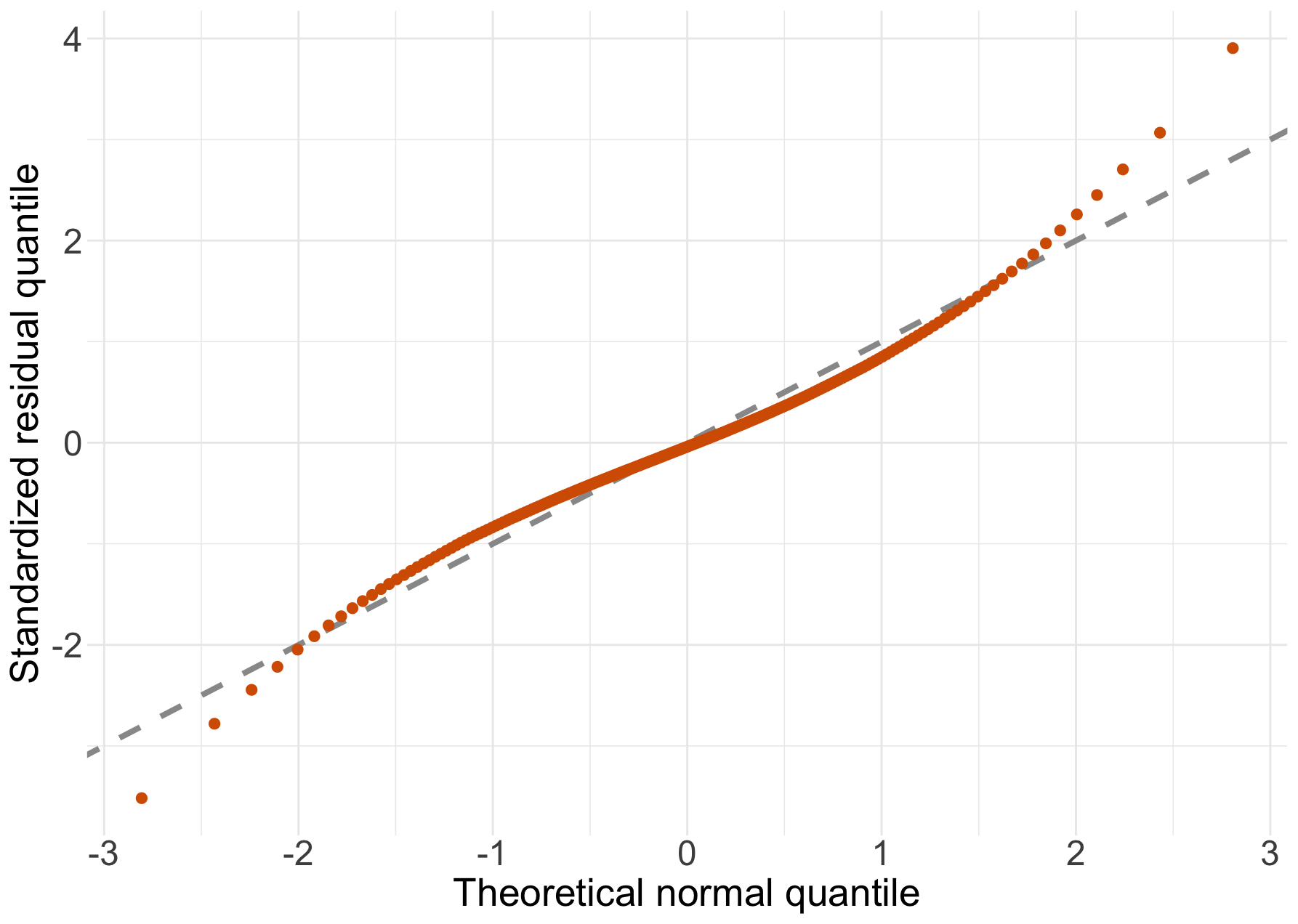}
    \label{fig:GOF3}
\end{figure}

Appendix Figure~\ref{fig:GOF3} evaluates the Gaussian error assumption by comparing the empirical distribution of the fitted residuals with a standard normal distribution. For each observed dyad and period, the residual is calculated as \(\widehat{\epsilon}_{ijt}=Y_{ijt}-\mathbf{x}_{ijt}^{\top}\widehat{\boldsymbol{\beta}}_t-\widehat{a}_{it}-\widehat{b}_{jt}-\widehat{\mathbf{u}}_{it}^{\top}\widehat{\mathbf{v}}_{jt}\). The residuals are pooled across dyads and periods, standardized using their overall mean and standard deviation, and evaluated at a common set of quantiles. The horizontal axis reports the theoretical quantiles of a standard normal distribution, while the vertical axis reports the corresponding quantiles of the standardized residuals. Under the Gaussian error assumption, the points should follow the dashed 45-degree line. The residual quantiles closely follow this line through the center of the distribution though depart from it in both tails. The fitted model therefore approximates the central part of the residual distribution reasonably well.

Common goodness-of-fit diagnostics for binary bipartite networks also include degree distributions and four-cycle counts. These statistics are not directly applicable here because the shipbuilding network is weighted rather than binary. Constructing them would require either imposing a threshold to convert the observed weights into binary ties or selecting a particular weighted definition of a network cycle. Because the resulting diagnostics could depend heavily on these choices, I instead evaluate model fit using statistics that retain the observed edge weights, including overall dispersion, heterogeneity in country and industry means, variation in profile correlations, and the distribution of the fitted residuals.

\bibliographystyle{chicago-apsr.bst}
\bibliography{sample}